\documentclass[sii]{ipart}
\RequirePackage[OT1]{fontenc}
\RequirePackage{amsthm,amsmath}
\RequirePackage[numbers,square]{natbib}
\RequirePackage[colorlinks]{hyperref}
\pdfoutput=1
\usepackage{subfigure}
\usepackage{graphicx}
\usepackage{amssymb}
\usepackage{amsfonts}
\usepackage{amsmath}
\usepackage{amsbsy}

\pubyear{2014}
\volume{0}
\issue{0}
\firstpage{1}
\lastpage{1}
\startlocaldefs
\theoremstyle{plain}
\newtheorem{theorem}{Theorem}[section]

\theoremstyle{definition}

\theoremstyle{remark}
\newtheorem{remark}[theorem]{Remark}

\endlocaldefs

\begin{document}

\begin{frontmatter}
\title{Detecting hidden periodicities for models with cyclical errors}
\runtitle{Detecting hidden periodicities for models with cyclical errors}

\begin{aug}

\author{\fnms{Mar\'{i}a Pilar}
\snm{Fr\'{i}as}\thanksref{t2}\ead[label=e1]{mpfrias@ujaen.es}},
\address{Department of Statistics
and Operations Research\\
University of Ja\'{e}n, Spain\\
\printead{e1}}
\author{\fnms{Alexander V.} \snm{Ivanov}\ead[label=e2]{alexntuu@gmail.com}}
\address{National Technical University\\
``Kyiv Polytechnic Institute" Ukraine\\
\printead{e2}}
\and
\author{\fnms{Nikolai} \snm{Leonenko}\thanksref{t1,t2,t3}
\ead[label=e3]{LeonenkoN@cardiff.ac.uk}}
\address{School of Mathematics \\
Cardiff University, United Kingdom\\
\printead{e3}}
\author{\fnms{Francisco} \snm{Mart\'{i}nez}
\ead[label=e4]{fmartin@ujaen.es}}
\address{Department of Computer Science\\
University of Ja\'{e}n, Spain\\
\printead{e4}}

\author{\fnms{Mar\'{i}a Dolores} \snm{Ruiz-Medina}\thanksref{t1,t2,t3}
\ead[label=e5]{mruiz@ugr.es}}
\address{Department of Statistics and Operations Research \\
University of Granada, Spain\\
\printead{e5}}

\thankstext{t1}{Partially supported by grant
of the European commission PIRSES-GA-2008-230804 (Marie Curie)}
\thankstext{t2}{Partially supported by projects MTM2012-32674 of the DGI}
\thankstext{t3}{Partially supported by Australian Research Council grants
A10024117 and DP 0345577}
\runauthor{M.P. Fr\'{i}as et al.}

%\affiliation{Some University and Another University}
\end{aug}
\received{\sday{XX} \smonth{12} \syear{2009}}

\begin{abstract}
In this paper, the estimation of parameters in the harmonic regression with cyclically dependent errors is addressed.  Asymptotic properties of the least-squares estimates are analyzed by simulation experiments. By numerical simulation, we prove that consistency and asymptotic normality of the least-squares parameter estimator studied holds under different scenarios, where theoretical results do not exist, and have yet to be proven. In particular, these two asymptotic properties are shown by simulations  for the least-squares parameter estimator in the non-linear regression model analyzed, when its error term  is defined as a non-linear transformation of a Gaussian random process displaying long-range dependence.
\end{abstract}



\end{frontmatter}

\section{Introduction}
\label{Introduction}

Classical models of ``hidden periodicities" have been widely studied and applied in natural sciences such as oceanography, astronomy, seismology and \newline medicine. Early work on the estimation of the parameters in the harmonic regression can be found in \cite{Schuster} which first introduced an estimation ``search" technique based on the periodogram. The first studies of the problem in a more formal treatment can be seen in \cite{Bartlett,Moran,Grenander}. Least-squares estimate (LSE) of the parameters in the trigonometric regression and their asymptotic covariance matrix is studied in \cite{Whittle}. This problem can be formulated in the following way. Consider regression model

\begin{equation}
x(t)=g(t,\theta )+\varepsilon (t),  \label{nonregr1}
\end{equation}%
\noindent where
\begin{equation}
g(t,\theta )=\sum_{k=1}^{N}\left( A_{k}\cos \varphi _{k}t+B_{k}\sin \varphi
_{k}t\right) ,  \label{nonregr2}
\end{equation}%
with $\theta =(\theta _{1},\theta _{2},\theta _{3},\dots ,\theta
_{3N-2},\theta _{3N-1},\theta _{3N})=(A_{1},B_{1},\varphi _{1},\ldots
,A_{N},B_{N},\varphi _{N})\in \mathbb{R}^{3N},%
\;C_{k}^{2}=A_{k}^{2}+B_{k}^{2}>0,\;k=1,\ldots ,N,\quad 0\leq \underline{%
\varphi }<\varphi _{1}<\cdots <\varphi _{N}<\overline{\varphi }<\infty,$ and $\{\varepsilon (t),\ t\in \mathbb{S}\}$, $\mathbb{S}=\mathbb{R}$ or $\mathbb{Z}$, is the
random noise process defining the error term through time. Process $%
\varepsilon $ is assumed to be a zero-mean stationary process.

The LSE, $\hat{\theta}_{T},$ of an unknown parameter $%
\theta \in \Theta $, obtained from the observations $\{x(t),t \in [0,T]\},$ or $\{x(t),t=1,\ldots,T\},$
is any random variable $\hat{\theta}_{T}\in \Theta ^{c},$ having the
property
\begin{equation}
Q_{T}(\hat{\theta}_{T})=\inf_{\tau \in \Theta ^{c}}Q_{T}(\tau ),\quad
Q_{T}(\tau )=\frac{1}{T}\int_{0}^{T}[x(t)-g(t,\tau )]^{2}\nu(dt),
\label{lse}
\end{equation}%
where $\Theta ^{c}$ is the closure of $\Theta $ and $\nu(dt)$ represents a counting measure in the case of discrete time (i.e., $\nu({t})=1, t \in \mathbb{Z}_{+}=\mathbb{N}$), and Lebesgue measure $dt$ in continuous time (i.e., $\nu(dt)=dt, t \in \mathbb{R}_{+}$).

Nonlinear regression models with independent or weakly dependent errors
have been extensively studied (see, for example, \cite{r27,r26,sko,pol},
and the references therein). The first results on nonlinear regression with errors having a slowly decreasing correlation function, i.e., with Long-Range Dependence (LRD) in discrete time were obtained by \cite{rob1,Koul2,Koul1}. The volume \cite{r17} presents a review of the most relevant applications of processes with LRD. The asymptotic theory of LSE in nonlinear regression with LRD has been
considered in \cite{Muk,r28,r29}. In
papers \cite{Iv2,Iv3}
asymptotic distributions of a class of M-estimates and L$_{p}$-estimates
($1<p<2$) in nonlinear regression model with LRD form were presented. The problem of the estimation of the unknown parameters
of the trigonometric regression with cyclical dependent stationary noise is studied in \cite{Iv2014}.  The authors derived LSE consistency and
asymptotic normality of the regression function (\ref{nonregr2}) parameters, and error term $\varepsilon $ being a zero-mean stationary process, generated by
nonlinear transformation of a stationary Gaussian process $\xi $ displaying
cyclical dependence. Specifically, for a stationary process  $\xi $ defined on a
complete probability space $(\Omega ,\mathbb{F},P):$
\begin{equation*}
\xi (t)=\xi (\omega ,t):\Omega \times \mathbb{S}\longrightarrow \mathbb{R}.
\end{equation*}%
Such a process is assumed to satisfy the following assumption.

\textbf{A1.} Random function $\xi$ is a
real-valued and measurable stationary mean-square continuous Gaussian
process with $\mathit{E}\xi (t)=0, $ and $\mathit{E}\xi ^{2}(t)=1.$ Its
covariance function (c.f.) is of the form:
\begin{equation}
B\left( t\right) =\mathit{E}[\xi (0)\xi (t)]=\sum_{j=0}^{\kappa
}D_{j}B_{\alpha _{j},\varkappa _{j}}\left( t\right),  \label{cov1}
\end{equation}%
\noindent  $t\in \mathbb{R}$, $\kappa \geq 0$, $\sum_{j=0}^{\kappa }D_{j}=1$, $D_{j}\geq 0,$
$j=0,\dots ,\kappa$ , where %
\begin{equation*}
B_{\alpha _{j},\varkappa _{j}}\left( t\right) =\frac{\cos \left( \varkappa
_{j}t\right) }{\left( 1+t^{2}\right) ^{\alpha _{j}/2}},
\end{equation*}

\noindent  $0\leq \varkappa_{0}<\varkappa _{1}<...<\varkappa _{\kappa }$, $\alpha _{j}>0$, $t\in
\mathbb{R}$, $j=0,\dots ,\kappa$.

Although \cite{Iv2014} dealts with the nonlinear
regression model (\ref{nonregr1}) with regression function (\ref{nonregr2}),
and cyclical dependent stationary noise with covariance function (\ref{cov1}%
), the results given in \cite{Iv2014} on linearization, and asymptotic
uniqueness, as well as on asymptotic normality hold for a more general class
of regression functions. The general class of non-linear
regression functions that could be considered includes the family of
functions $g$ such that, the family of matrix-valued measures, defined by
\begin{equation*}
\boldsymbol{\mu }_{T}(d\lambda )=(\mu _{T}^{jl}(d\lambda ,\theta
))_{j,l=1}^{q},\end{equation*}
\begin{equation*}
\mu _{T}^{jl}(d\lambda ,\theta )=\frac{%
g_{T}^{j}(\lambda ,\theta )\overline{g_{T}^{l}(\lambda ,\theta )})d\lambda }{%
\left( \int\limits_{\mathbb{R}}\left\vert g_{T}^{j}(\lambda ,\theta
)\right\vert ^{2}d\lambda \int\limits_{\mathbb{R}}\left\vert
g_{T}^{l}(\lambda ,\theta )\right\vert ^{2}d\lambda \right) ^{\frac{1}{2}}}%
,  \label{regmes1}
\end{equation*}%
$T>0,$
\begin{equation*}
g_{T}^{j}(\lambda ,\theta )=\int\limits_{0}^{T}e^{it\lambda }\frac{\partial
}{\partial \theta _{j}}g(t,\theta )dt,  \label{regmes2}
\end{equation*}%
\noindent $j=1,\dots ,q$, $\lambda \in
\mathbb{R}$, $\theta \in \Theta$, weakly converges, as $T \rightarrow \infty,$ to an atomic spectral measure of regression function $\boldsymbol{%
\mu }$ with atoms $\Xi _{regr}=\{\delta _{1},\dots ,\delta _{n}\}.$

Limit theorems for non-linear transformations of Gaussian stationary processes were considered.
In the derivation of these limit results, the above mentioned
weak-convergence to the spectral measure of regression function and the diagram formulae were applied. In the discrete case this
phenomenon was discussed in \cite{r52,r53} for some other regression
scheme.

Although the model definition included possible LRD in the error term, this property has not been considered to show the asymptotic properties of the LSE. That is, for $\alpha m>1,$ $\alpha=\min_{j=0,\ldots,\kappa} \alpha_j,$ $m$ is Hermite rank of $G$ (see below), the consistency and limiting Gaussian distribution of the LSE for general regression function are proven in \cite{Iv2014}. Using limit theorems of \cite{Iv2013} it can be seen that these results hold for $\alpha>1/2$ and $m=1$. In this article, the statements of the papers \cite{Iv2013,Iv2014} for the trigonometric regression function (\ref{nonregr2}) for $\alpha>1/2$ are confirmed by simulation. In addition, for $\alpha<1/2$ these results are unknown, but in this paper we show that they are correct also
by simulation, at least for non overlapping spectra as is explained in Section 2.1.

The outline of the paper is the following: a review of principal results concerning the asymptotic normality and consistency of LSE in regression model (\ref{nonregr1}) is done in Section 2. A simulation study to prove the previous results is illustrated in Section 3. Also, some remarks on the asymptotic properties of the LSE considering a broader range of values of model parameters that define the noise process are set out in this section. Section 4 provides the  final comments and conclusions.

\section{Consistency and asymptotic normality of the LSE of the parameters of
trigonometric regression}

In this section a review of the published work regarding consistency and asymptotic normality of the LSE of the parameters of
trigonometric regression with cyclically dependent errors is carried out. The assumptions made on the Gaussian process $\xi $
generating the random noise $\varepsilon ,$ representing the time-dependent
error term in the regression model (\ref{nonregr1}) are summarized below.

Random process $\xi$ is assumed to satisfy condition  \textbf{A1}. Therefore, the covariance function (\ref{cov1}) admits the following spectral
representation:
\begin{equation*}
B(t)=\int\limits_{\mathbb{R}}e^{i\lambda t}f(\lambda )d\lambda ,\quad t\in
\mathbb{R},
\end{equation*}%
\noindent where the spectral density (s.d.) is of the form:
\begin{equation*}
f\left( \lambda \right) =\sum_{j=0}^{\kappa }D_{j}f_{\alpha _{j},\varkappa
_{j}}\left( \lambda \right) ,\quad \lambda \in \mathbb{R},
\end{equation*}%
\noindent with, $f_{\alpha
_{j},\varkappa_{j}}\left( \lambda \right) $ being defined by
\begin{eqnarray*}
f_{\alpha _{j},\varkappa _{j}}\left( \lambda \right) &=&\frac{c_{1}\left(
\alpha _{j}\right) }{2}\left[ K_{\frac{\alpha _{j}-1}{2}}\left( \left\vert
\lambda+\varkappa _{j}\right\vert \right) \left\vert \lambda +\varkappa
_{j}\right\vert ^{\frac{\alpha _{j}-1}{2}}\right.\\&&\left.+K_{\frac{\alpha_{j}-1}{2}}\left(
\left\vert \lambda -\varkappa _{j}\right\vert \right) \left\vert \lambda
-\varkappa _{j}\right\vert ^{\frac{\alpha _{j}-1}{2}}\right] ,
\end{eqnarray*}%
\noindent $\lambda \in \mathbb{R}$, and
\begin{equation*}
c_{1}\left( \alpha_{j} \right) =\frac{2^{\left( 1-\alpha_{j} \right) /2}}{%
\sqrt{\pi }\,\Gamma \left( \frac{\alpha_{j}}{2}\right) }.
\end{equation*}

\bigskip \noindent Here,
\begin{equation*}
K_{\nu }\left( z\right) =\frac{1}{2}\int_{0}^{\infty }s^{\nu -1}\exp \left\{
-\frac{1}{2}\left( s+\frac{1}{s}\right) z\right\} ds,
\end{equation*}%
\noindent $z\geq 0$, $\nu
\in \mathbb{R}$, is the modified Bessel function of the third kind and order $\nu $
or McDonald's function.

The following asymptotic expansions are known (see, i.e.,
\cite{GR}, formulae 8.485, 8.445 and 8.446): if $\nu
\notin \mathbb{Z}$,

\begin{eqnarray*}
K_{-\nu }\left( z\right) &=&K_{\nu }\left( z\right) \\
&=&\frac{\pi }{2\sin (\pi \nu )}\left\{ \sum_{j=0}^{\infty }\frac{%
(z/2)^{2j-\nu }}{j!\Gamma (j+1-\nu )}\right.\\&&\left.-\sum_{j=0}^{\infty }\frac{%
(z/2)^{2j+\nu }}{j!\Gamma (j+1+\nu )}\right\} ,
\end{eqnarray*}%
while if $\nu =\pm m,$ where $m$ is a nonnegative integer,%
\begin{eqnarray*}
K_{\nu }\left( z\right) &=&\frac{1}{2}\sum_{j=0}^{m-1}\frac{(-1)^{j}(m-j-1)!}{%
j!}\left( \frac{z}{2}\right) ^{2j-m}\\&&
+(-1)^{m+1}\sum_{j=0}^{\infty }\frac{(z/2)^{m+2j}}{j!(m+j)!}\left\{ \ln
\frac{z}{2}\right.\\&&\left.-\frac{1}{2}\Psi (j+1)-\frac{1}{2}\Psi (j+m+1)\right\} ,
\end{eqnarray*}%
where $\Psi (z)=(\frac{d}{dz}\Gamma (z))/\Gamma (z)$ is the logarithm
derivative of the Gamma function.

\noindent We have: for $\alpha _{j}>1$
\begin{equation*}
\lim_{\lambda \rightarrow 0}f_{\alpha _{j},0}\left( \lambda \right) =\frac{%
\Gamma \left(\frac{\alpha _{j}-1}{2}\right)}{\left[ 2\sqrt{\pi }\Gamma (%
\frac{\alpha _{j}}{2})\right]} ,
\end{equation*}%
for $\alpha _{j}=1,$ and $\lambda \rightarrow 0$%
\begin{equation*}
f_{\alpha _{j},0}\left( \lambda \right) \sim \frac{1}{\pi }\left\{ \ln
\left\vert \lambda \right\vert +\ln 2+\Psi (1)\right\} ,
\end{equation*}%
where $\Psi (1)=-\gamma ,\gamma $ is the Euler constant.

For $0<\alpha _{j}<1,$ and $\lambda \rightarrow 0$
\begin{equation*}
f_{\alpha _{j},0}\left( \lambda \right) =c_{2}\left( \alpha _{j}\right)
\frac{1}{\left\vert \lambda \right\vert ^{1-\alpha _{j}}}(1-h_{j}(\left\vert
\lambda \right\vert )),
\end{equation*}%
\noindent where $c_{2}(\alpha _{j})=[2\Gamma (\alpha _{j})\cos \frac{\alpha
_{j}\pi }{2}]^{-1},$ and%
\begin{eqnarray*}
h_{j}\left( \left\vert \lambda \right\vert \right) &=&\frac{\Gamma \left(
\frac{\alpha _{j}+1}{2}\right) }{\Gamma \left( \frac{3-\alpha _{j}}{2}%
\right) }\left\vert \frac{\lambda }{2}\right\vert ^{1-\alpha _{j}}+\frac{%
\Gamma \left( \frac{\alpha _{j}+1}{2}\right) }{4\Gamma \left( \frac{3+\alpha
_{j}}{2}\right) }\left\vert \frac{\lambda }{2}\right\vert ^{2}\\&&+o\left(
\left\vert \lambda \right\vert ^{2}\right) .  \label{hf}
\end{eqnarray*}

\noindent Thus, for $j=0,\dots ,\kappa ,$ $0<\alpha _{j}<1$
\begin{eqnarray*}
f_{\alpha _{j},\varkappa _{j}}\left( \lambda \right) &=&\frac{c_{2}\left(
\alpha _{j}\right) }{2}\left[ \left\vert \lambda +\varkappa _{j}\right\vert
^{\alpha _{j}-1}\left( 1-h_{j}\left( \left\vert \lambda +\varkappa
_{j}\right\vert \right) \right) \right.\\&&\left.+\left\vert \lambda -\varkappa
_{j}\right\vert ^{\alpha _{j}-1}\left( 1-h_{j}\left( \left\vert \lambda
-\varkappa _{j}\right\vert \right) \right) \right] .  \label{f1}
\end{eqnarray*}%
Therefore, the s.d. $f$ has $2\kappa +2$ different singular points \newline $\left\{
-\varkappa _{\kappa },-\varkappa _{\kappa -1},..,-\varkappa _{1},-\varkappa
_{0},\varkappa _{0},\varkappa _{1},...,\varkappa _{\kappa }\right\} $ under
condition \textbf{A1}, when $\varkappa _{0}\neq 0,$ and $0<\alpha _{j}<1,$ $%
j=0,\dots ,\kappa .$ If $\varkappa _{0}=0,$ the s.d. $f$ has $2\kappa +1$
different singular points.

For $\alpha _{j}=1$ and $\lambda \rightarrow \pm \varkappa _{j}:$
\begin{equation*}
f_{1,\varkappa _{j}}\left( \lambda \right) \sim \frac{c_{1}\left( \alpha
_{j}\right) }{2}K_{0}\left( \left\vert 2\varkappa _{j}\right\vert \right) +%
\frac{1}{2\pi}\left\{ \ln \left\vert \lambda \mp \varkappa
_{j}\right\vert +\ln 2+\Psi (1)\right\} ,
\end{equation*}%
while for $\alpha _{j}>1$ and $\lambda \rightarrow \pm \varkappa _{j}:$%
\begin{equation*}
f_{\alpha _{j},\varkappa _{j}}\left( \lambda \right) \rightarrow \frac{%
c_{1}\left( \alpha _{j}\right) }{2}K_{\frac{\alpha _{j}-1}{2}}\left(
\left\vert 2\varkappa _{j}\right\vert \right) +\frac{1}{2}\frac{\Gamma (%
\frac{\alpha _{j}-1}{2})}{\left[ 2\sqrt{\pi }\Gamma (\frac{\alpha _{j}}{2})%
\right] }.
\end{equation*}

\textbf{A2}. The stochastic process $\varepsilon $
is given by $\varepsilon (t)=G(\xi (t)),$ $t\in \mathbb{R},$ with $\xi (t)$
satisfying condition \textbf{A1}, and $G:\mathbb{R}\longrightarrow \mathbb{R}
$ being a non-random measurable function such that $\mathit{E}G(\xi (0))=0,$
and $\mathit{E}G^{2}(\xi (0))<\infty .$

Under condition \textbf{A2}, function $G\in L_{2}(\mathbb{R},\varphi (x)dx),$
with $\varphi (x)=\frac{1}{\sqrt{2\pi }}e^{-\frac{x^{2}}{2}},$ $x\in \mathbb{%
R},$ being the standard Gaussian density, and
\begin{equation}
G(x)=\sum_{k=1}^{\infty }\frac{C_{k}}{k!}H_{k}(x),\quad \sum_{k=1}^{\infty }%
\frac{C_{k}^{2}}{k!}=\mathit{E}[G^{2}(\xi (0))]<\infty ,  \label{3.1}
\end{equation}%
\noindent where%
\begin{equation*}
C_{k}=\int_{\mathbb{R}}G(x)H_{k}(x)\varphi (x)dx.
\end{equation*}%
Here, the Hermite polynomials
\begin{equation}
H_{k}(x)=(-1)^{k}e^{\frac{x^{2}}{2}}\frac{d^{k}}{%
dx^{k}}e^{-\frac{x^{2}}{2}},\quad k=0,1,2,\ldots ,\label{Hm}
\end{equation}%
\noindent constitute a complete orthogonal system in the Hilbert space $%
L_{2}(\mathbb{R},\varphi (x)dx).$

\textbf{A3}. We assume that the function $G$ has Hermite rank $Hrank(G)=m,$ that is, either $C_{1}\neq 0$ and $m=1,$ or, for some $m\geq 2,$ $%
C_{1}=\cdots =C_{m-1}=0,\ C_{m}\neq 0.$

Under conditions \textbf{A1}-\textbf{A3}, the process $\{\varepsilon
(t)=G(\xi (t)),\ t\in \mathbb{R}\},$ admits a Hermite series expansion in
the Hilbert space $L_{2}(\Omega ,\mathbb{F} ,P):$%
\begin{equation*}
\varepsilon (t)=G(\xi (t))=\sum_{k=m}^{\infty }\frac{C_{k}}{k!}H_{k}(\xi
(t)).  \label{3.3}
\end{equation*}

In the following modification of the LSE proposed in \cite{WAL} is used (see,
also \cite{r25,Iv1}). Consider a monotone non-decreasing system of open
sets $S_{T}\subset S(\underline{\varphi },\overline{\varphi }),\ T>T_{0}>0,$
given by the condition that the true value of unknown parameter $\varphi $
belongs to $S_{T},$ and
\begin{equation*}
\lim_{T\rightarrow \infty }\inf_{1\leq j<k\leq N,\ \varphi \in
S_{T}}T(\varphi _{k}-\varphi _{j})=+\infty ,
\end{equation*}%
\begin{equation}
\lim_{T\rightarrow \infty
}\inf_{\varphi \in S_{T}}T\varphi _{1}=+\infty ,  \label{3.6}
\end{equation}
\noindent where
\begin{equation*}
S(\underline{\varphi },\overline{\varphi })=\left\{ 0\leq \underline{\varphi
}<\varphi _{1}<\dots <\varphi _{N}<\overline{\varphi }<\infty \right\} .
\end{equation*}

The LSE \ $\widehat{\theta }_{T}$ in the Walker sense of unknown parameter \newline $%
\theta =(A_{1},B_{1},\varphi _{1},\ldots ,A_{N},B_{N},\varphi _{N}$ ) in the
model (\ref{nonregr1}) with nonlinear regression function (\ref{nonregr2})
is said to be any random vector $\hat{\theta}_{T}\in \Theta _{T}$ having the
property:%
\begin{equation*}
Q_{T}(\hat{\theta}_{T})=\inf_{\tau \in \Theta _{T}}Q_{T}(\tau ),  \label{3.4}
\end{equation*}%
\noindent where $Q_{T}(\tau )$ is defined in (\ref{lse}), and $\Theta_{T}
\subset \mathbb{R}^{3N}$ is such that $A_{k}\in \mathbb{R},\ B_{k}\in
\mathbb{R},\ k=1,\ldots ,N$, and $\varphi \in S_{T}^{c},$ the closure in $%
\mathbb{R}^{N}$ of the set $S_{T}.$

\begin{remark}
The 2nd condition (\ref{3.6}) is satisfied if $\underline{\varphi }>0.$ If $%
S_{T}\subset S(\underline{\varphi },\overline{\varphi }),$ the relations
given in (\ref{3.6}) are, for example, satisfied for a parametric set $S_{T}$%
, such that
\begin{equation*}
\inf_{1\leq j<k\leq N,\ \varphi \in S_{T}}(\varphi _{k}-\varphi
_{j})=T^{-1/2},\quad \inf_{\varphi \in S_{T}}\varphi _{1}=T^{-1/2}.
\end{equation*}
\end{remark}
\setcounter{theorem}{0}
\begin{theorem} \cite{Iv2014} \label{th1}
Under conditions \textbf{A1} and \textbf{A2}, the LSE in the Walker sense
\begin{equation*}
\hat{\theta}_{T}=(\hat{A}_{1T},\hat{B}_{1T},\hat{\varphi}_{1T},\ldots ,\hat{A%
}_{NT},\hat{B}_{NT},\hat{\varphi}_{NT})
\end{equation*}%
\noindent of the unknown parameter $$\theta =(A_{1},B_{1},\varphi _{1},\ldots
,A_{N},B_{N},\varphi _{N})$$ of the regression function (\ref{nonregr2}) is
weakly consistent as $T\rightarrow \infty $, that is,
\begin{equation*}
\hat{A}_{k_{T}}\overset{P}{\longrightarrow }A_{k},\ \hat{B}_{k_{T}}\overset{P%
}{\longrightarrow }B_{k},\ T(\hat{\varphi}_{k_{T}}-\varphi _{k})\overset{P}{%
\longrightarrow }0,
\end{equation*}%
\noindent  $k=1,\ldots ,N$, where $\overset{P}{\longrightarrow }$ stands for the convergence
in probability.
\end{theorem}

The following condition is needed to prove the limiting normal distribution of the LSE
of the parameters in the trigonometric regression with cyclically dependent errors.
This constraint on $\alpha$ is opposed to the presence of LRD in noise process.

\textbf{A4}. Either 1) Hrank$(G)=1,$ $\alpha >1;$ or 2)
Hrank$(G)=m\geq 2,$ $\alpha m>1;$ where $\alpha =\min_{j=0,1,\ldots
,\kappa}\alpha _{j}.$

 The asymptotic convergence to the Gaussian distribution of the LSE in the Walker
sense of the function (\ref{nonregr2}) is obtained in Theorem \ref{th2} for certain ranges of the parameters defining the spectral singularities of $\xi$. Specifically, assumption \textbf{A4} defines parameter range $\alpha =\min_{j=0,\dots ,\kappa}\alpha_{j}>1/m.$ Here this condition is rewritten to include some differences and extension due to the consideration of a more general class of models.  Simulations in Section \ref{sec.NR} show that the Gaussian limit results hold for $\alpha_{j} > 0,$ $j=0,\ldots,\kappa.$  A new condition is formulated, \textbf{A5}, where the limit regression spectral measure and the spectrum of the Gaussian random process generating the error term  may not be overlapped .

\textbf{A4'}. Hrank$(G)=1,$ $0<\alpha <\frac{1}{2};$ where $\alpha =\min_{j=0,1,\ldots
,\kappa}\alpha _{j}.$

\textbf{A5}. The singular points in the spectrum of noise, denoted as $\Xi _{noise}=\{\pm \varkappa_{0},\ldots,\pm \varkappa_{\kappa}\},$ $0\leq \varkappa_{0}<\varkappa _{1}<\ldots<\varkappa _{\kappa }$ and spectral measure atoms $\Xi_{regr} = \{\delta_{1},\ldots, \delta_{n}\}$ may not be overlapped. That is, $\Xi _{noise}
\cap \Xi _{regr} = \emptyset.$

The asymptotic Gaussian distribution of the LSE in the Walker sense of the
regression function (\ref{nonregr2}) is established in the following result.

\begin{theorem} \cite{Iv2014}
\label{th2} Under conditions \textbf{A1}-\textbf{A4}, the LSE in the Walker
sense of the function (\ref{nonregr2}) of unknown parameter is
asymptotically normal, that is, the vector
\begin{equation*}
\left( T^{1/2}(\hat{A}_{kT}-A),T^{1/2}(\hat{B}_{kT}-B),T^{3/2}(\hat{\varphi}%
_{kT}-\varphi )\right),
\end{equation*}
\noindent $k=1,\ldots ,N$ converges weakly to the multidimensional normal vector $%
N_{3N}(0,\Gamma ),$ where the matrix $\Gamma >0$ is of the form $\Gamma
=diag\left( \Gamma _{k}\right) _{k=1}^{N},$ with
\begin{eqnarray*}
\Gamma _{k}&=&\frac{4\pi }{A_{k}^{2}+B_{k}^{2}}\sum_{j=m}^{\infty }\frac{%
C_{j}^{2}}{j!}f^{(\ast j)}(\varphi _{k})\\&&\left(
\begin{array}{ccc}
A_{k}^{2}+B_{k}^{2} & -3A_{k}B_{k} & -6B_{k} \\
-3A_{k}B_{k} & A_{k}^{2}+B_{k}^{2} & 6A_{k} \\
-6B_{k} & 6A_{k} & 12%
\end{array}%
\right) .  \label{matrixgamma}
\end{eqnarray*}
\noindent Here, $f^{(\ast j)}(\lambda ),$ $\lambda \in \mathbb{R},$ is the $%
j $-th convolution of the s.d. given under assumption {\bfseries %
A1}.
\end{theorem}
Theorem \ref{th2} follows Theorem 5  in \cite{Iv2014} by direct computations. In this theorem the limiting distribution of the LSE estimators is obtained for a more general class of $g$ functions.

\section{Numerical results}\label{sec.NR}
Let us consider the following model:
\begin{equation}
x(t)=g(t,\theta)+ G(\xi(t)),\quad t\in \mathbb{S}_{+}, \quad \mathbb{S}_{+}=\mathbb{R}_{+}\quad or \quad \mathbb{N},\label{mod1}
\end{equation}
with nonlinear regression function,
\begin{equation}
g(t,\theta)=A \cos(\varphi t)+B \sin (\varphi t), \label{g2}
\end{equation}
where $\theta=(A,B,\varphi)$, $C=A^2+B^2>0$, $\varphi<\infty$. Consider $g_{i}(t,\theta)=(\partial/\partial \theta_{i})g(t,\theta)$, $i=1,2,3$, such that,
\begin{equation*}
d_{iT}^{2}=\int_{0}^{T}\left[g_{i}(t,\theta)\right]^{2} \nu(dt)<\infty, \quad T>0,\quad i=1,2,3.
\end{equation*}
Let $\nabla g(t,\theta)=\left(g_{1}(t,\theta),g_{2}(t,\theta),g_{3}(t,\theta)\right)'$ be the column vector gradient of the function $g(t,\theta)$. We use the notation $d_{T}^{2}(\theta)=\mbox{diag}(d_{iT}^{2})_{i=1}^{3}$. In the theory of statistical estimation of unknown parameter $\theta\in \Theta \subset \mathbb{R}^{3}$ for (\ref{mod1}), the asymptotic behavior, as $T\to \infty,$ of the functional
\begin{equation}
\zeta_{T}=d_{T}^{-1}(\theta)\int_{0}^{T}\nabla g(t,\theta)G(\xi(t))\nu(dt), \label{functional}
\end{equation}
plays a crucial role, since, under certain conditions, the asymptotic distribution of the normalized LSE $d_{T}(\theta)(\hat{\theta}_{T}-\theta)$, and properly normalized functional (\ref{functional}) coincide, as $T\to \infty$; see \cite{r27,r28}.

For model (\ref{mod1}) we have,
\begin{eqnarray*}
g_{1}(t,\theta)&=&\frac{\partial}{\partial A}g(t,\theta)=\cos(\varphi t),\\
g_{2}(t,\theta)&=&\frac{\partial}{\partial B}g(t,\theta)=\sin(\varphi t),\\
g_{3}(t,\theta)&=&\frac{\partial}{\partial \varphi}g(t,\theta)=-A t \sin(\varphi t)+B t \cos(\varphi t).\\
\end{eqnarray*}

%
%\begin{equation}
%x(t)=A\cos(\varphi t)+B\sin(\varphi t)+G(\xi(t)), \qquad t=0,\ldots,T, \label{modsim1}
%\end{equation}

In the capacity of function $G$, we take the first few Hermite polynomials (see equation (\ref{Hm})):
\begin{enumerate}
\item[]\textbf{Case H1:} $G(u)=u$,
\item[]\textbf{Case H2:} $G(u)=u^{2} -1$,
\item[]\textbf{Case H3:} $G(u)=u^{3} -3u$,
\item[]\textbf{Case H4:} $G(u)=u^{4}-6u^{2} +3$.

\end{enumerate}

\noindent These four cases are under conditions \textbf{A2} - \textbf{A3}.   Random function $\xi$ is a
real-valued and measurable stationary mean-square continuous Gaussian
process with $\mathit{E}\xi (t)=0, $ and $\mathit{E}\xi ^{2}(t)=1,$ and covariance function:

    \begin{equation}
    B(t)=\frac{\cos(\varkappa t)}{(1+t^{2})^{\alpha/2}},\quad t\in \mathbb{S}_{+}, \quad \mathbb{S}_{+}=\mathbb{R}_{+}\quad or \quad \mathbb{N}.\label{covA}
    \end{equation}

In this section, numerical results show the consistency for particular cases under conditions \textbf{A1}, \textbf{A2}, \textbf{A4'} and \textbf{A5}. Specifically,  the LSE in the Walker sense
\begin{equation*}
\hat{\theta}_{T}=(\hat{A}_{T},\hat{B}_{T},\hat{\varphi}_{T})
\end{equation*}%
\noindent of the unknown parameter $\theta =(A,B,\varphi)$ of the regression function (\ref{g2}) seems to be
weakly consistent as $T\rightarrow \infty $, under the assumption of LRD in process $\xi$. Moreover, the limiting distribution of the LSE in the Walker sense of the function (\ref{g2}) parameters for special cases under conditions \textbf{A1},\textbf{A2},\textbf{A3},\textbf{A4'} and \textbf{A5}  seems to be normally distributed.

The paper \cite{Iv2014} provides an analysis of the most relevant results concerning asymptotic normality of the LSE of trigonometric
regression parameters. Here, we address the numerical problem of the estimation of the unknown
parameter from the observation of random process $\{x(t),t=1,\ldots,T]\}$ defined in (\ref{nonregr1}), when $T\rightarrow \infty, $ and under the hypothesis of Theorems \ref{th1} and \ref{th2}. The asymptotic normality of the LSE of parameters of model (\ref{nonregr1}) is studied using simulated data. The numerical experiments have been conducted for different assumptions on noise distributions and values of the covariance function (\ref{cov1}).

We simulate the process (\ref{mod1}) with different $T$ values, $A=1$, $B=1$, $\varphi=0.6$. The set $S_{T}$ in equation (\ref{3.6}) is chosen as $S_{T}=(1/\sqrt{T},1)$. The generation of the random vectors $\{\xi(t),t=0,\ldots,T\}$ has been done from a multivariate normal distribution with zero mean vector, and covariance matrix (\ref{covA}). The values of the parameters used to simulate the error term are $\varkappa=0.5,$ $\alpha=0.85,1.50, 2.50$ for simulation experiment 1 and $\alpha=0.25, 0.45$ for simulation experiment 2.  For each combination of $\varkappa$, $\alpha$ and $G$, we generate 1000 different data sets from (\ref{mod1}) using different sequences of $\xi(t).$

\begin{figure}[!hhh]
\begin{center}
\subfigure[Case H1]{\includegraphics[width=0.35\columnwidth]{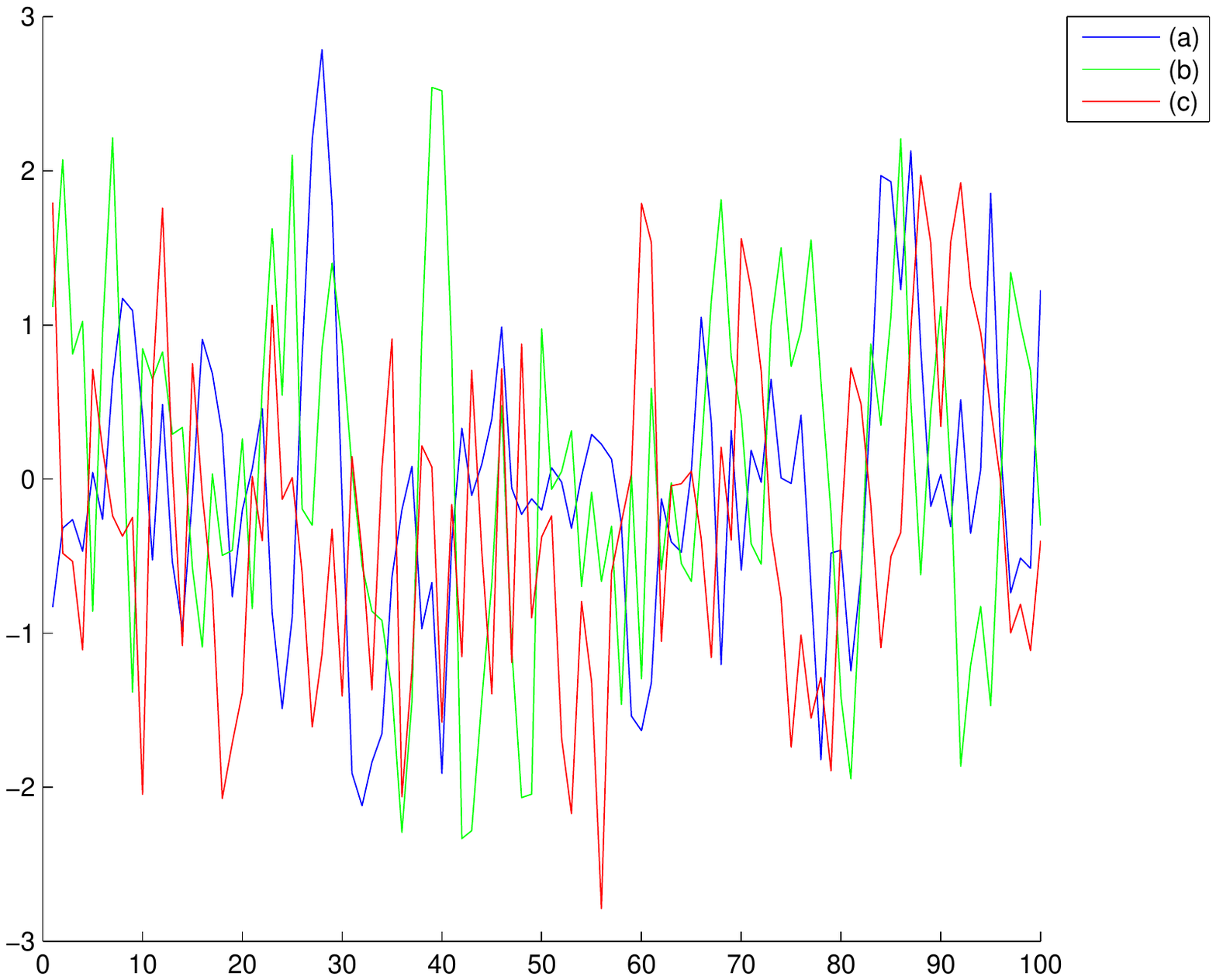}}
\hspace{0.1cm}
\subfigure[Case H2]{\includegraphics[width=0.35\columnwidth]{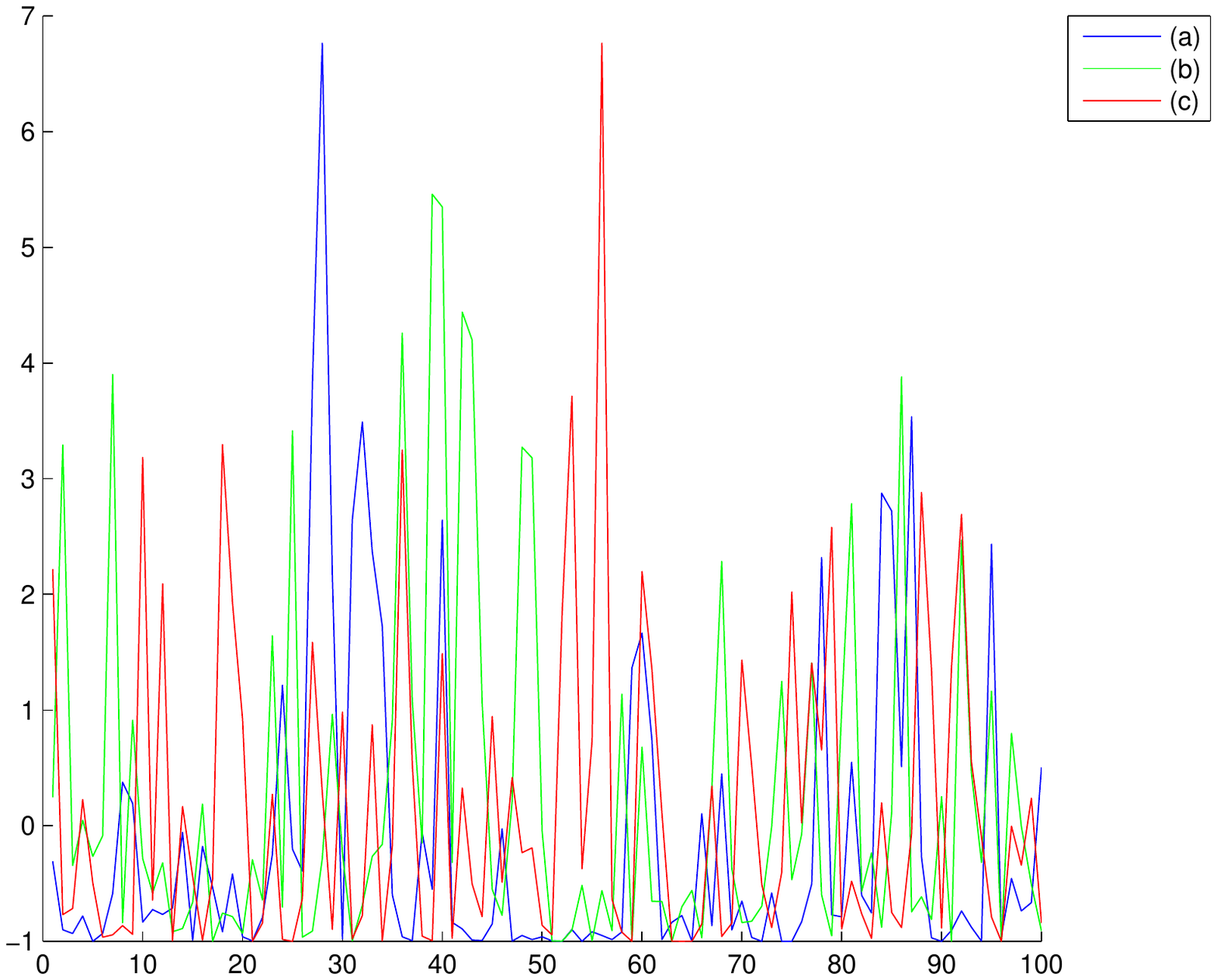}}
\hspace{0.1cm}\subfigure[Case H3]{\includegraphics[width=0.35\columnwidth]{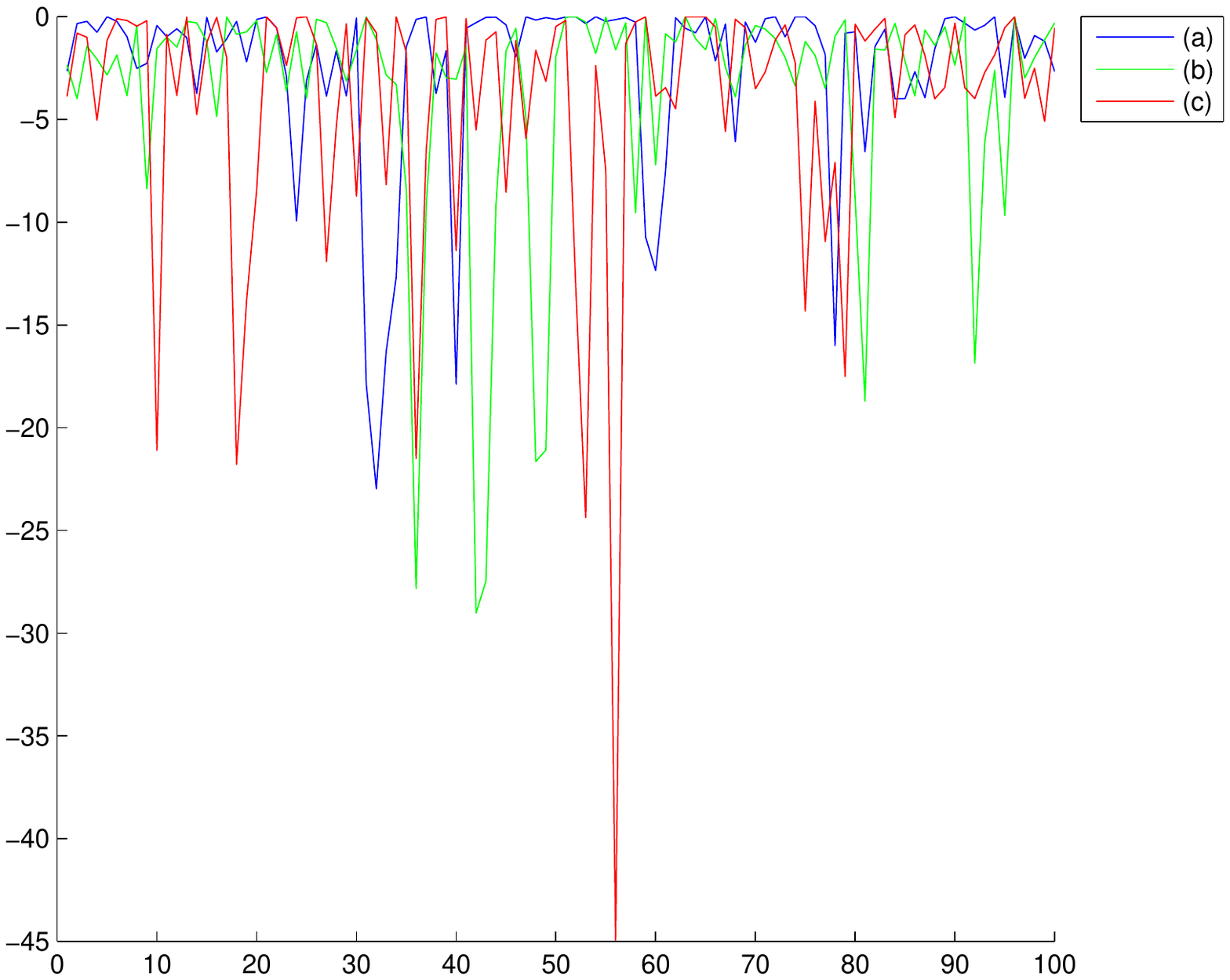}}
\hspace{0.1cm}\subfigure[Case H4]{\includegraphics[width=0.35\columnwidth]{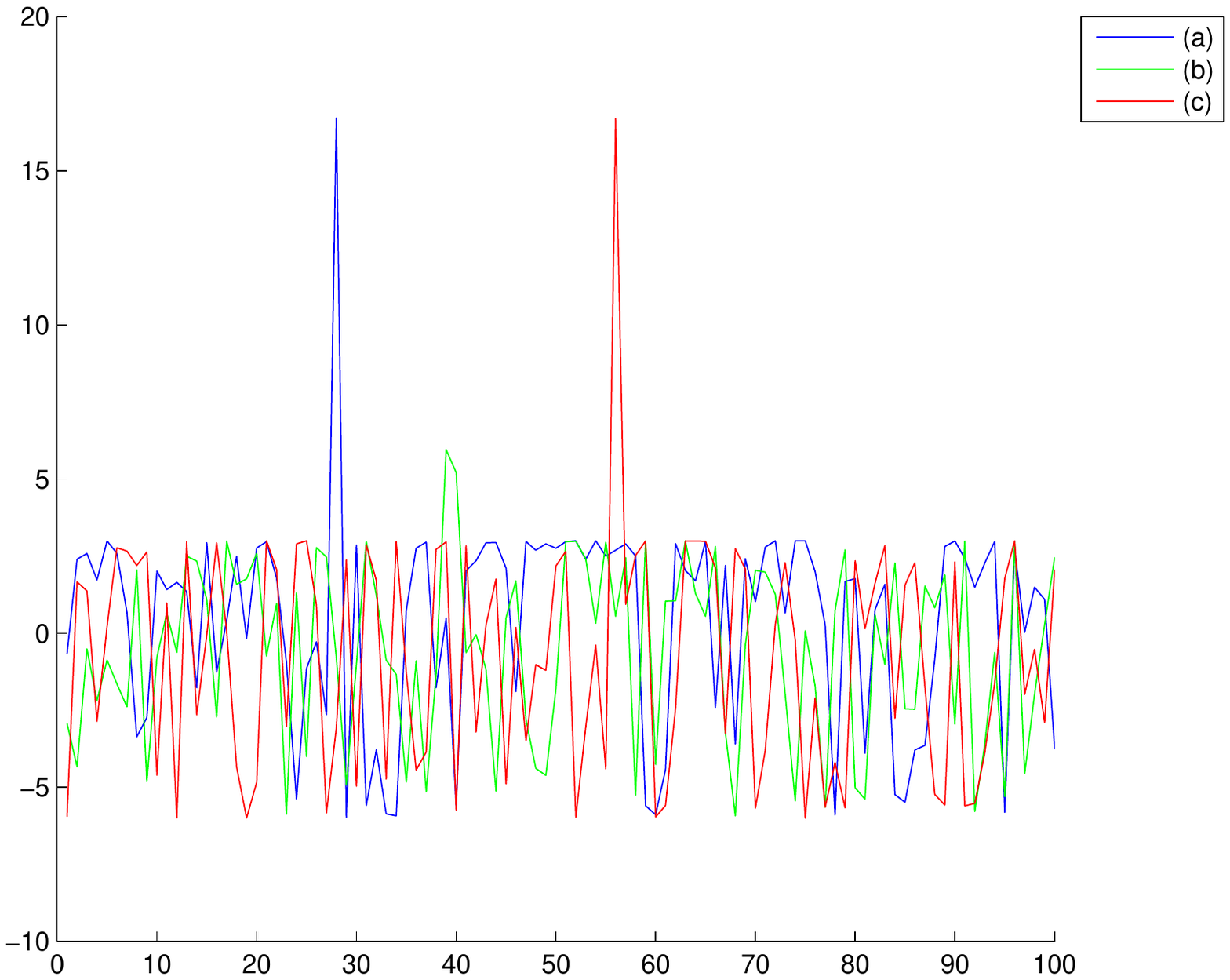}}
\subfigure[Case H1]{\includegraphics[width=0.35\columnwidth]{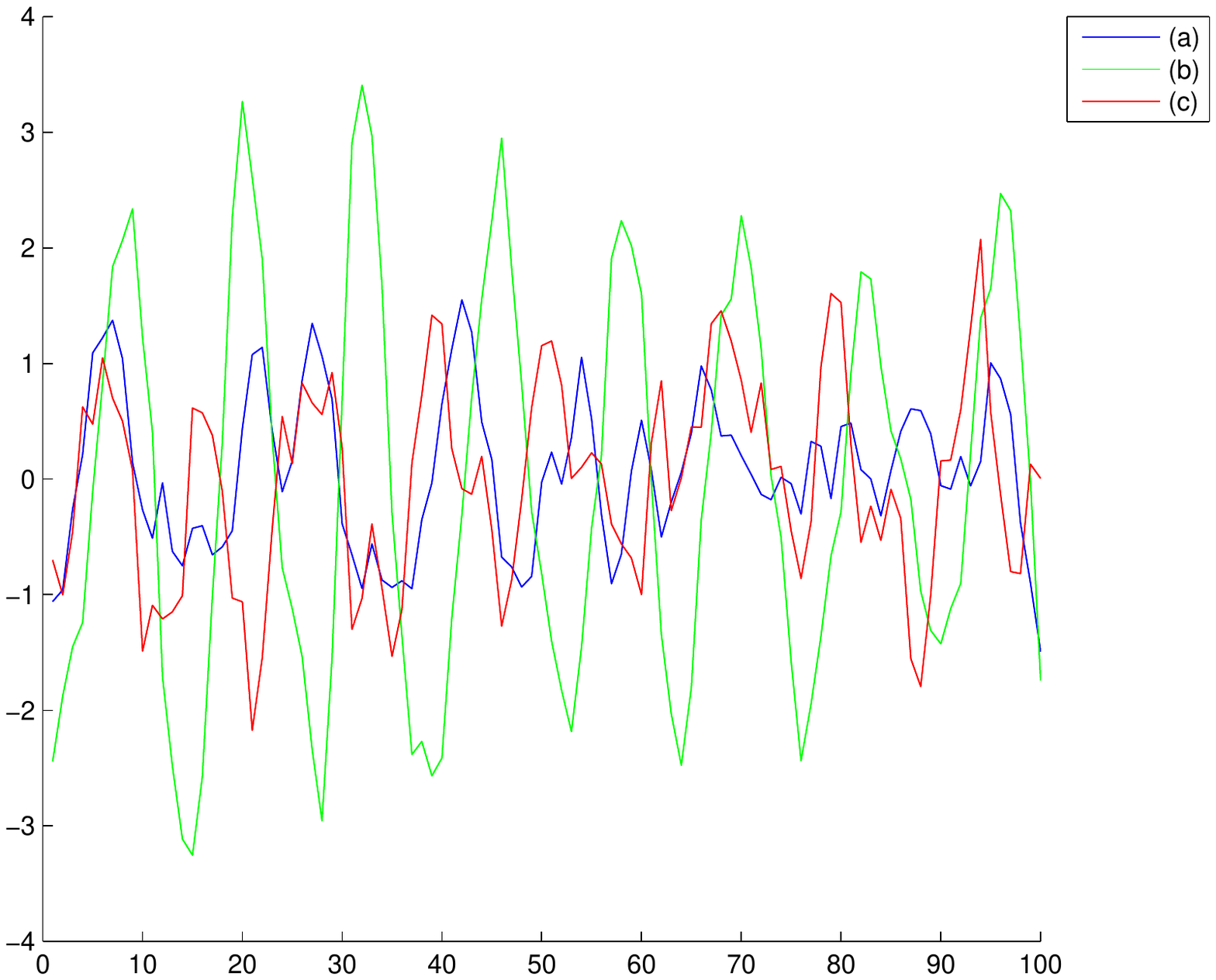}}
\hspace{0.1cm}
\subfigure[Case H2]{\includegraphics[width=0.35\columnwidth]{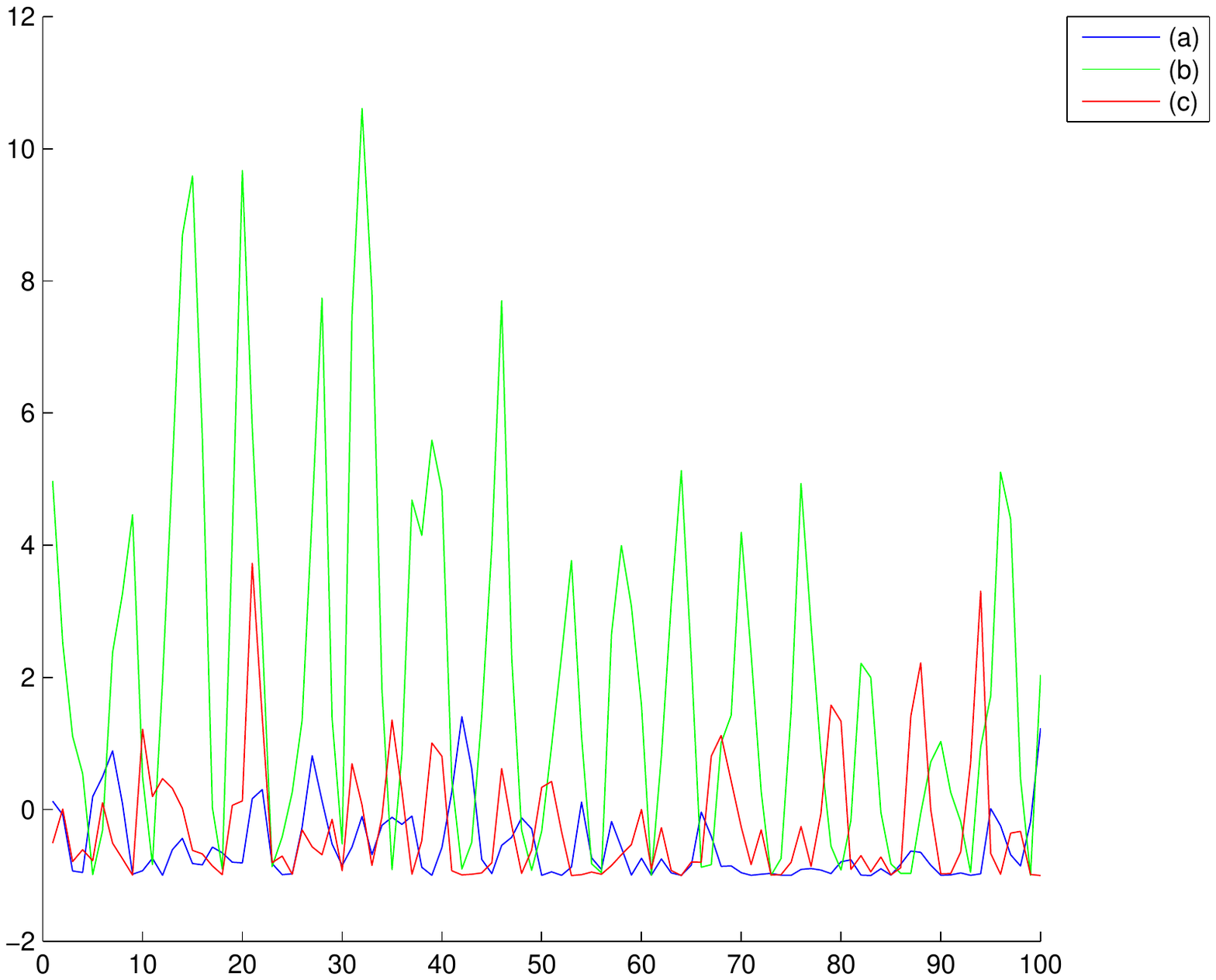}}
\hspace{0.1cm}\subfigure[Case H3]{\includegraphics[width=0.35\columnwidth]{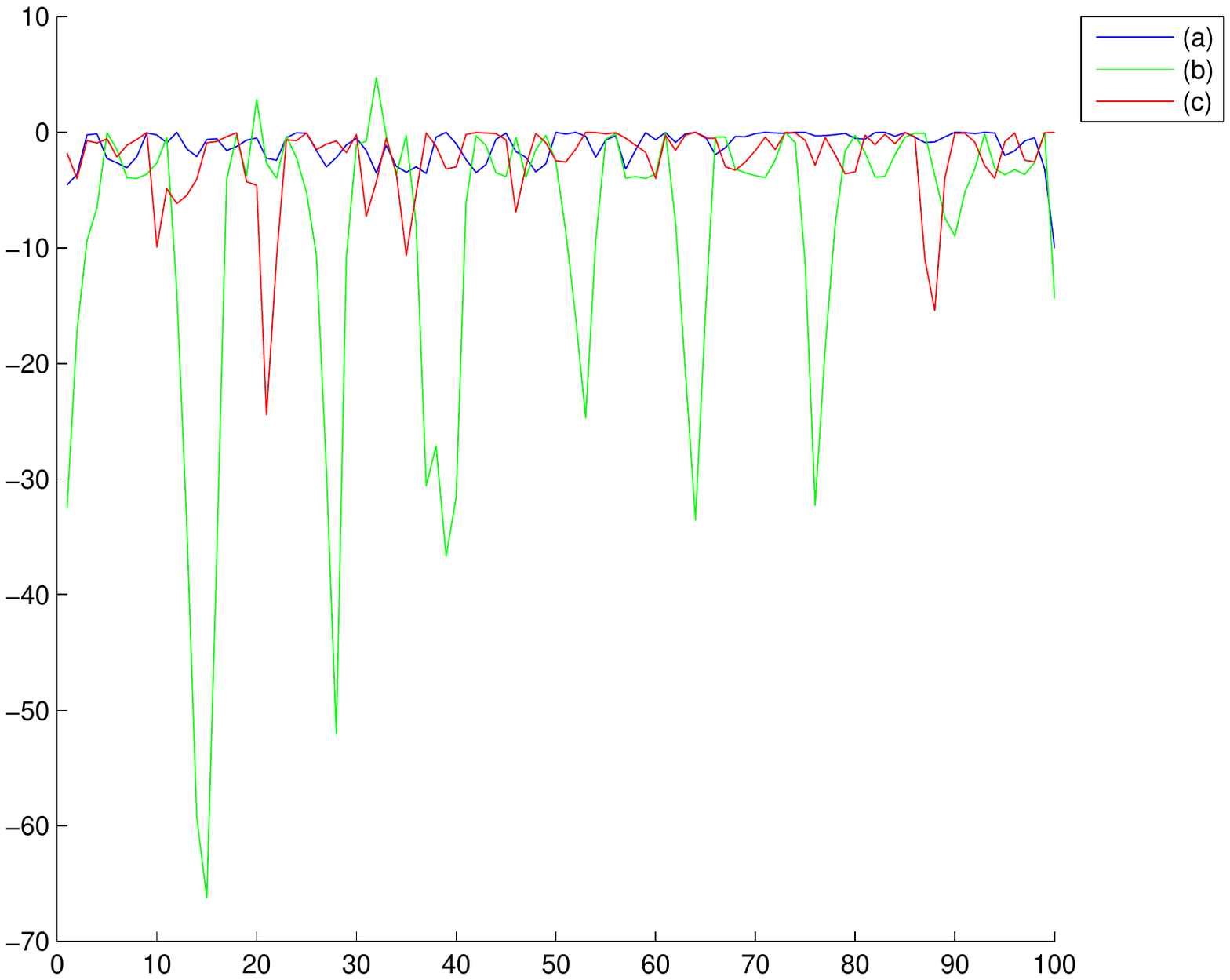}}
\hspace{0.1cm}\subfigure[Case H4]{\includegraphics[width=0.35\columnwidth]{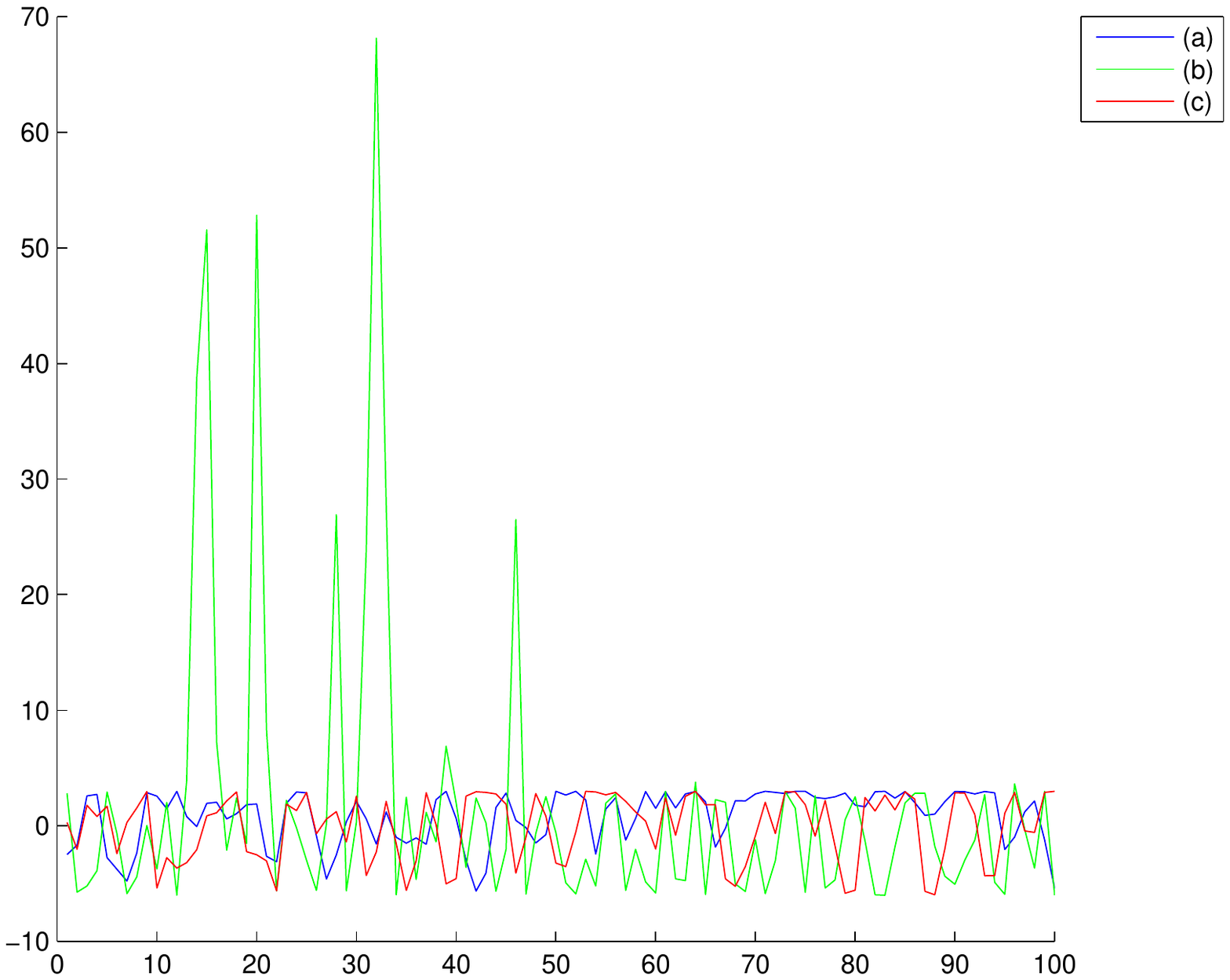}}
\end{center}
\caption{Simulated $\varepsilon $ process for cases: (a) $\alpha=0.85$, (b) $\alpha=1.50$, (c) $\alpha=2.50$ (top) and (a) $\alpha=0.25$, (b) $\alpha=0.45$ (bottom).}\label{fig.path1}
\end{figure}

To illustrate the simulation method, we can write the random vector $\mathbf{V}=\left(\xi(1),\ldots,\xi(T)\right)'$, as
 \begin{equation}\mathbf{V}=L\eta,\label{9}\end{equation}
where $\eta$ is an independent standard normal vector of dimension $T$ and $L$ is a lower triangular Cholesky factor of $\Sigma,$ so $\Sigma =LL'$, with

$$\Sigma =\left(\begin{array}{cccc} B(0)&B(1)&\cdots&B(T-1)\\
                                   B(1)&B(0)&\cdots&B(T-2)\\
                                   \vdots&\vdots&\ddots&\vdots\\
                                   B(T-1)&B(T-2)&\cdots&B(0)\end{array}\right).$$
Firstly, vector $\eta$ is generated as an independent zero-mean Gaussian random vector. Secondly, we apply equation (\ref{9}). In Figure \ref{fig.path1}, realizations of random vector $V$ are shown for different values of $\alpha.$ Moreover, realizations of the error term $\varepsilon $ generated by nonlinear transformation of  stationary Gaussian process $\xi $ can be seen in Figure \ref{fig.path1} for above mentioned functions $G$.

\subsection{Simulation experiment 1}

Here, the results of the papers \cite{Iv2013,Iv2014} concerning the asymptotic normality and consistency of LSE in regression model (\ref{nonregr1}) for the trigonometric regression function are confirmed by simulation. The values of the parameters used to simulate the error term are $\varkappa=0.5,$ $\alpha=0.85,1.50, 2.50,$ (see, Figure 1).

  The convergence to the Gaussian distribution of the LSE of $\theta$ in model (\ref{mod1}) is checked by simulations studying of the behavior of
\begin{eqnarray*}
\hat{\zeta}_{T}^{A}&=&\frac{1}{W_{A}(T)}\sum_{t=1}^{T}W_{A}(t) G(\xi(t)),\\
W_{A}^{2}(T) &=& \sum_{t=1}^{T} (\cos \varphi t)^2,\quad W_{A}(t)=\cos (\varphi t),\\
\hat{\zeta}_{T}^{B}&=&\frac{1}{W_{B}(T)}\sum_{t=1}^{T}W_{B}(t) G(\xi(t)),\\
W_{B}^{2}(T) &=& \sum_{t=1}^{T} (\sin \varphi t)^2,\quad W_{B}(t)= \sin( \varphi t),
\end{eqnarray*}
and
\begin{eqnarray*}
\hat{\zeta}_{T}^{\varphi}&=&\frac{1}{W_{\varphi}(T)}\sum_{t=1}^{T} W_{\varphi}(t)G(\xi(t)),\\
W^{2}(T) &=& \sum_{t=1}^{T}t^{2}[-A\sin \varphi t+ B\cos \varphi t]^{2},\\
W_{\varphi}(t)&=&t[-A\sin \varphi t+B\cos \varphi t] .
\end{eqnarray*}
for increasing values of T.

Three statistical tests, Henze-Zirkler's \cite{Henze,TrujilloHz}, Doornik-Hansen Omnibus \cite{Doornik,TrujilloDorH} and the Chi-square plot \cite{Mardia,TrujilloMardia}, are applied to the simulated random vectors ($\hat{\zeta}_{T}^{A}$, $\hat{\zeta}_{T}^{B}$, $\hat{\zeta}_{T}^{\varphi}$) to evaluate whether the data belongs to a
multivariate normal distribution (MVN) or not. Although we can find many tests for MVN in the literature, the uniformly most powerful
test does not exist and it is recommended to perform several tests to evaluate the belonging to MVN. These three tests are known to have good overall power
against alternatives to normality (see, for example,  \cite{Mecklin}).

\begin{figure}[hptb]
\begin{center}
\subfigure[$\alpha=0.85$, H1]{\includegraphics[width=0.3\columnwidth]{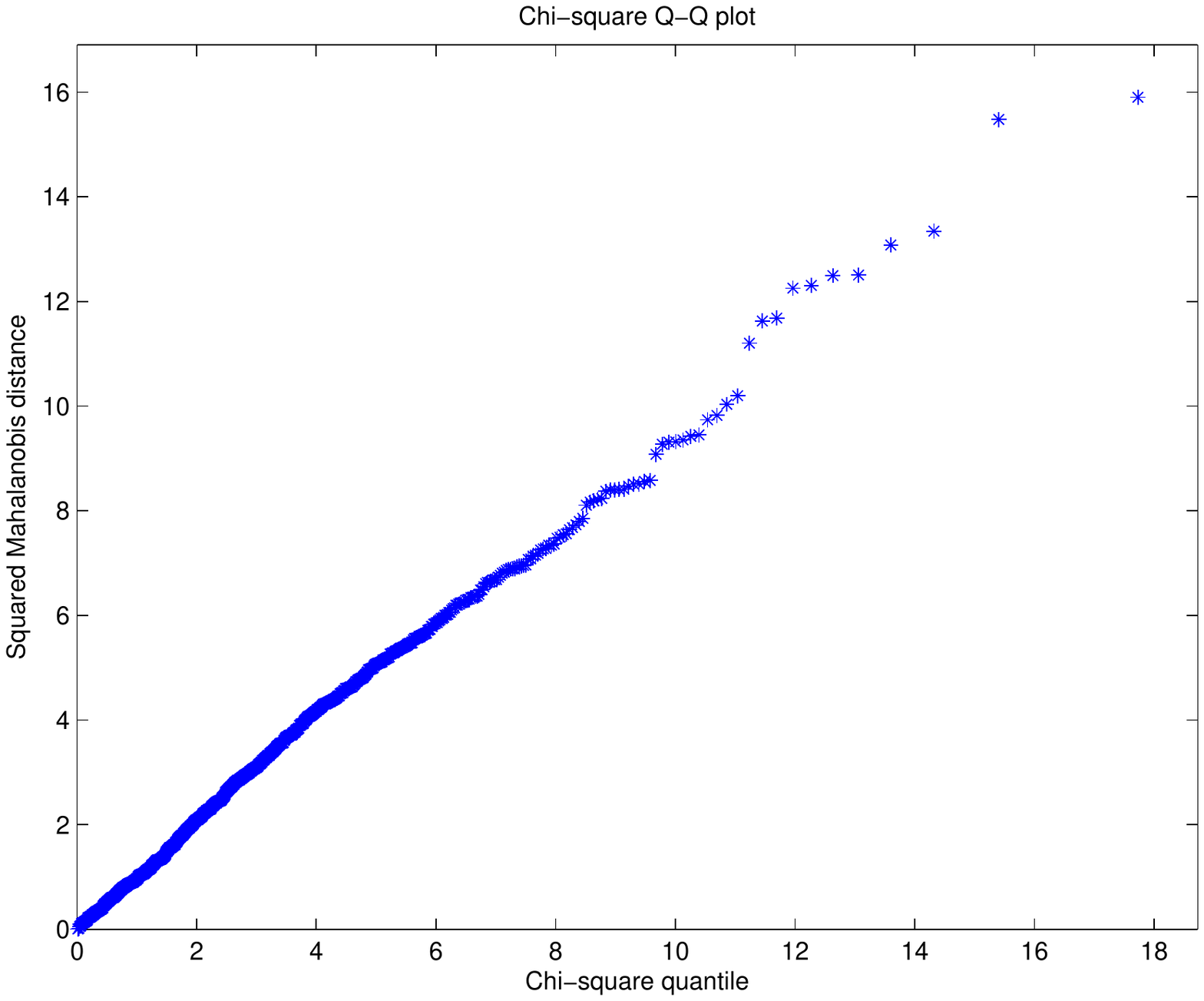}}
%\hspace{0.1cm}
\subfigure[$\alpha=0.85$, H2]{\includegraphics[width=0.3\columnwidth]{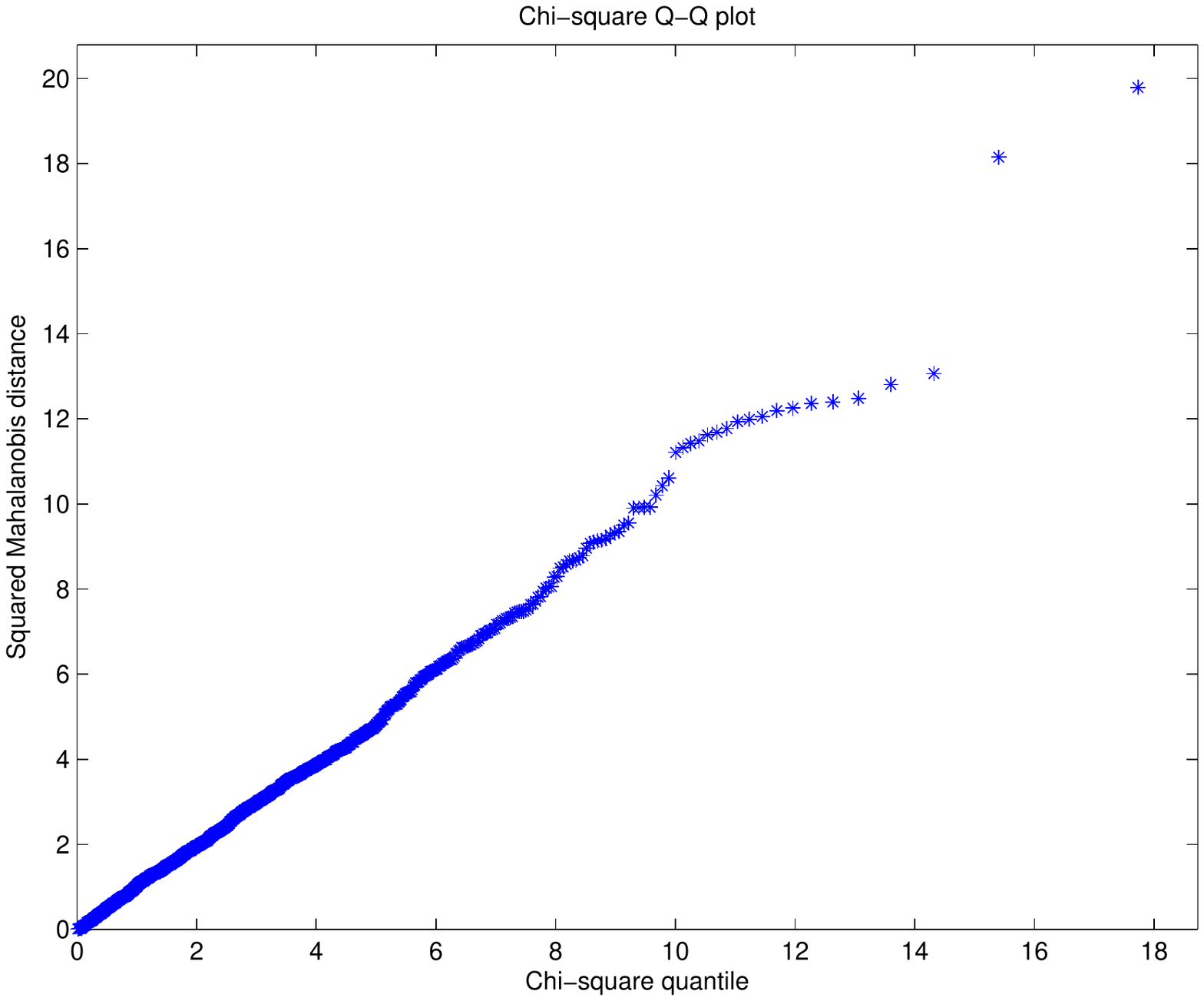}}
%\hspace{0.1cm}
\subfigure[$\alpha=0.85$, H3]{\includegraphics[width=0.3\columnwidth]{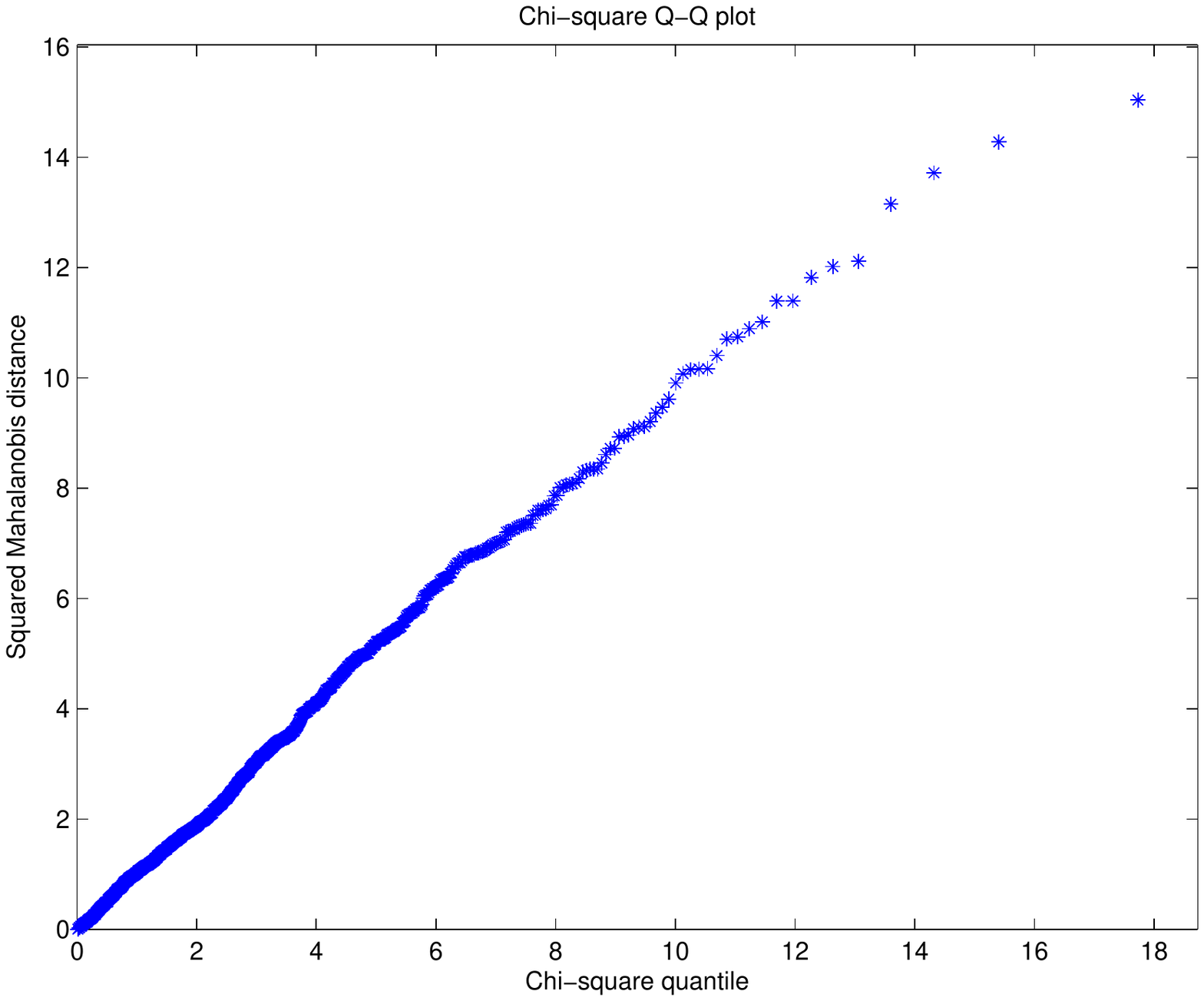}}
%\hspace{0.1cm}
\subfigure[$\alpha=0.85$, H4]{\includegraphics[width=0.3\columnwidth]{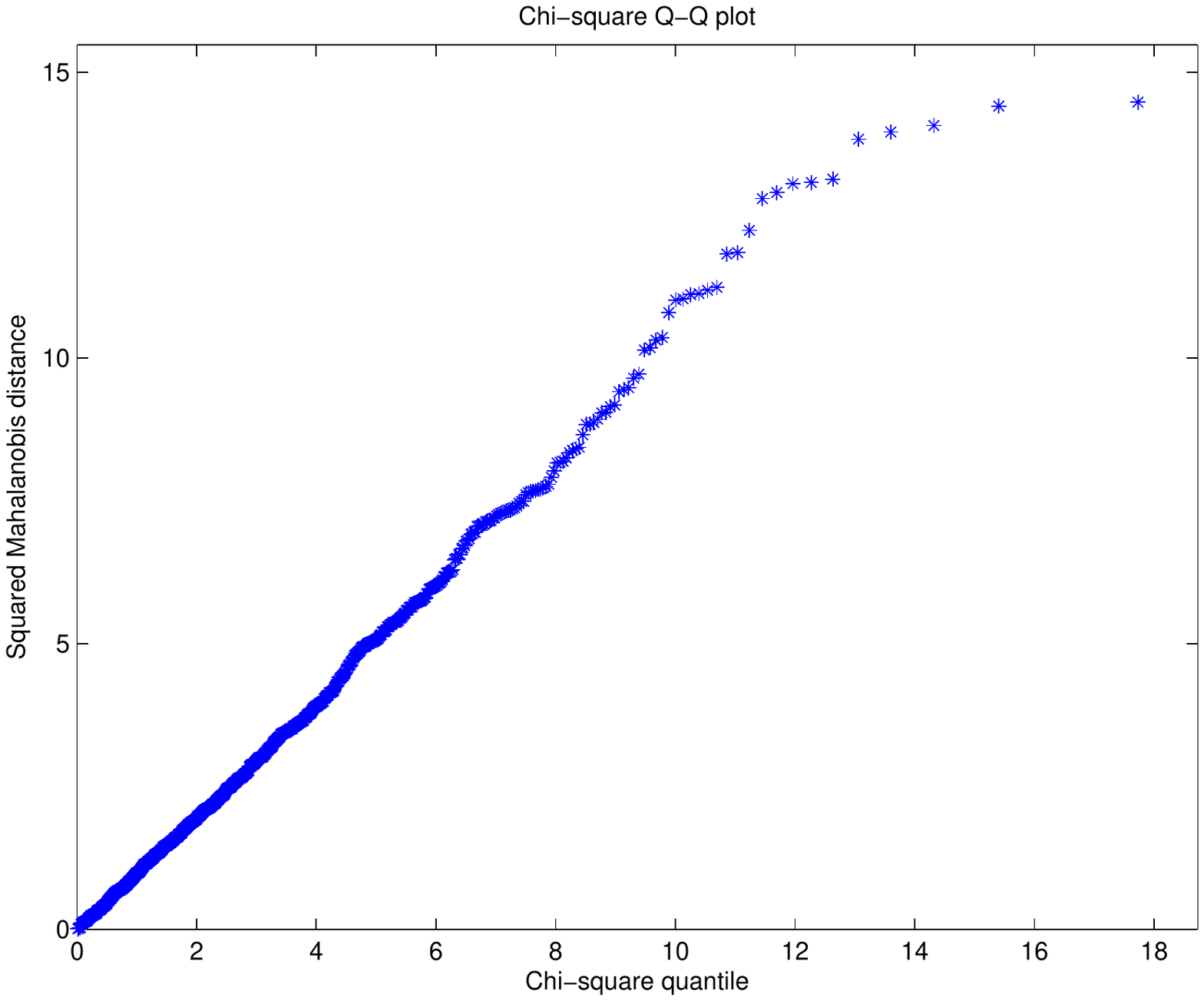}}
\subfigure[$\alpha=1.50$, H1]{\includegraphics[width=0.3\columnwidth]{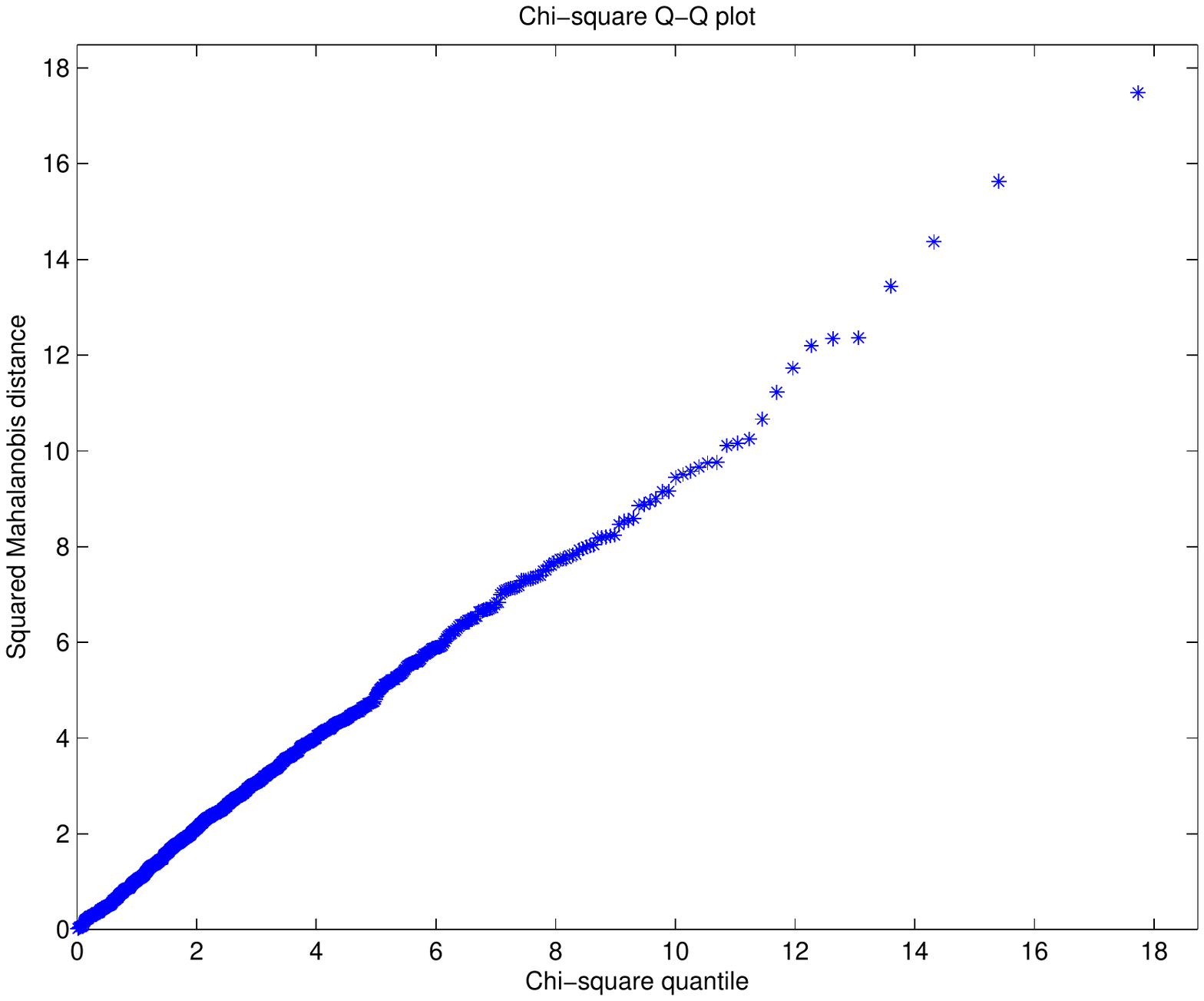}}
%\hspace{0.1cm}
\subfigure[$\alpha=1.50$, H2]{\includegraphics[width=0.3\columnwidth]{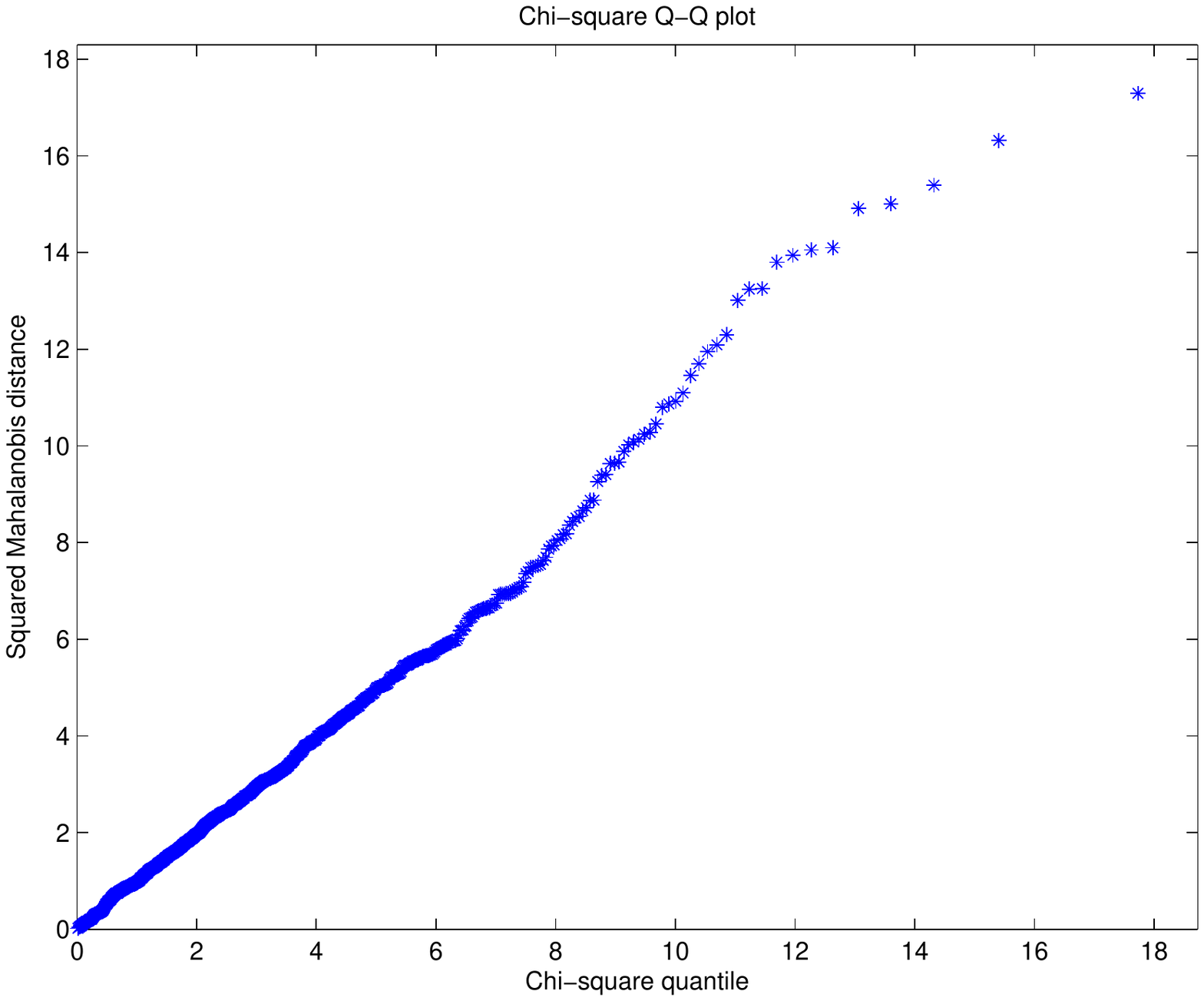}}
%\hspace{0.1cm}
\subfigure[$\alpha=1.50$, H3]{\includegraphics[width=0.3\columnwidth]{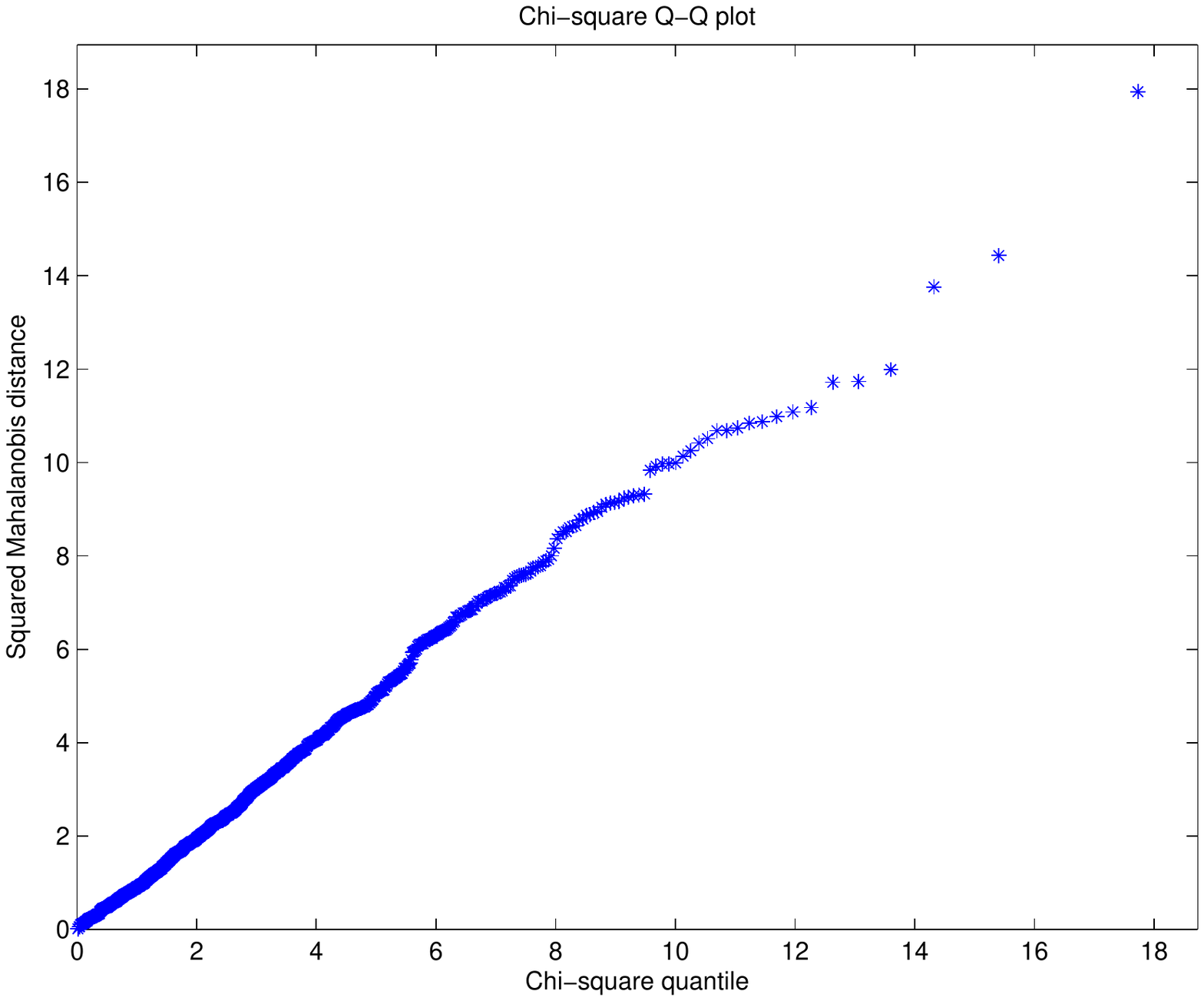}}
%\vspace{0.1cm}
\subfigure[$\alpha=1.5$, H4]{\includegraphics[width=0.3\columnwidth]{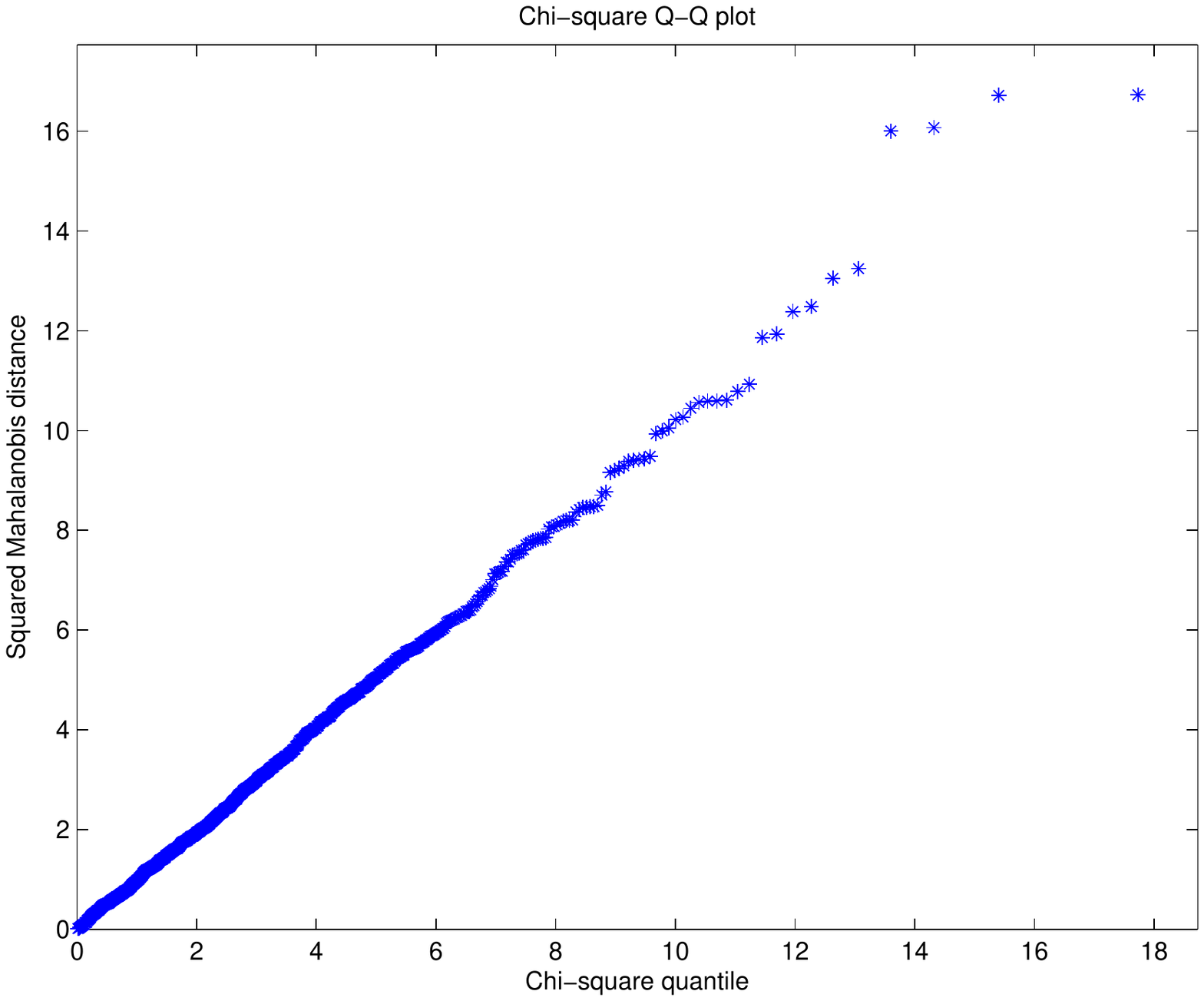}}
\subfigure[$\alpha=2.50$, H1]{\includegraphics[width=0.3\columnwidth]{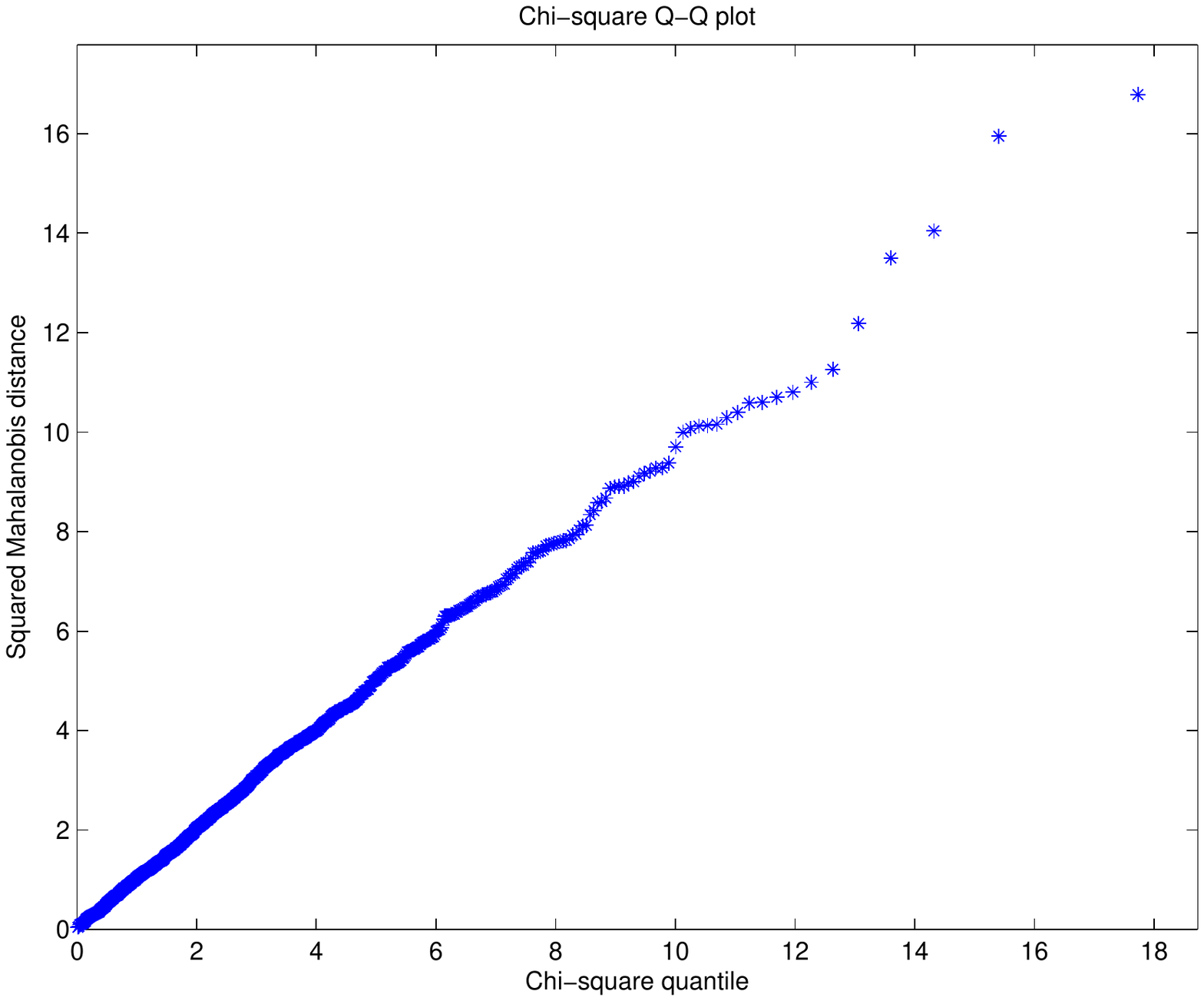}}
%\hspace{0.1cm}
\subfigure[$\alpha=2.50$, H2]{\includegraphics[width=0.3\columnwidth]{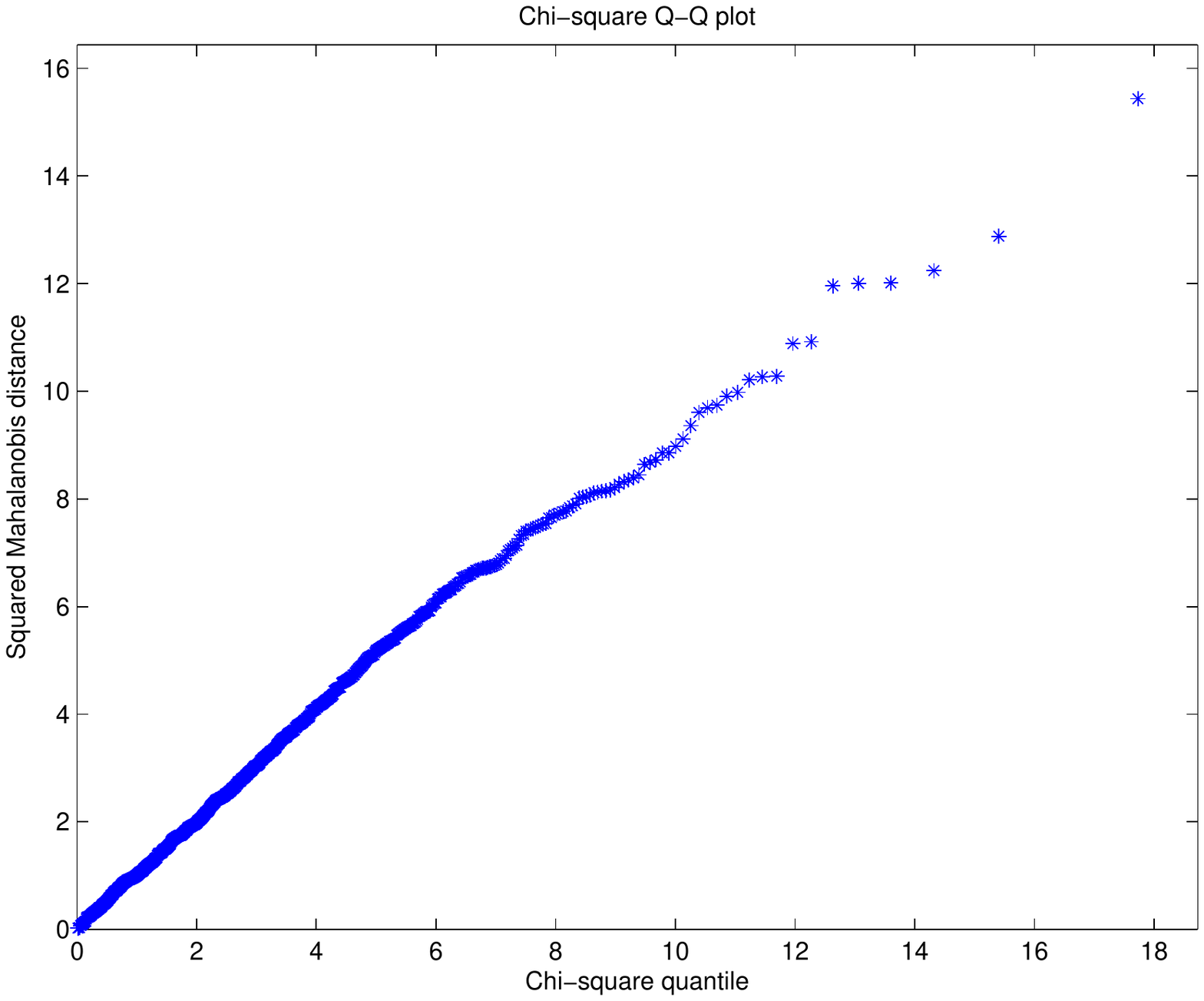}}
%\hspace{0.1cm}
\subfigure[$\alpha=2.50$, H3]{\includegraphics[width=0.3\columnwidth]{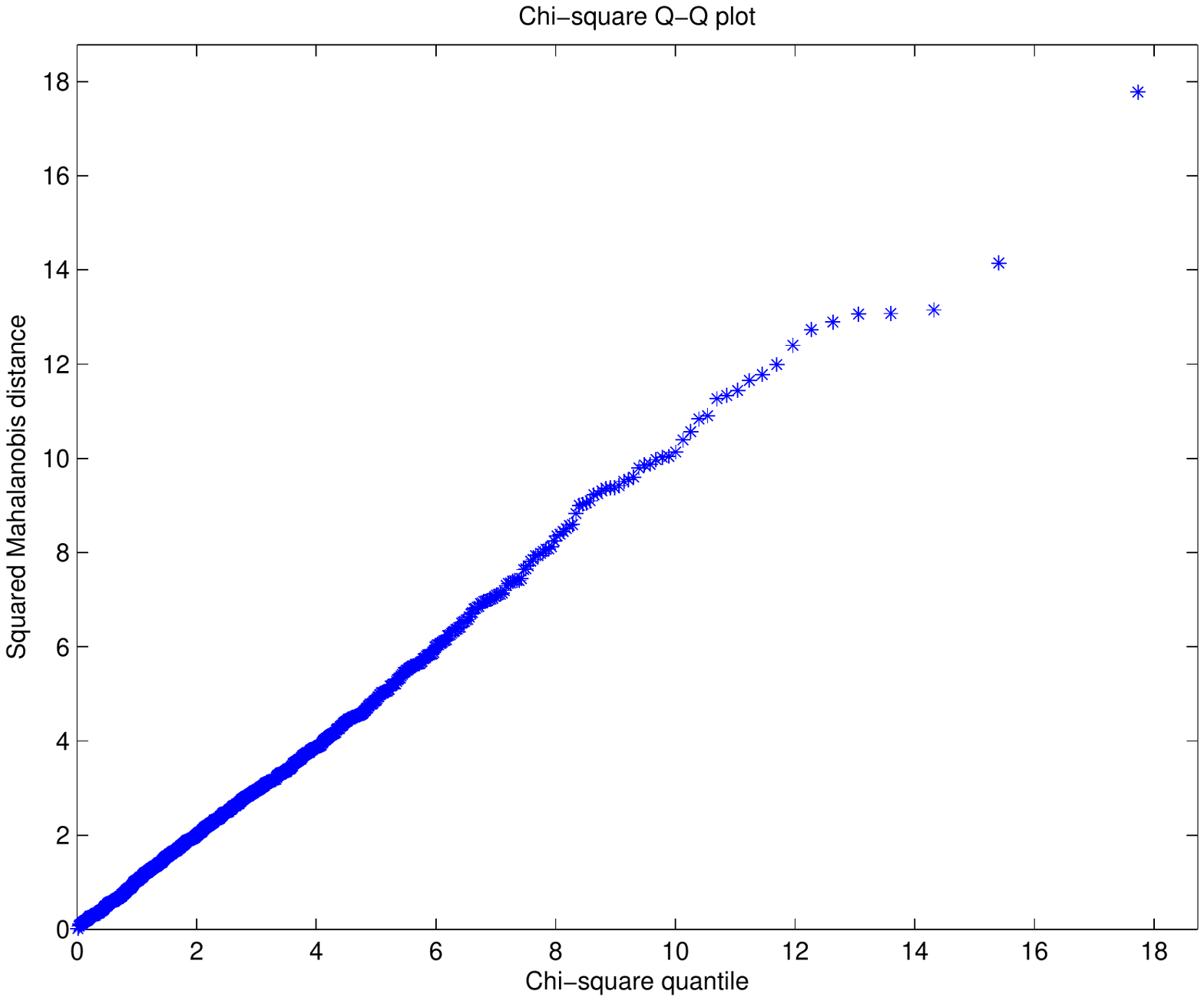}}
%\hspace{0.1cm}
\subfigure[$\alpha=2.50$, H4]{\includegraphics[width=0.3\columnwidth]{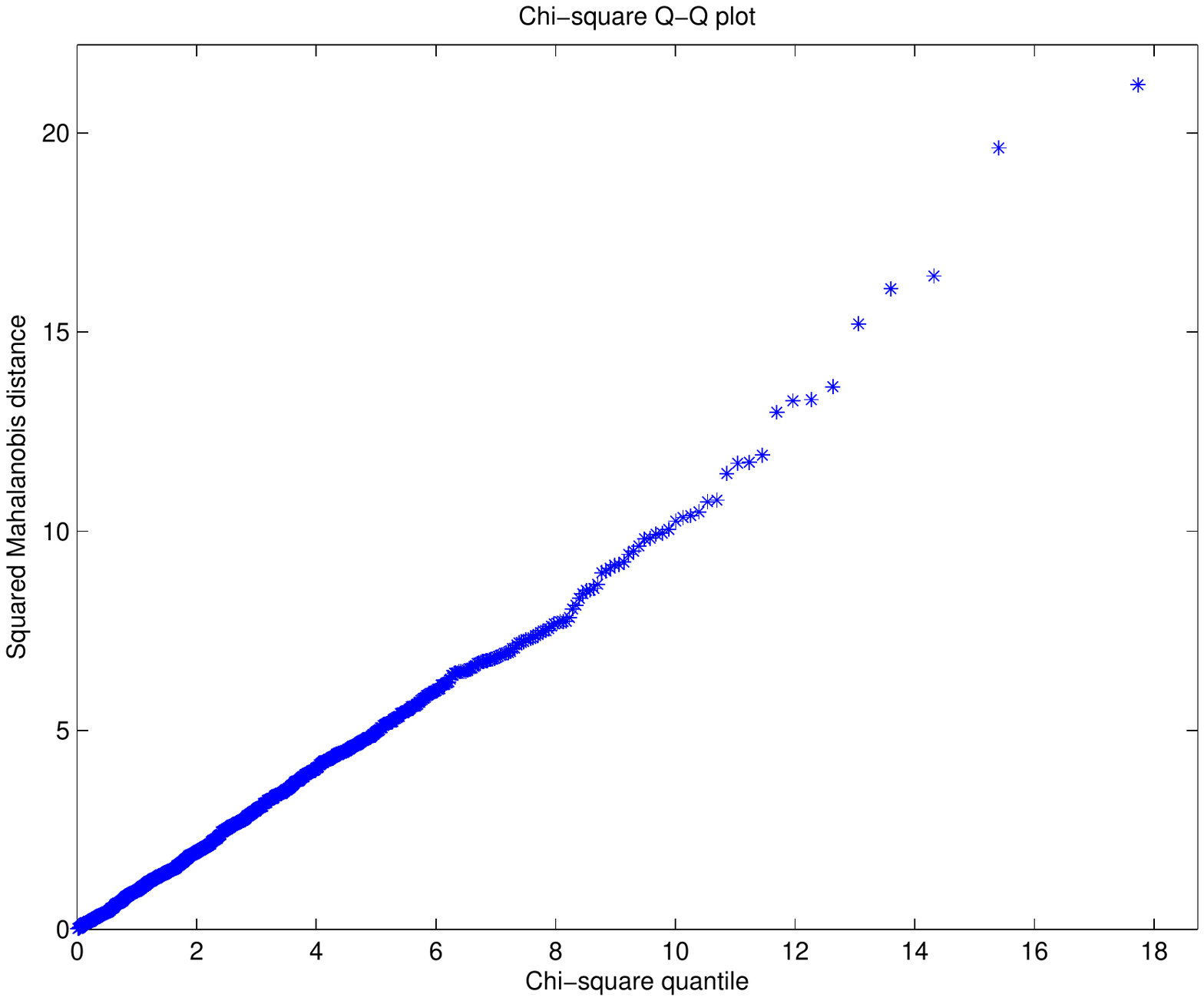}}

\end{center}
\caption{Chi-square Q-Q plot of the squared Mahalanobis distances of the simulated random vector ($\hat{\zeta}_{T}^{A}$, $\hat{\zeta}_{T}^{B}$,  $\hat{\zeta}_{T}^{\varphi}$), $T=30000$.}\label{Fig.1}
\end{figure}

 Figure \ref{Fig.1} shows the Chi-square quantile-quantile (Q-Q) plot. The graphs display the squared Mahalanobis distances of ($\hat{\zeta}_{T}^{A}$, $\hat{\zeta}_{T}^{B}$,  $\hat{\zeta}_{T}^{\varphi}$), $T=30000$,  versus quantiles of the Chi-square distribution with $d$  degrees of freedom ($d=3,$ number of variables). The squared Mahalanobis distance has an approximate Chi-squared distribution when the data are MVN. The interpretation is similar to the normal Q-Q plot, that is, if the graph is not linear, it can not ensure the multivariate normal distribution. The linear plot of data for the cases displayed in Figure \ref{Fig.1}, suggests that the asymptotic multivariate normality can be proved. In Table \ref{Tabla.1}, rejection rates of Henze-Zirkler's and Doornik-Hansen MVN tests applied to the simulated random vectors ($\hat{\zeta}_{T}^{A}$, $\hat{\zeta}_{T}^{B}$ $\hat{\zeta}_{T}^{\varphi}$) for different $T$ values are shown. These rates are calculated for significance level \% 1, using 50 sets of simulated random vectors ($\hat{\zeta}_{T}^{A}$, $\hat{\zeta}_{T}^{B}$,  $\hat{\zeta}_{T}^{\varphi}$), with $T=1000;5000;10000;15000;20000;30000$. Each set is composed of 1000 replications of ($\hat{\zeta}_{T}^{A}$, $\hat{\zeta}_{T}^{B}$,  $\hat{\zeta}_{T}^{\varphi}$). It is noted that the rate of rejection decreases as $T$ increases. For cases where the rate of rejection is higher, the value of $T$ needs to be increased to ensure the MVN in the random vector ($\hat{\zeta}_{T}^{A}$, $\hat{\zeta}_{T}^{B}$,  $\hat{\zeta}_{T}^{\varphi}$).

  Also, consistency of the LSEs can be verified with the results in figures \ref{Fig.5}-\ref{Fig.6}. Specifically, the small variance obtained suggests this property of the estimators under the hypothesis \textbf{A1}-\textbf{A3}. The LSEs $(\hat{A}_{T},\hat{B}_{T},\hat{\varphi}_{T}),$
  $T=[1000,5000]$, with discretization step size 250, of the parameters $(A,B,\varphi)$ of the regression function (\ref{g2}),  are computed numerically with Matlab function \verb"lsqnonlin" based on the  Levenberg-Marquardt algorithm (see, \cite{More}).

\begin{figure}[hptb]
\begin{center}
\subfigure[$\alpha=0.85$, H1]{\includegraphics[width=0.3\columnwidth]{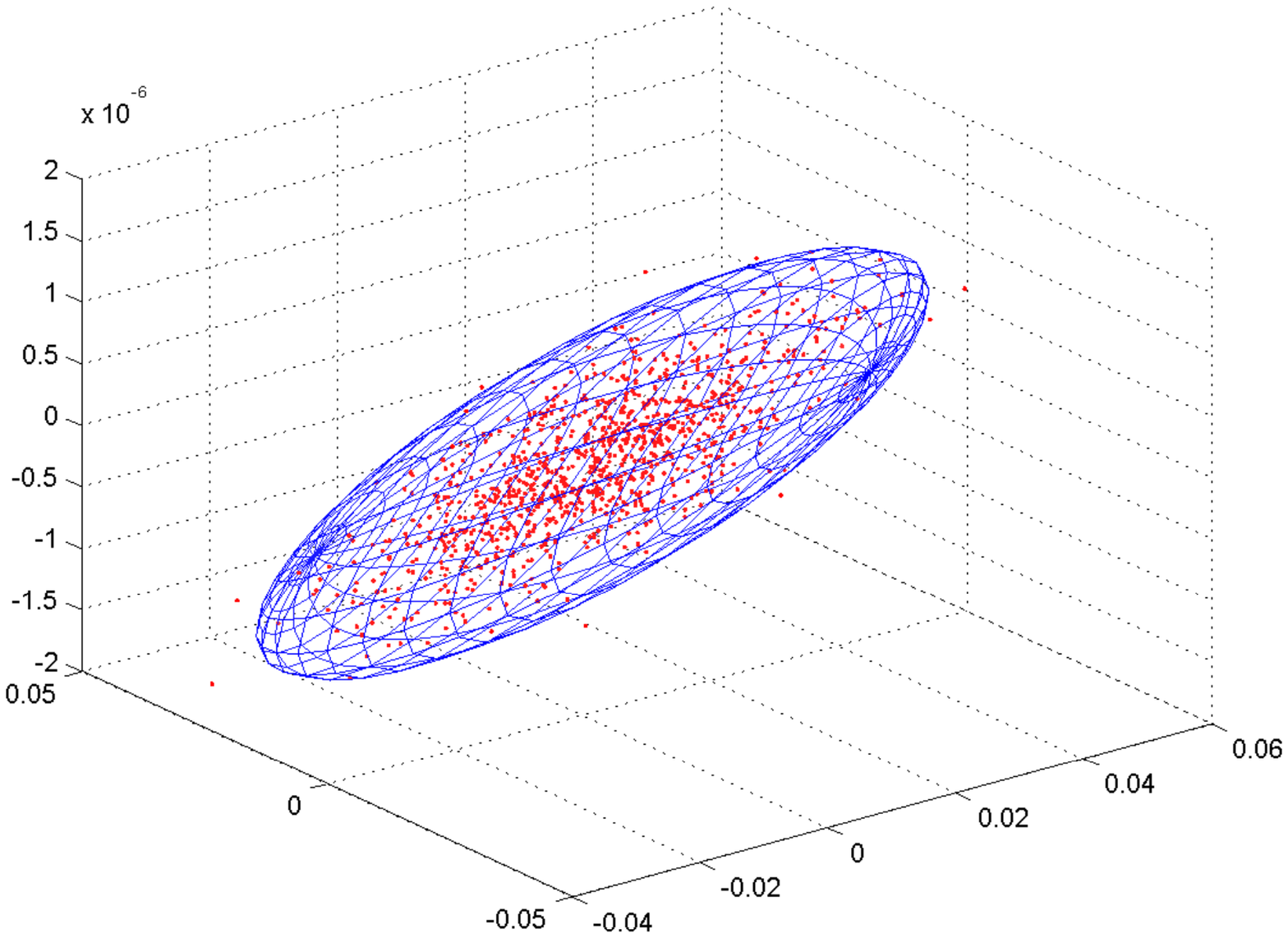}}
%\hspace{0.1cm}
\subfigure[$\alpha=0.85$, H2]{\includegraphics[width=0.3\columnwidth]{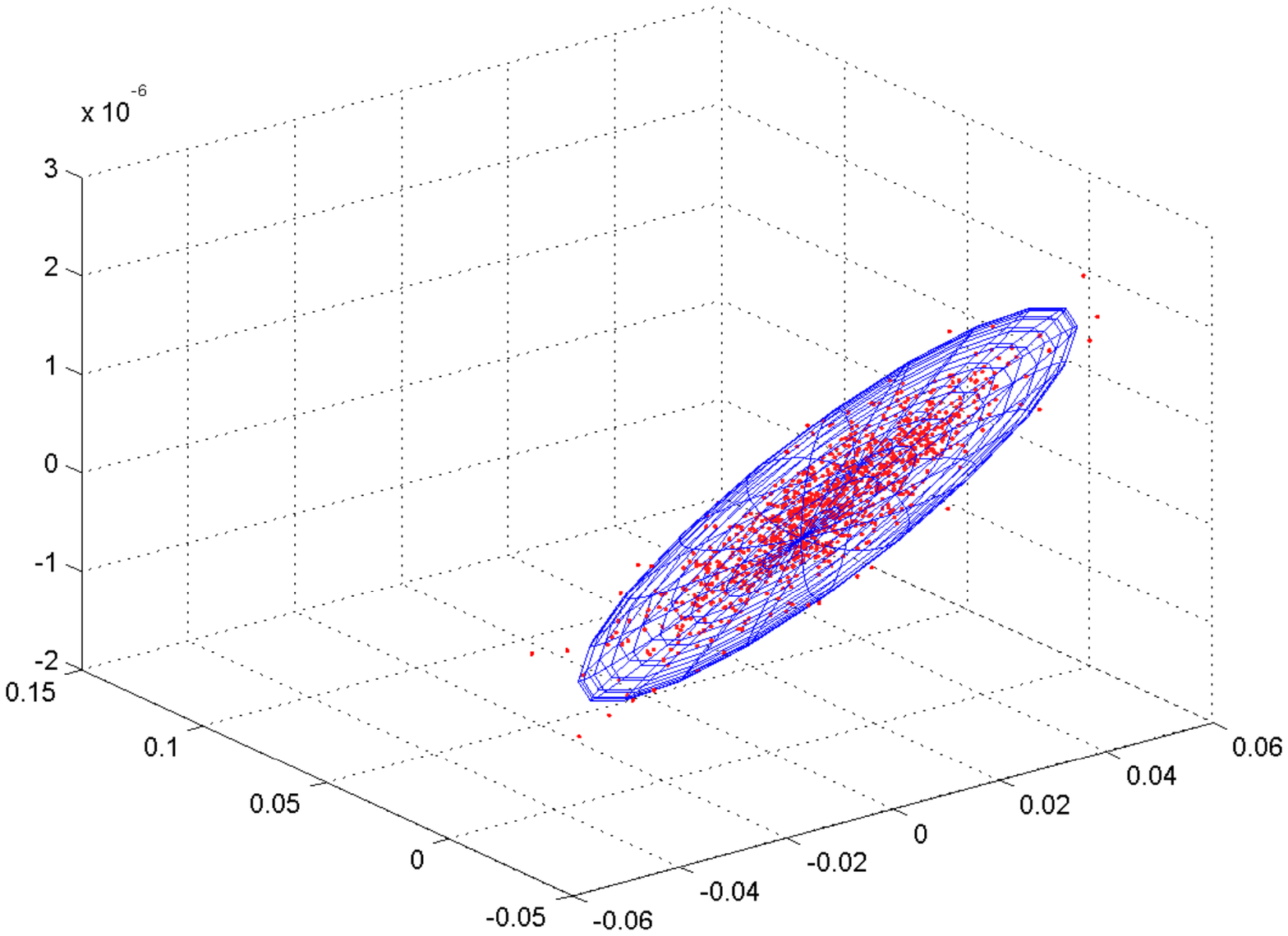}}
%\hspace{0.1cm}
\subfigure[$\alpha=0.85$, H3]{\includegraphics[width=0.3\columnwidth]{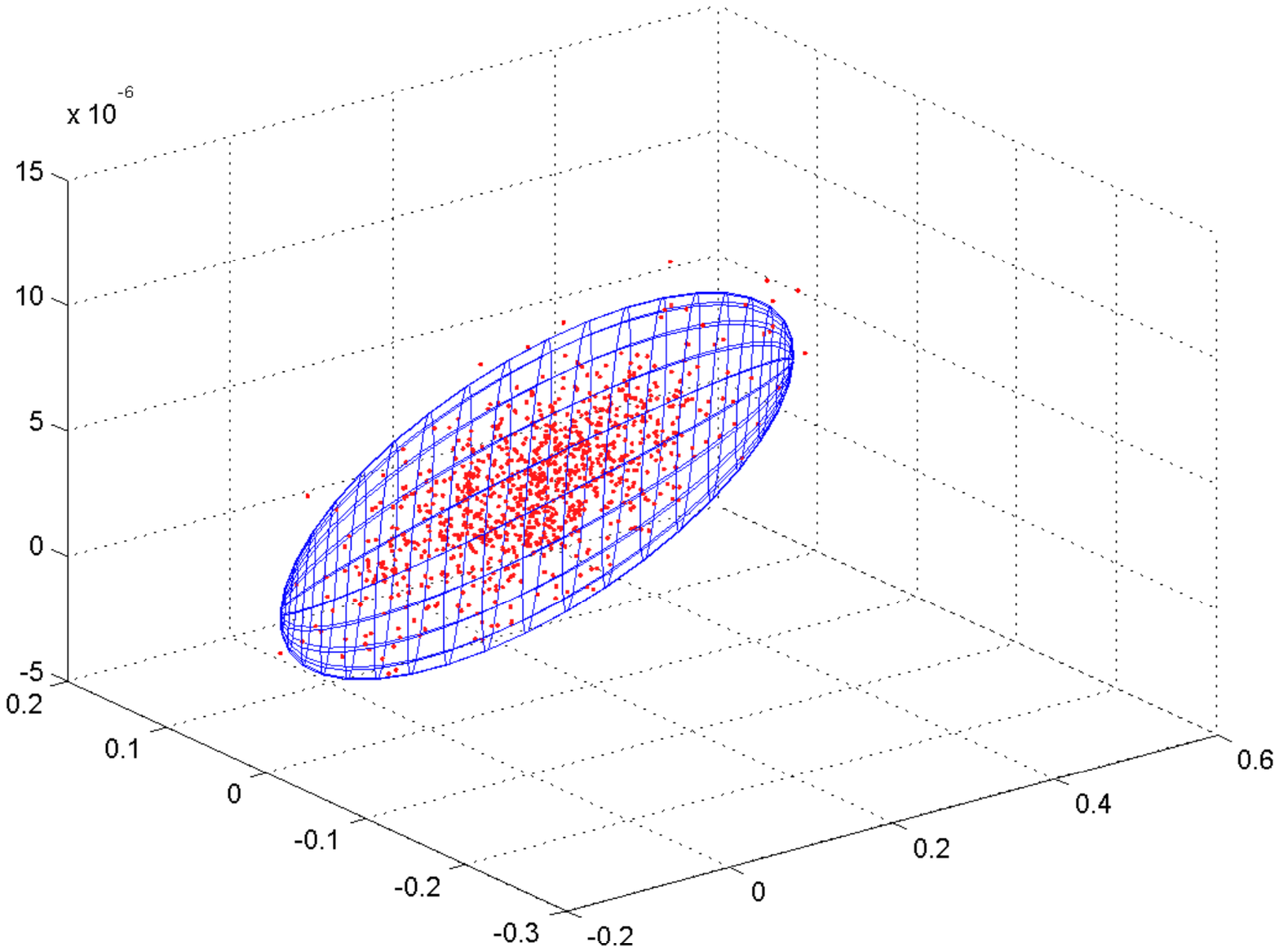}}
%\hspace{0.1cm}
\subfigure[$\alpha=0.85$, H4]{\includegraphics[width=0.3\columnwidth]{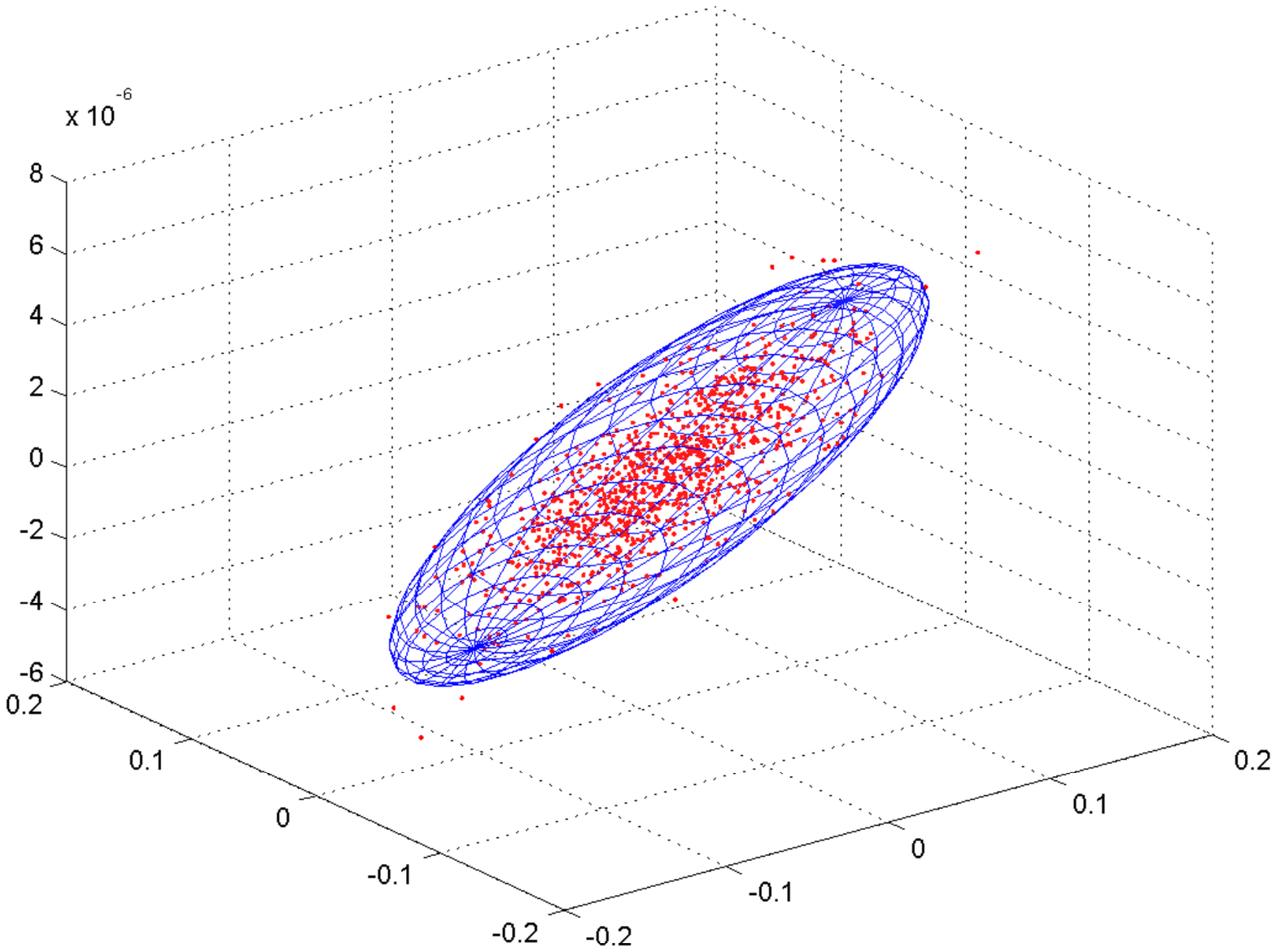}}
\subfigure[$\alpha=1.50$, H1]{\includegraphics[width=0.3\columnwidth]{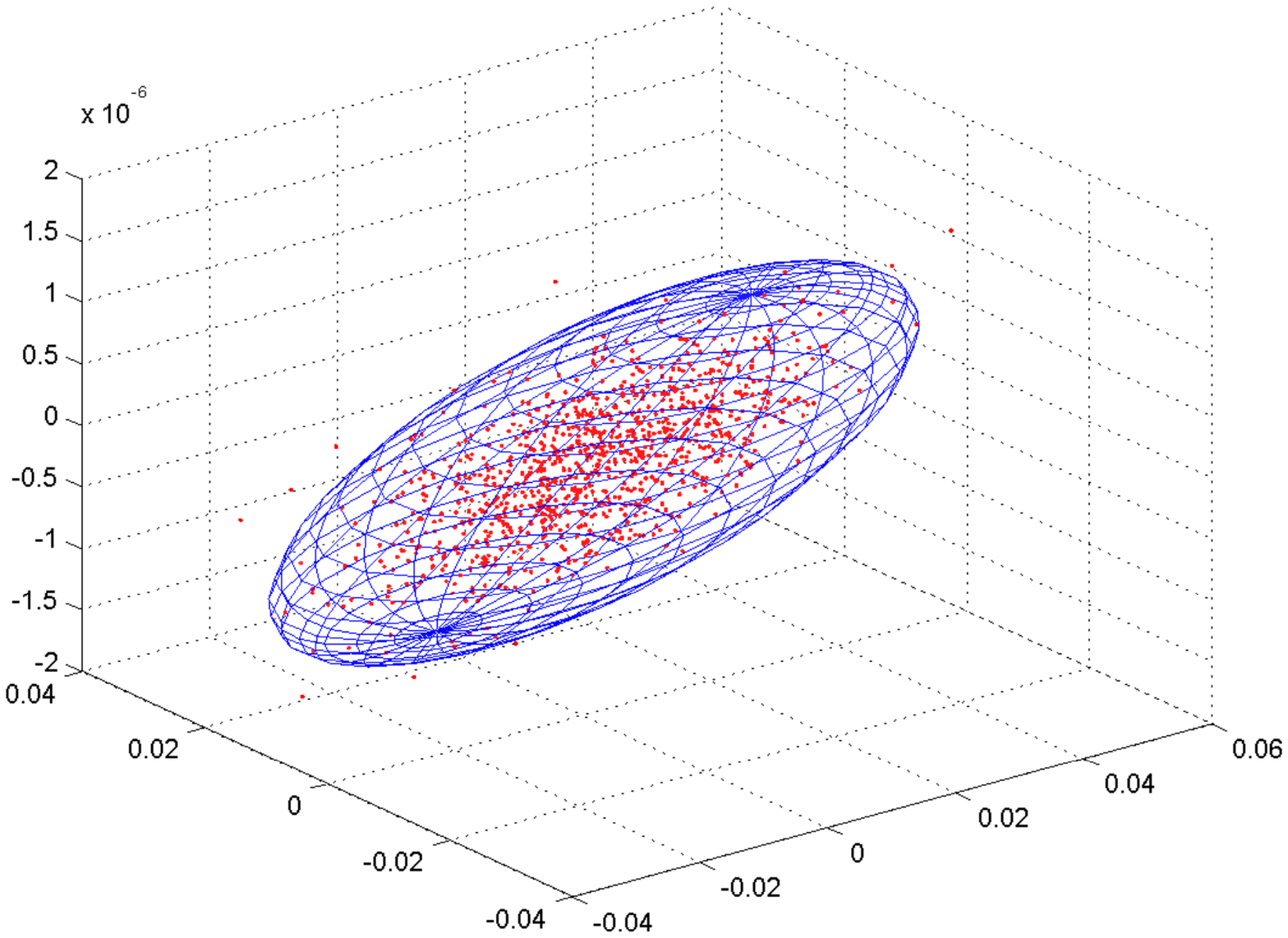}}
%\hspace{0.1cm}
\subfigure[$\alpha=1.50$, H2]{\includegraphics[width=0.3\columnwidth]{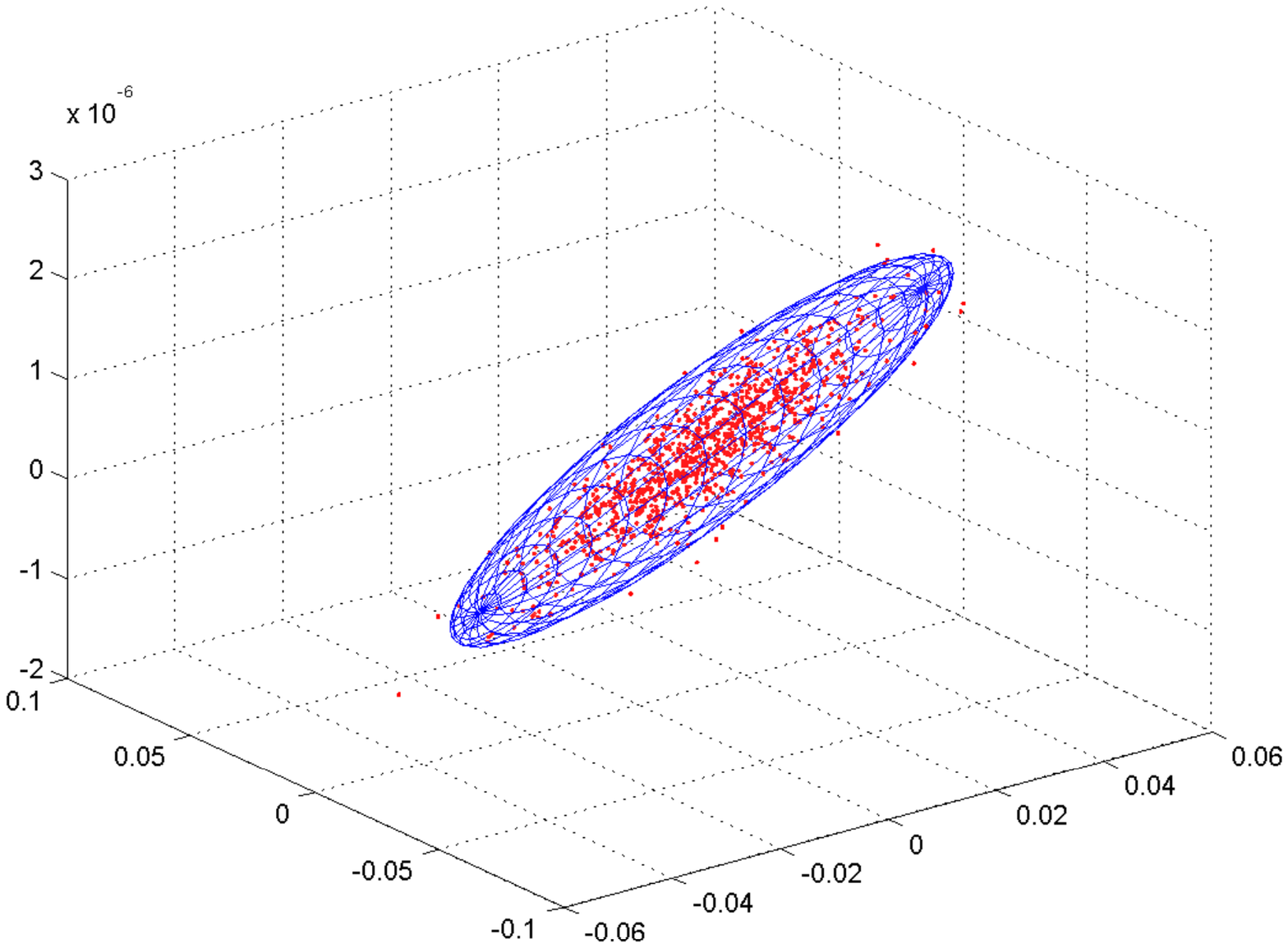}}
%\hspace{0.1cm}
\subfigure[$\alpha=1.50$, H3]{\includegraphics[width=0.3\columnwidth]{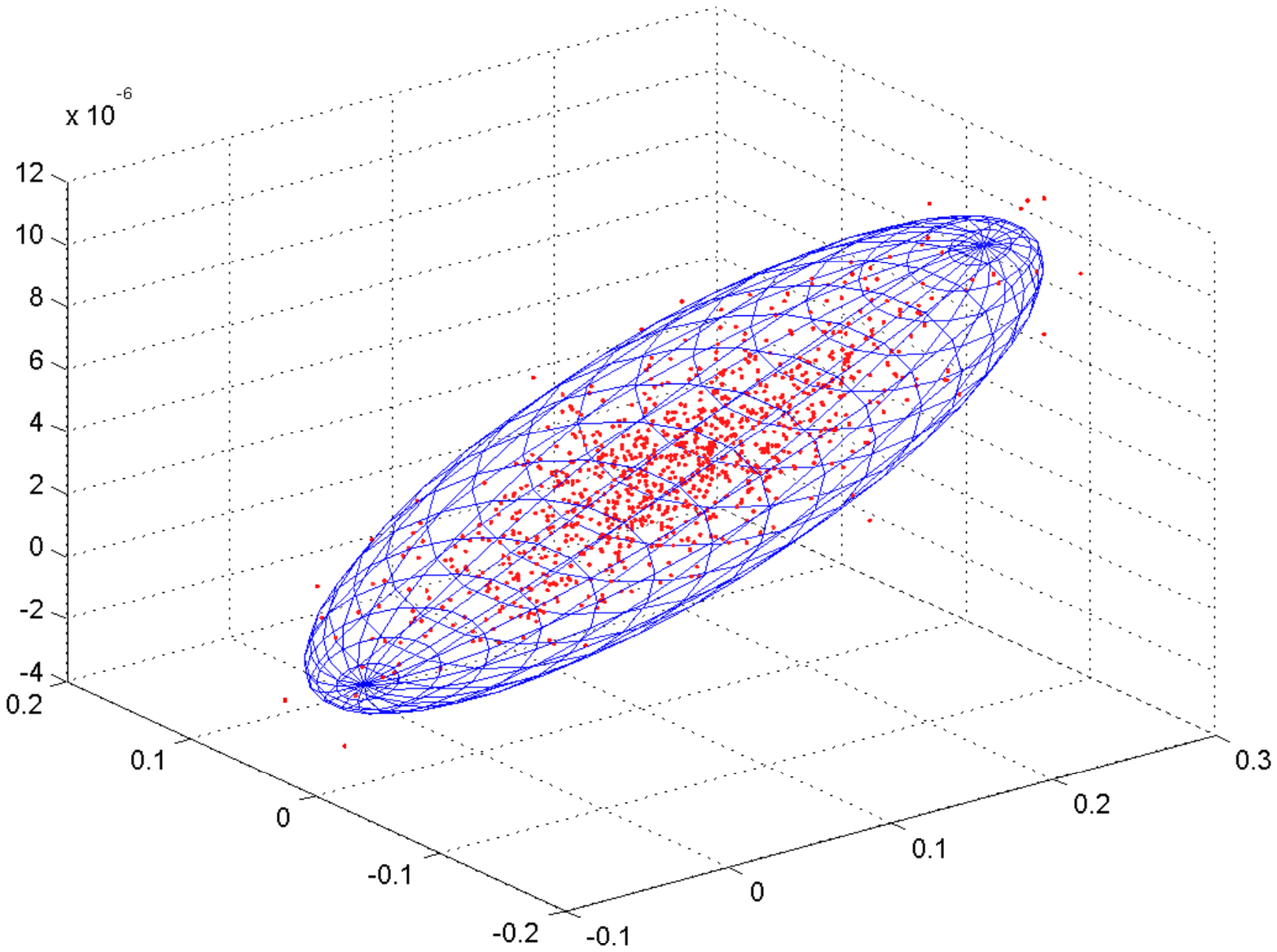}}
%\vspace{0.1cm}
\subfigure[$\alpha=1.50$, H4]{\includegraphics[width=0.3\columnwidth]{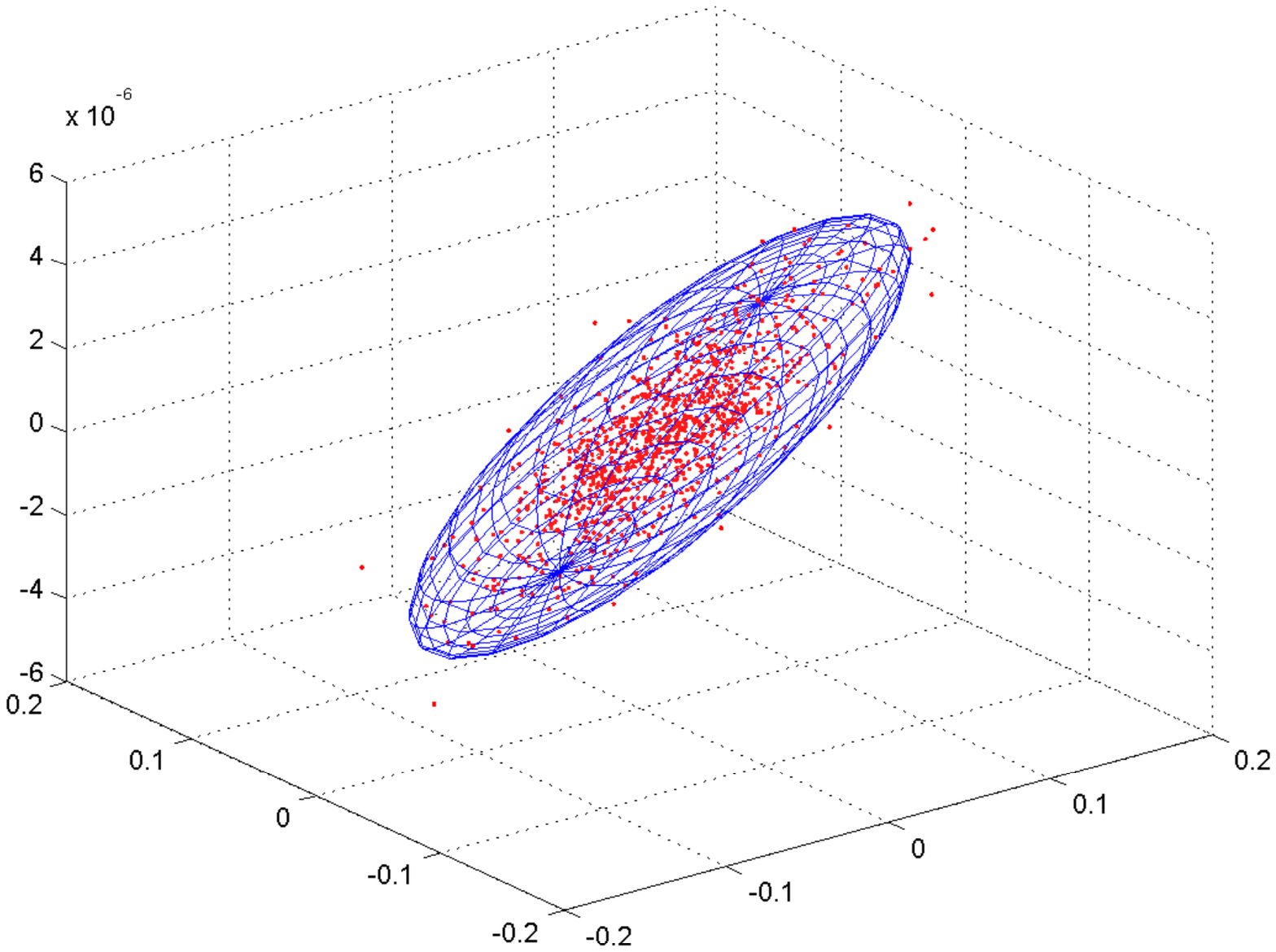}}
\subfigure[$\alpha=2.50$, H1]{\includegraphics[width=0.3\columnwidth]{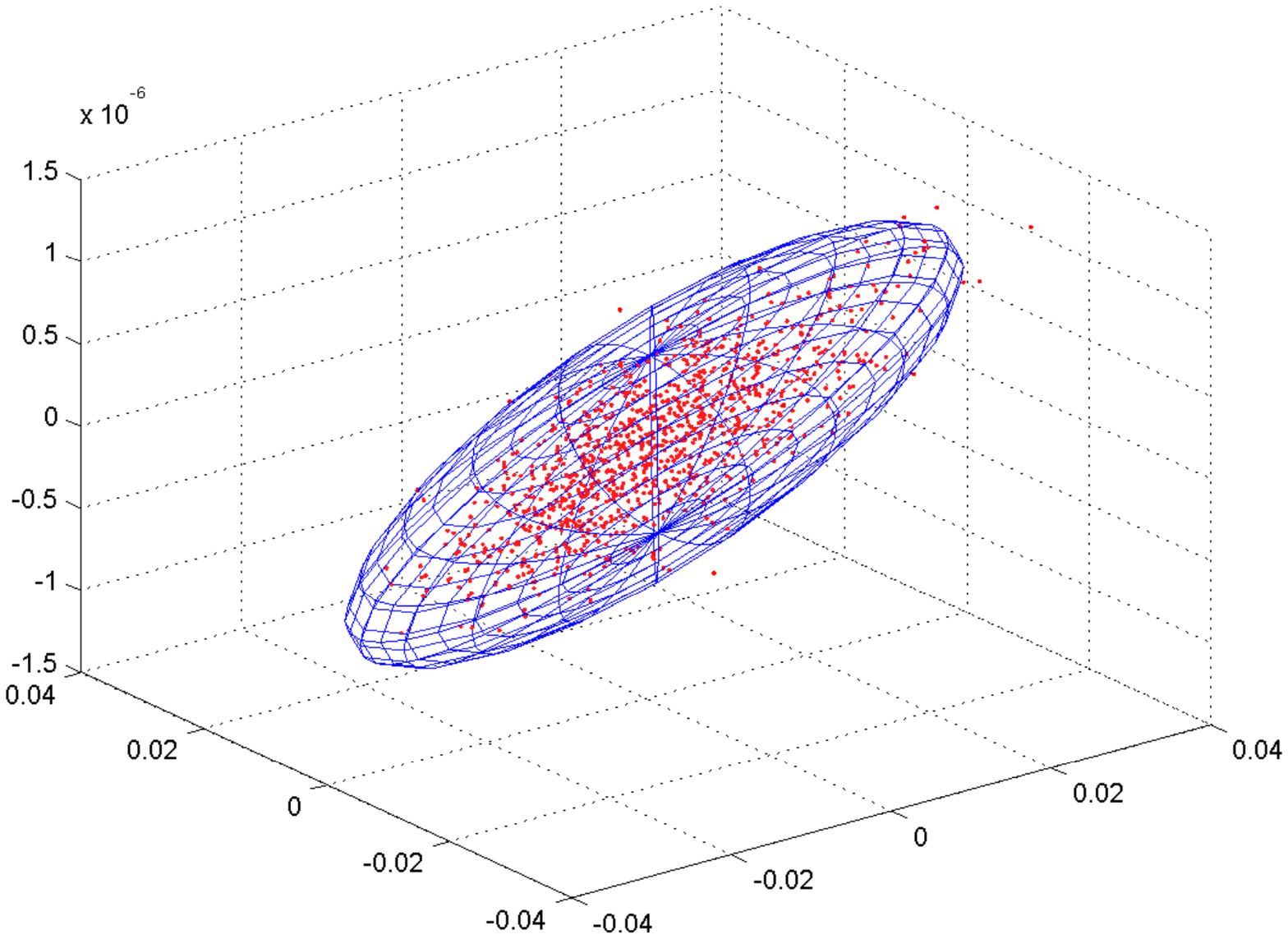}}
%\hspace{0.1cm}
\subfigure[$\alpha=2.50$, H2]{\includegraphics[width=0.3\columnwidth]{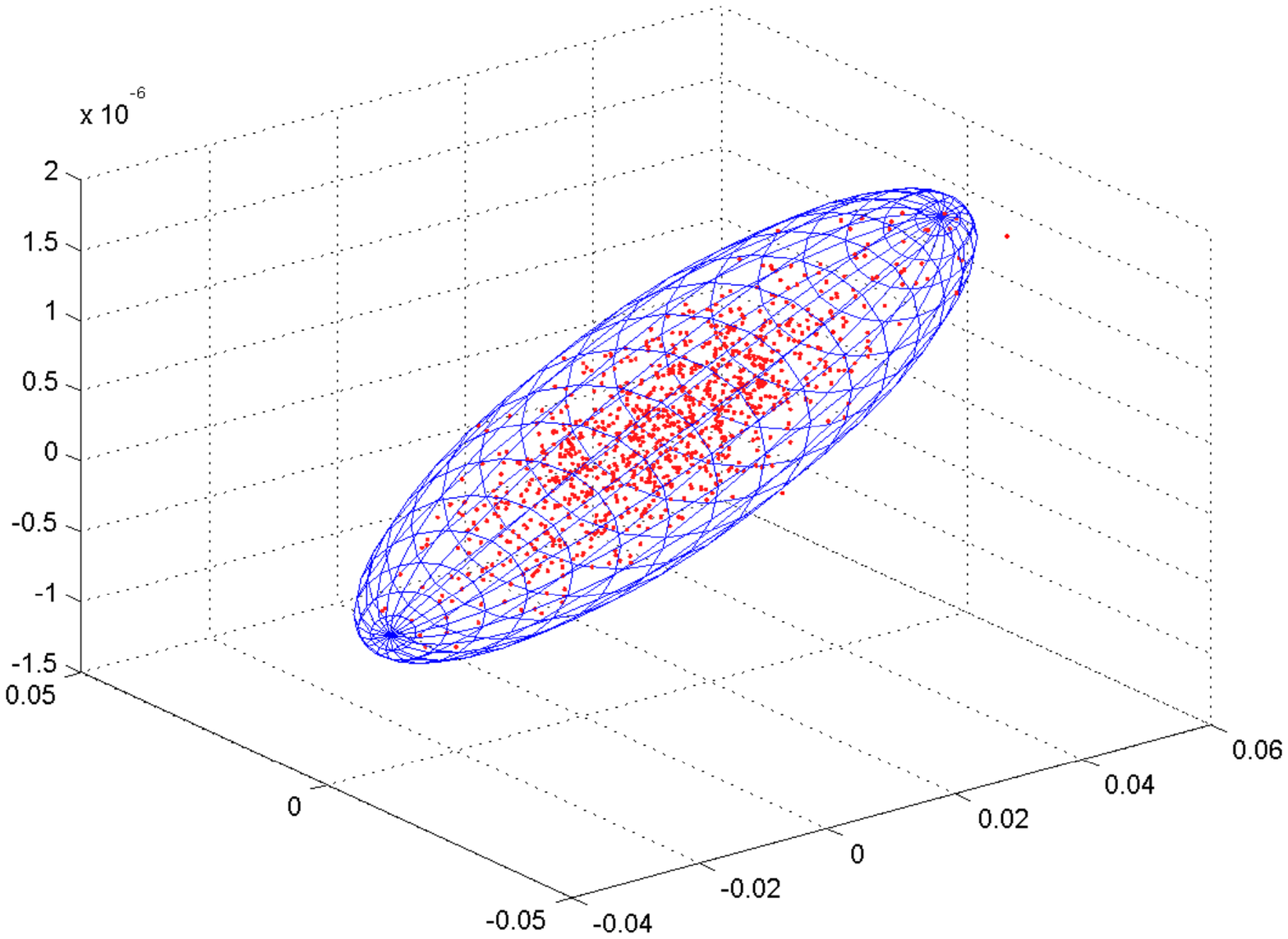}}
%\hspace{0.1cm}
\subfigure[$\alpha=2.50$, H3]{\includegraphics[width=0.3\columnwidth]{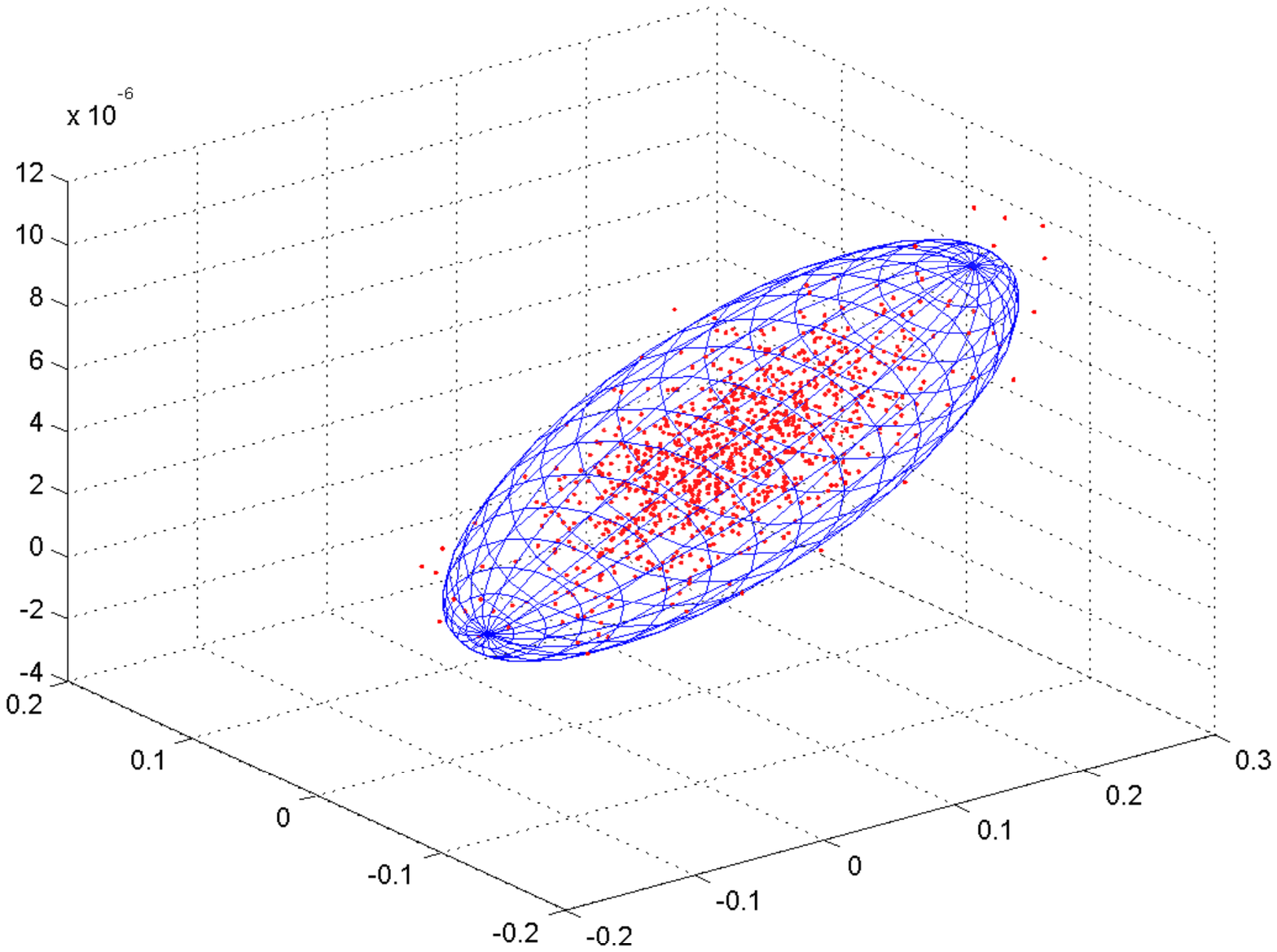}}
%\hspace{0.1cm}
\subfigure[$\alpha=2.50$, H4]{\includegraphics[width=0.3\columnwidth]{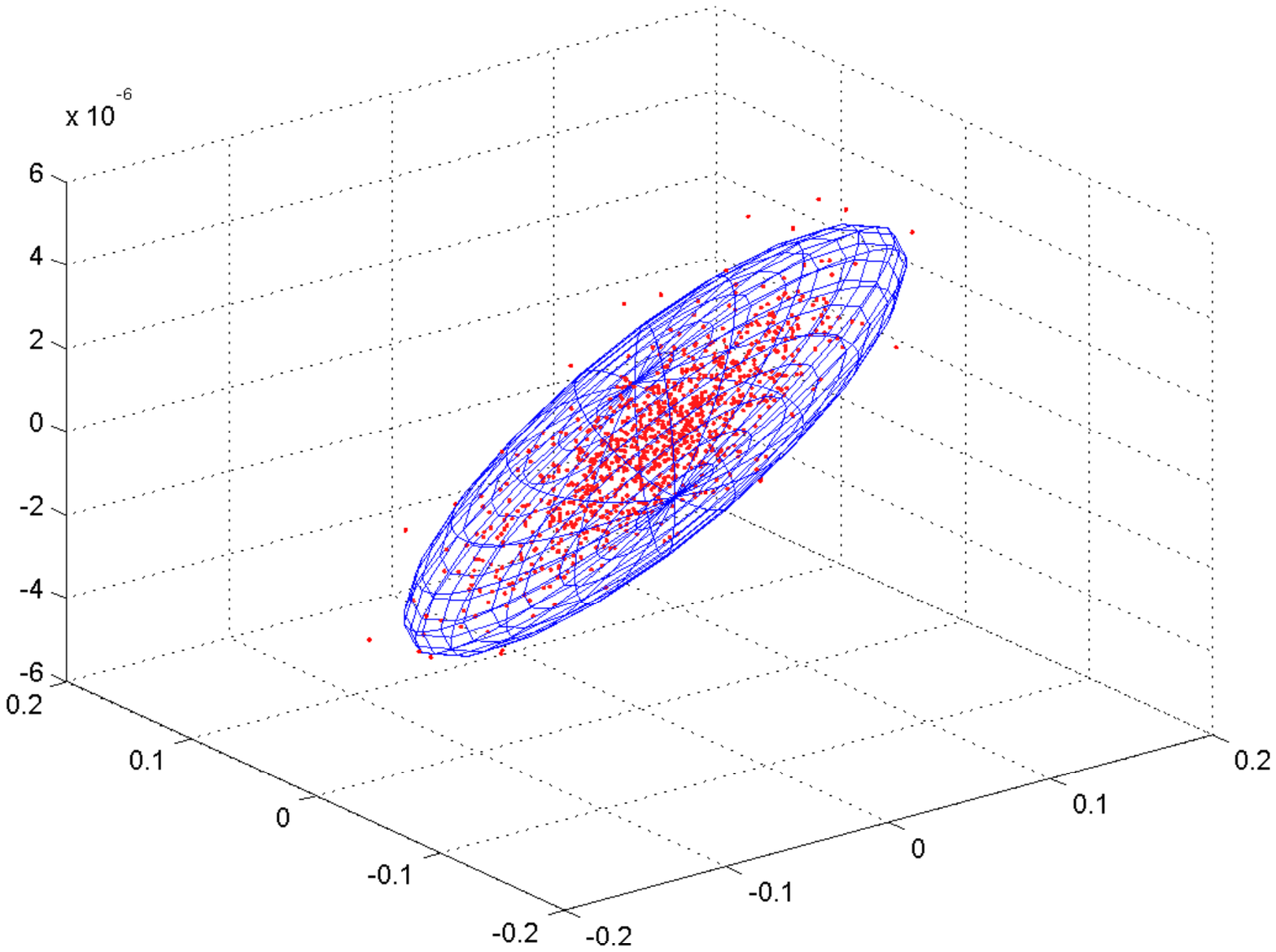}}
\end{center}
\caption{97\% Probability Contours of MVN and scatter plot for ($\hat{\zeta}_{T}^{A}$, $\hat{\zeta}_{T}^{B}$,  $\hat{\zeta}_{T}^{\varphi}$) values, $T=30000$.}\label{Fig.2}
\end{figure}

Another significant graph for random vectors with MVN, $%
N_{d}(\boldsymbol{\mu},\tilde{C} ),$ is the constant Probability Contours. This graph is represented as a ellipsoid formed for all $\mathbf{x}$ satisfying equation
$(\mathbf{x}-\boldsymbol{\mu})'\tilde{C}^{-1}(\mathbf{x}-\boldsymbol{\mu})=c^{2}, $
with a constant $c$. The axes of the ellipsoid are $\boldsymbol{\mu}\pm c \sqrt{\lambda_{i}} e_{i}$, where $e_{i}$ and $\lambda_{i}$ are the $i^{th}$ eigenvectors and eigenvalues of $\tilde{C}$, \cite{Rencher}. In Figure \ref{Fig.2}, the constant Probability Contours are calculated for each one of the  cases studied, where $\tilde{C}$ is estimated from the simulated sample of  ($\hat{\zeta}_{T}^{A}$, $\hat{\zeta}_{T}^{B}$,  $\hat{\zeta}_{T}^{\varphi}$) values and $c=3$, that is, if data are multivariate normally distributed, then 97\% of the data should be inside the ellipsoid. In the same graph the scatter plot of the ($\hat{\zeta}_{T}^{A}$, $\hat{\zeta}_{T}^{B}$,  $\hat{\zeta}_{T}^{\varphi}$) values is represented. In most cases, the simulated values (shown as red dots) are within the ellipsoid of the theoretical distribution (shown in blue).

\begin{figure}[hptb]
\begin{center}
\includegraphics[width=0.35\columnwidth]{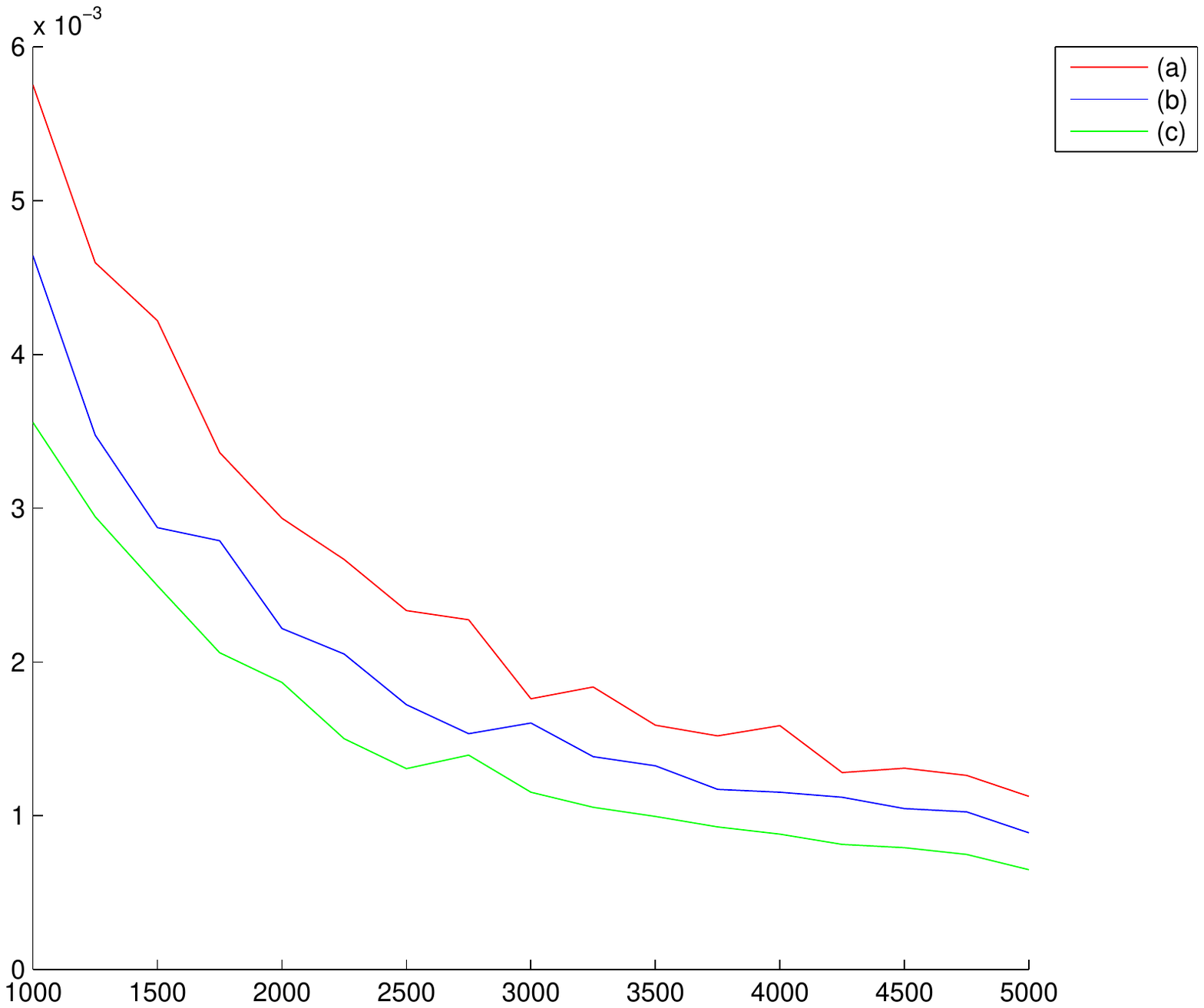}
\hspace{0.1cm}
\includegraphics[width=0.35\columnwidth]{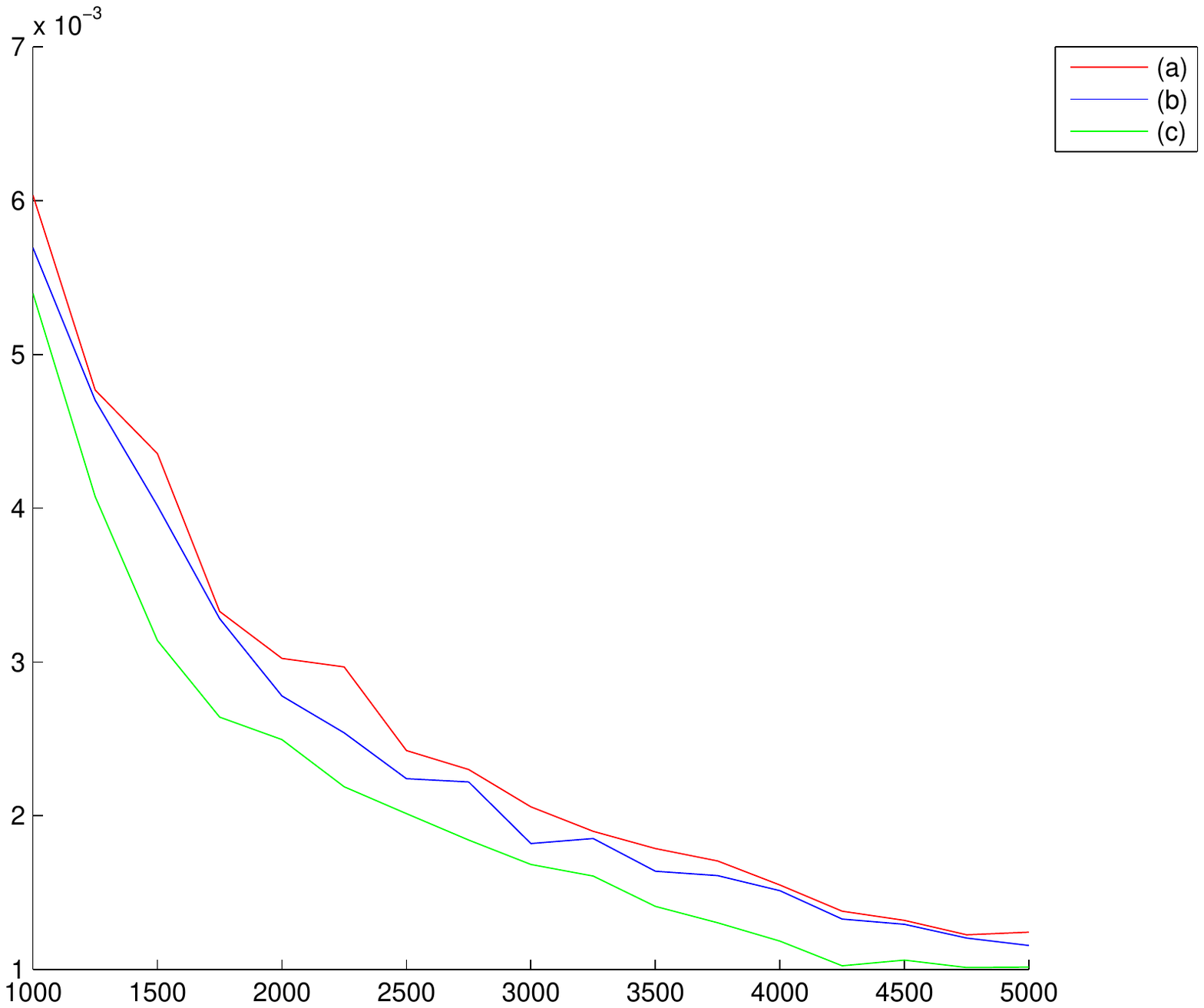}
\hspace{0.1cm}
\includegraphics[width=0.35\columnwidth]{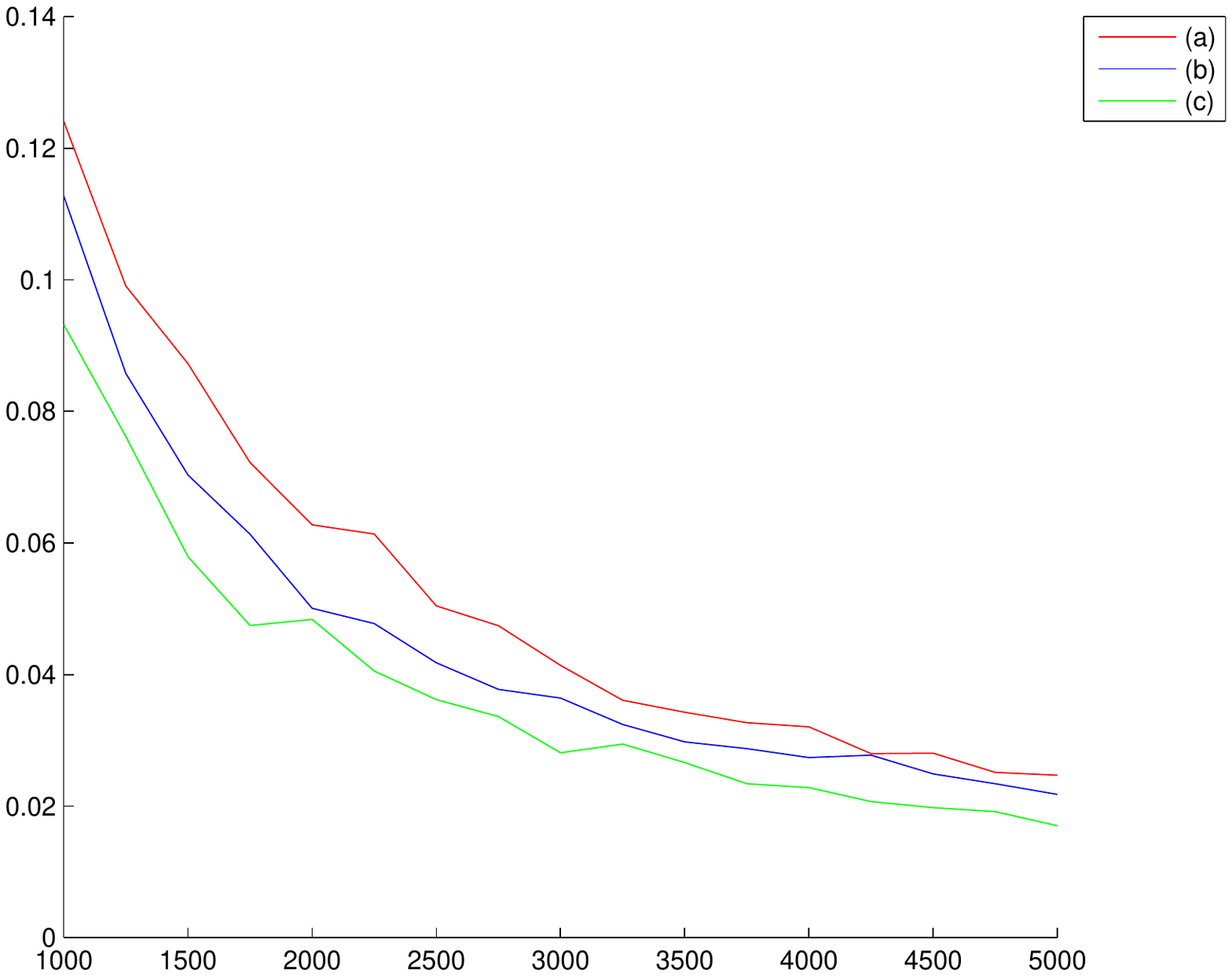}
\hspace{0.1cm}
\includegraphics[width=0.35\columnwidth]{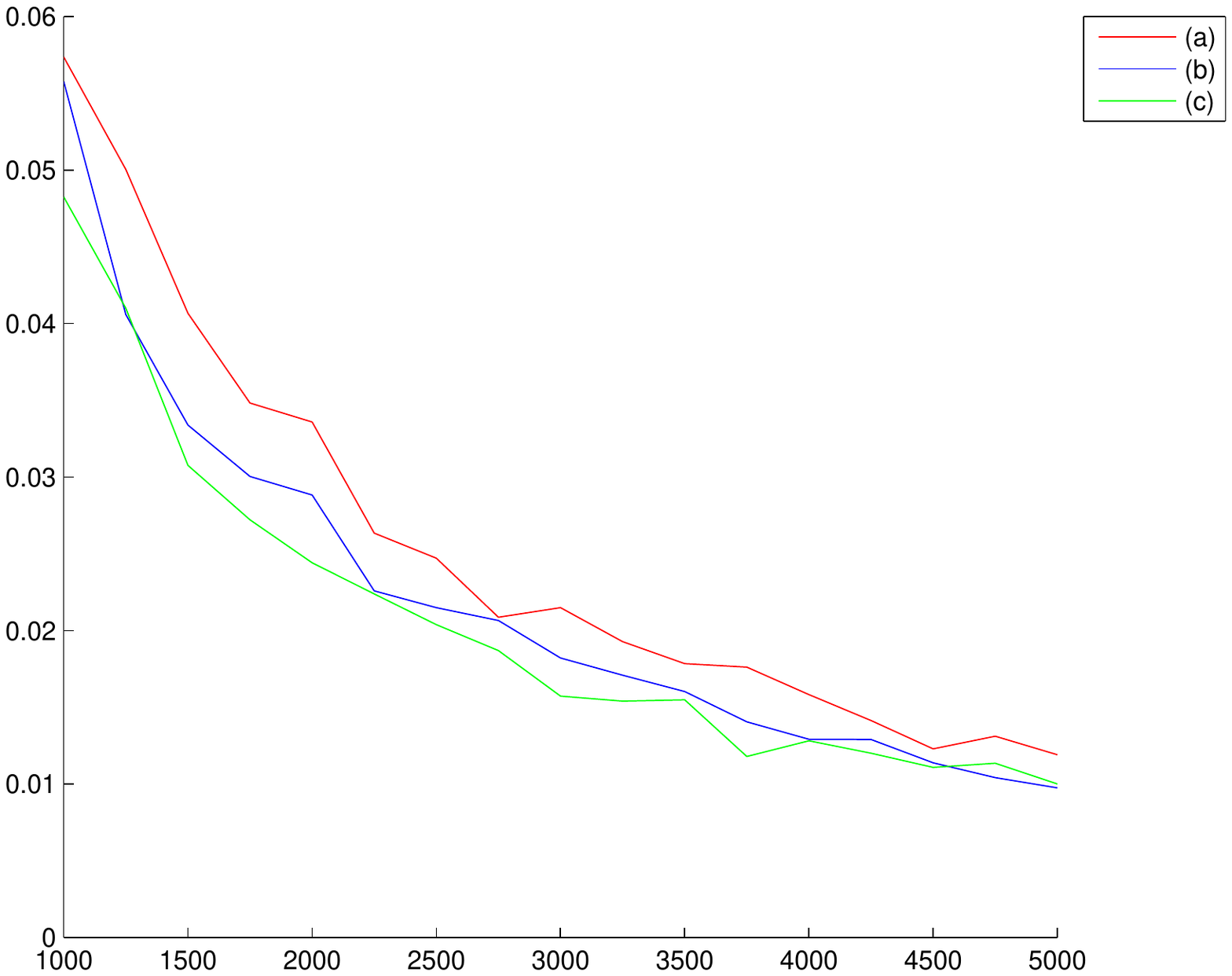}
\hspace{0.1cm}
\end{center}
\caption{Variance of $\hat{A}_{T}$, $T \in [1000, 5000]$, with discretization step size 250, for cases: (a) $\alpha=0.85$, (b) $\alpha=1.50$, (c) $\alpha=2.50$, case H1 (top-left), case H2 (top-right), case H3 (bottom-left) and case H4 (bottom-right).}\label{Fig.4}
\end{figure}

\begin{figure}[hptb]
\begin{center}
\includegraphics[width=0.35\columnwidth]{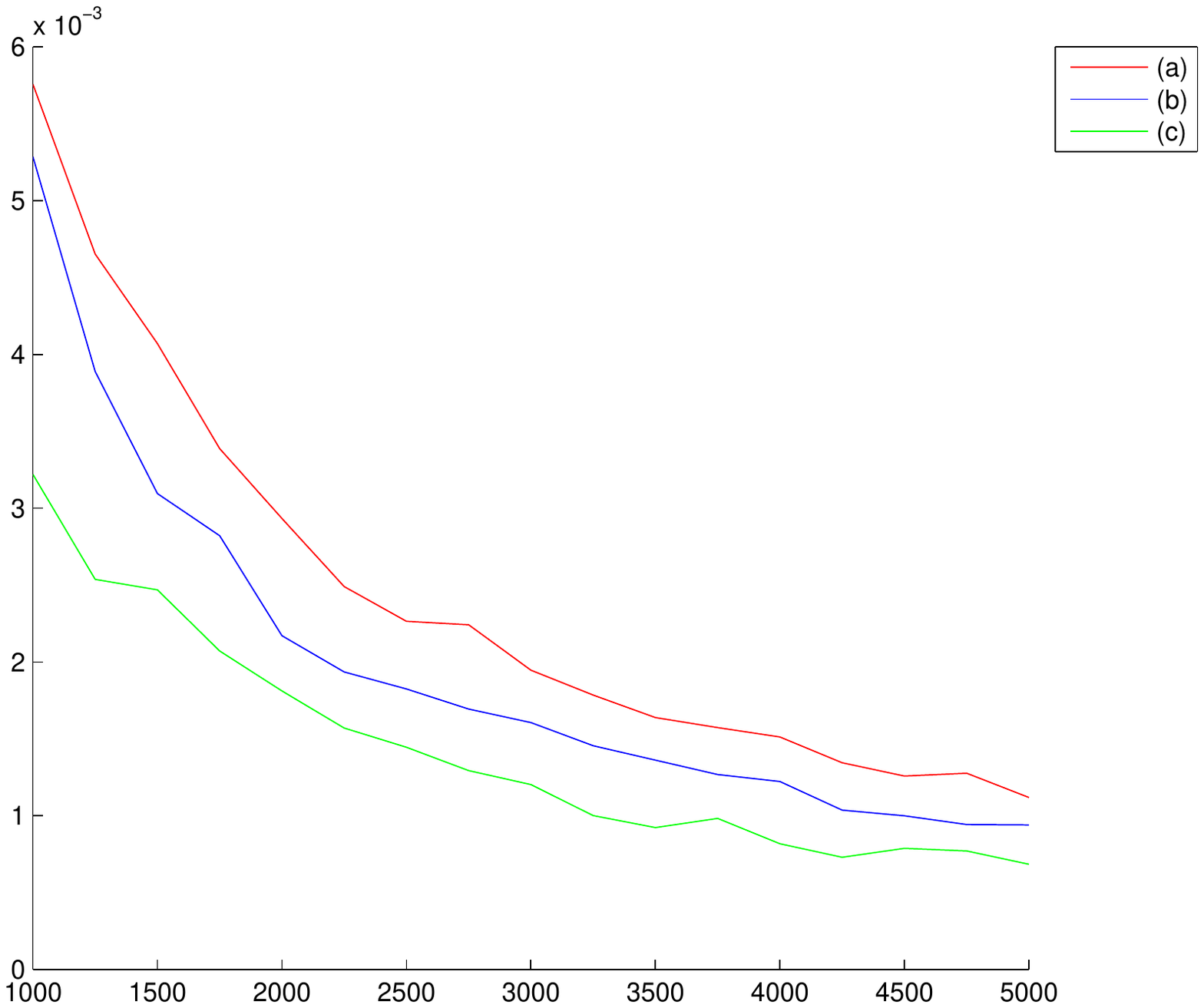}
\hspace{0.1cm}
\includegraphics[width=0.35\columnwidth]{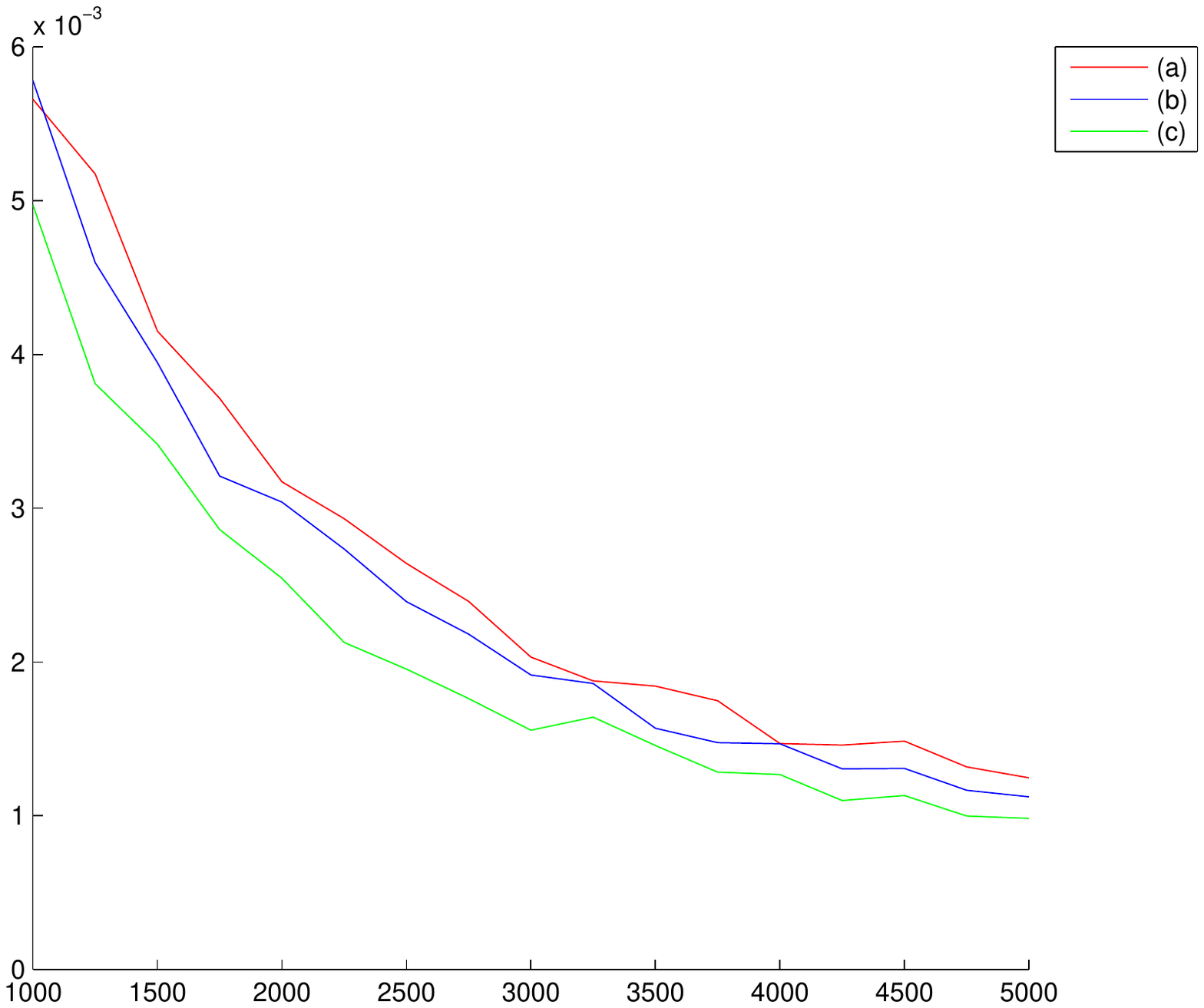}
\hspace{0.1cm}
\includegraphics[width=0.35\columnwidth]{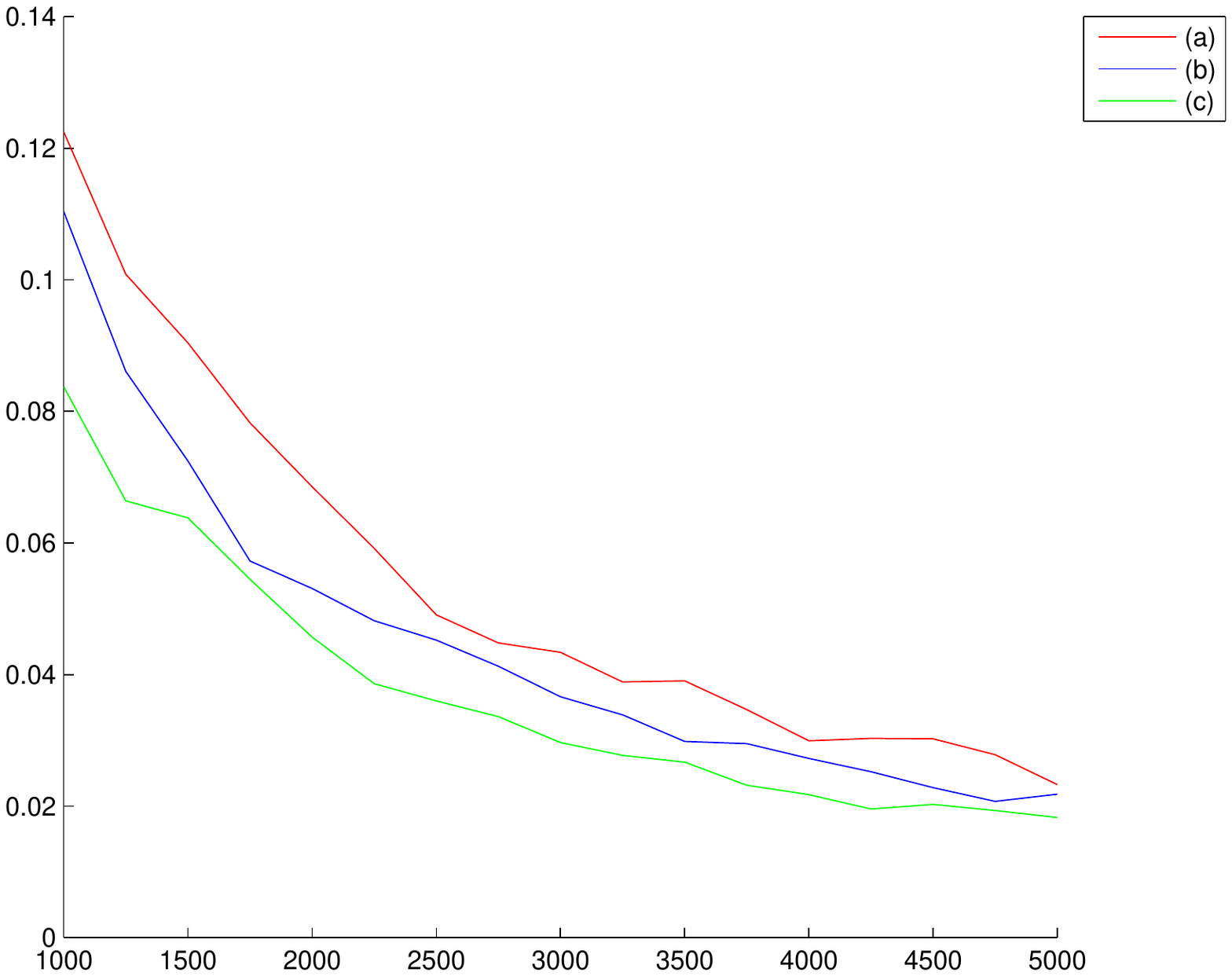}
\hspace{0.1cm}
\includegraphics[width=0.35\columnwidth]{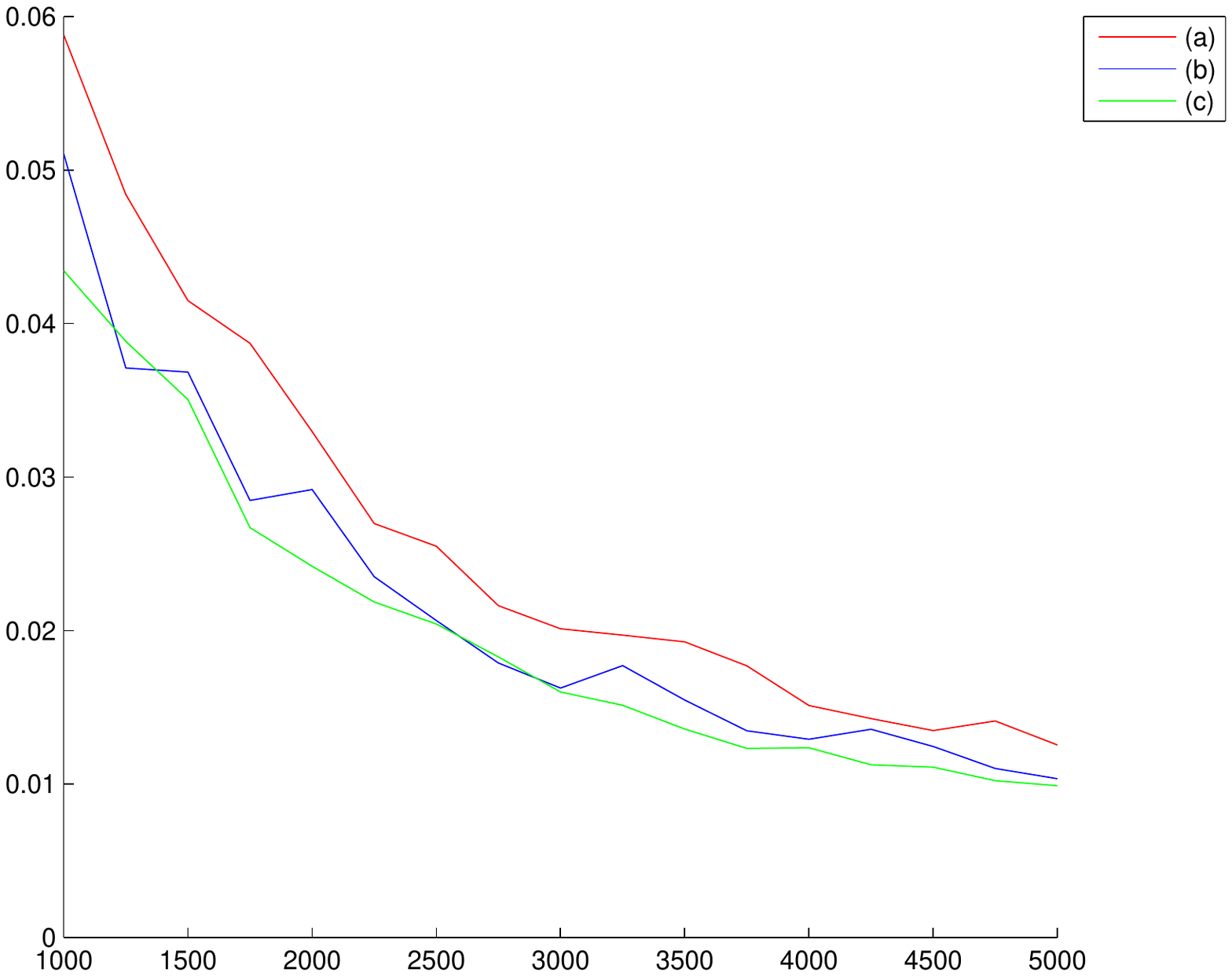}
\hspace{0.1cm}
\end{center}
\caption{Variance of  $\hat{B}_{T}$, $T \in [1000, 5000]$, with discretization step size 250, for cases: (a) $\alpha=0.85$, (b) $\alpha=1.50$, (c) $\alpha=2.50$, case H1 (top-left), case H2 (top-right), case H3 (bottom-left) and case H4 (bottom-right).}\label{Fig.5}
\end{figure}

\begin{figure}[hptb]
\begin{center}
\includegraphics[width=0.35\columnwidth]{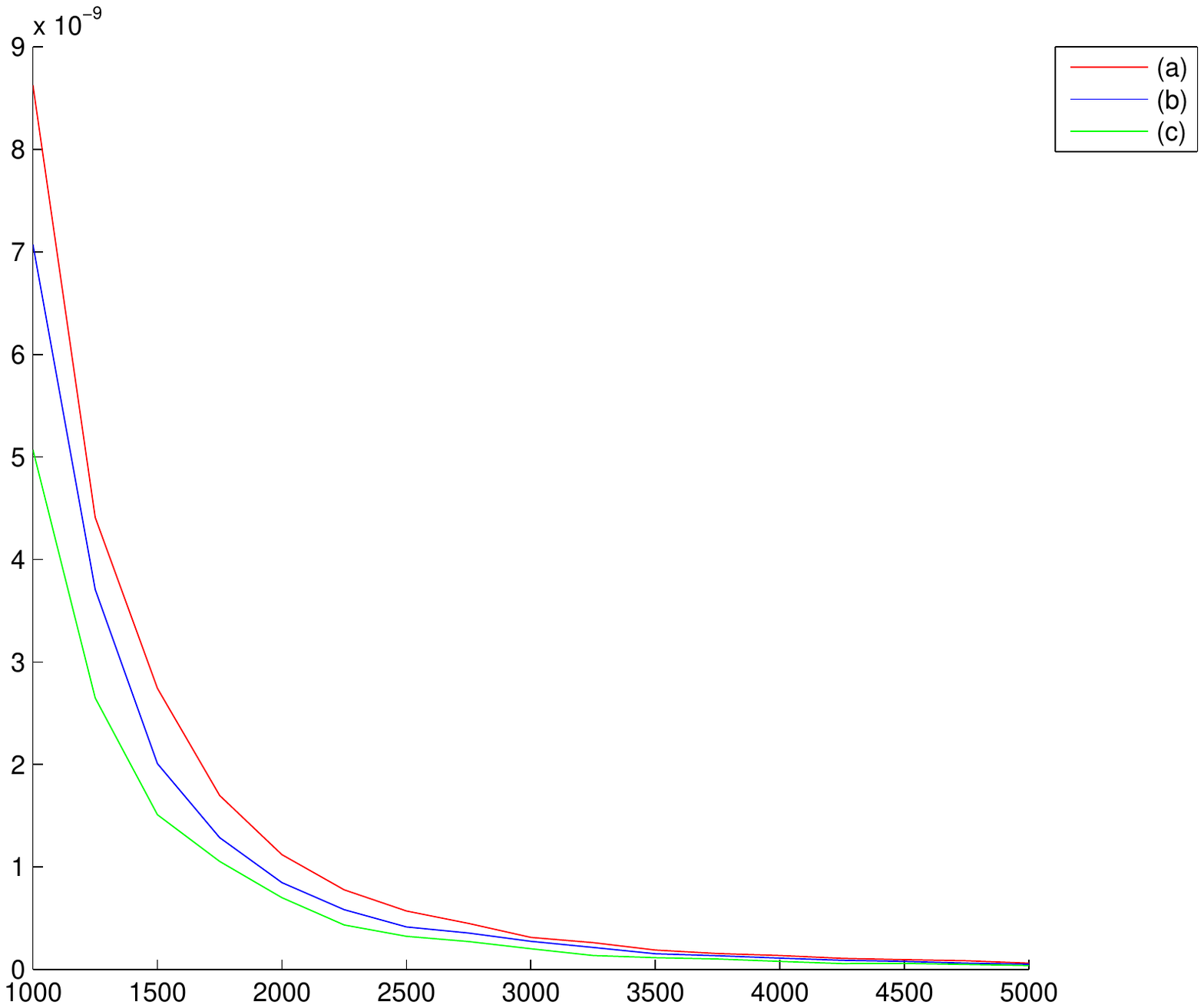}
\hspace{0.1cm}
\includegraphics[width=0.35\columnwidth]{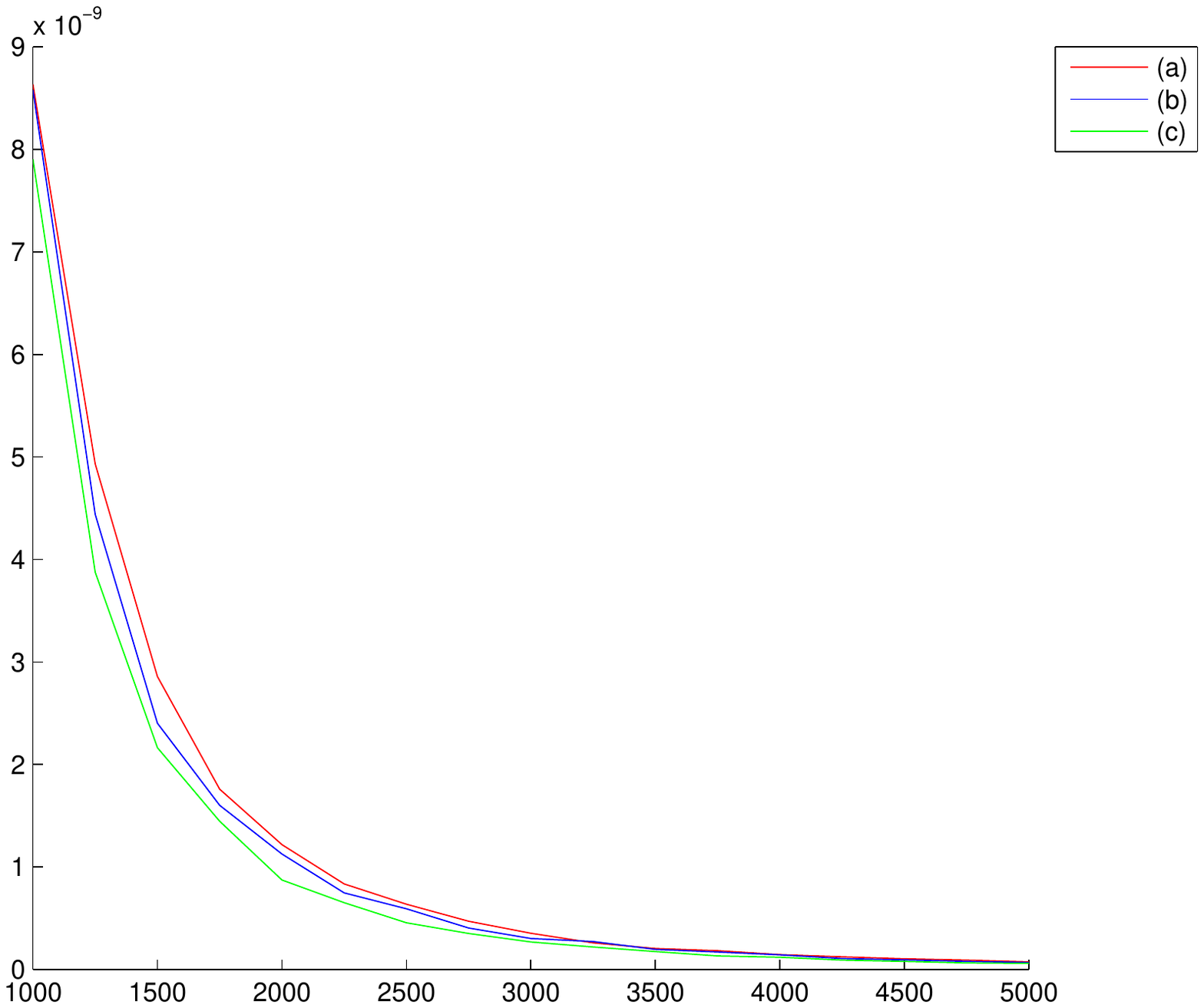}
\hspace{0.1cm}
\includegraphics[width=0.35\columnwidth]{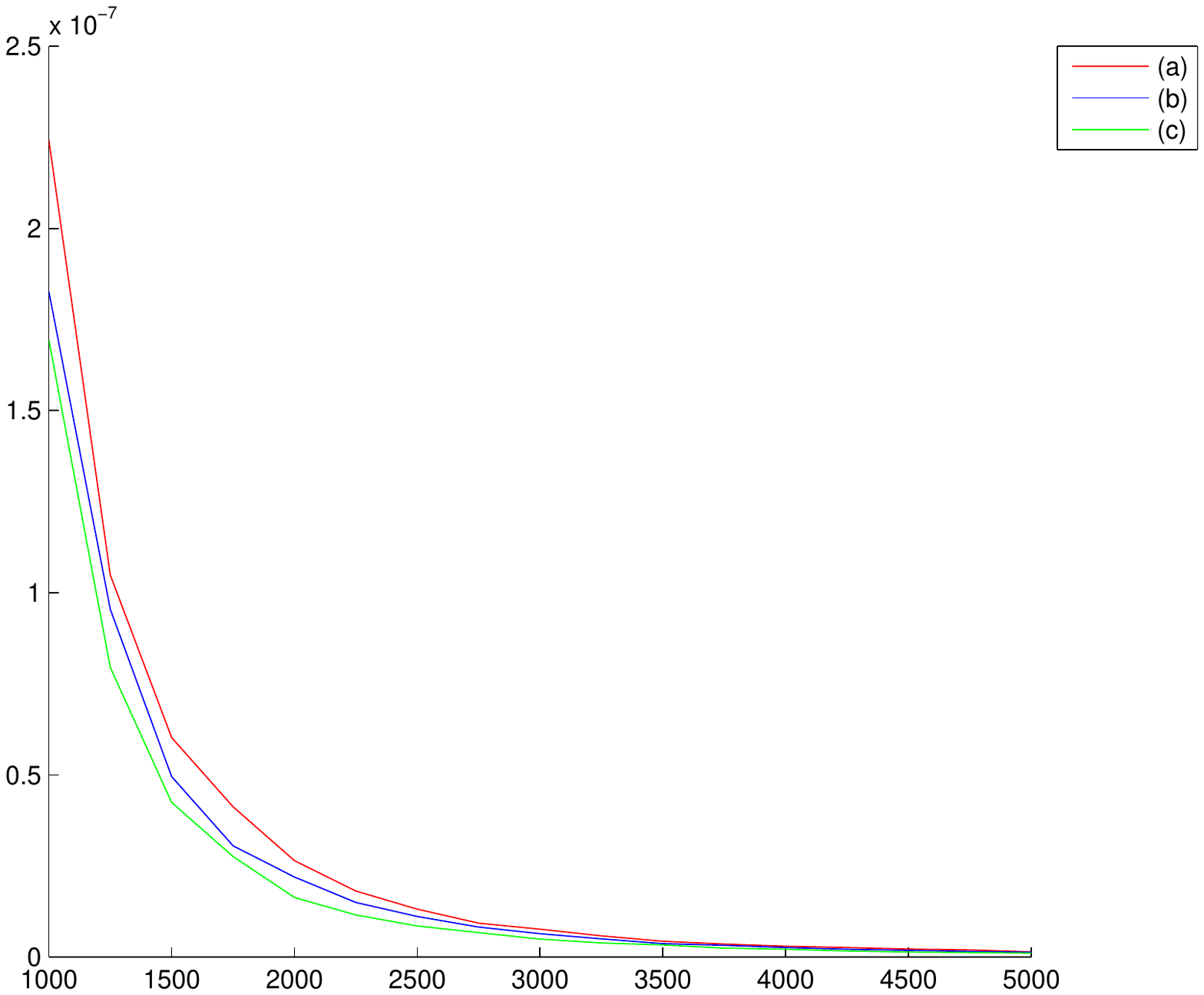}
\hspace{0.1cm}
\includegraphics[width=0.35\columnwidth]{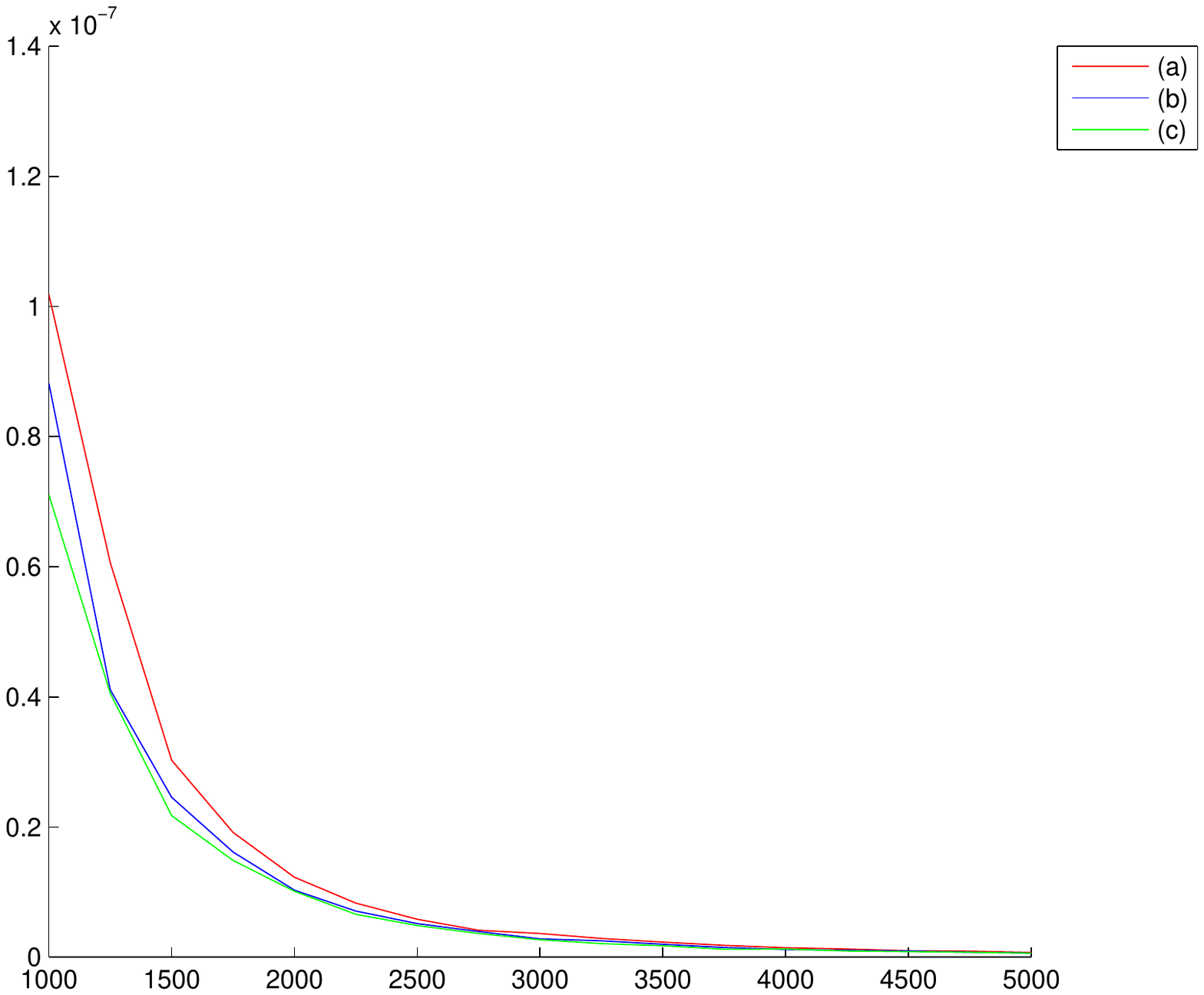}
\hspace{0.1cm}
\end{center}
\caption{Variance of $\hat{\varphi}_{T}$, $T \in [1000, 5000]$, with discretization step size 250, for cases: (a) $\alpha=0.85$, (b) $\alpha=1.50$, (c) $\alpha=2.50$, case H1 (top-left), case H2 (top-right), case H3 (bottom-left) and case H4 (bottom-right).}\label{Fig.6}
\end{figure}

\begin{table*}
\centering
\caption{Rejection rates of Henze-Zirkler's (T1) and Doornik-Hansen (T2) MVN tests applied to the simulated random vectors ($\hat{\zeta}_{T}^{A}$, $\hat{\zeta}_{T}^{B}$ $\hat{\zeta}_{T}^{\varphi}$) for different $T$ values. }
\label{Tabla.1}
\begin{tabular*}{\textwidth}{@{\extracolsep{\fill}}ccccccccc}
\hline

&&&\multicolumn{6}{c}{$T$} \\\cline{4-9}
$\alpha$&Case&Test &$1000$&$5000$& $10000$&$15000$&$20000$& $30000$ \\\hline

0.85&H1&T1&0.00&0.04&0.00&0.00&0.00&0.00\\
      &&T2&0.02&0.00&0.00&0.00&0.00&0.00\\
    &H2&T1&0.02&0.06&0.00&0.00&0.00&0.00\\
      &&T2&0.06&0.02&0.00&0.00&0.00&0.03\\
    &H3&T1&0.06&0.00&0.00&0.00&0.00&0.00\\
      &&T2&0.44&0.04&0.00&0.00&0.00&0.00\\
    &H4&T1&1.00&0.34&0.10&0.07&0.05&0.03\\
      &&T2&1.00&0.82&0.40&0.27&0.15&0.03\\\hline
1.50&H1&T1&0.00&0.02&0.00&0.00&0.00&0.00\\
      &&T2&0.00&0.00&0.00&0.00&0.00&0.00\\
    &H2&T1&0.02&0.00&0.00&0.00&0.00&0.00\\
      &&T2&0.14&0.02&0.00&0.00&0.00&0.00\\
    &H3&T1&0.08&0.04&0.00&0.00&0.05&0.00\\
      &&T2&0.30&0.00&0.00&0.00&0.00&0.03\\
    &H4&T1&1.00&0.22&0.00&0.07&0.00&0.00\\
      &&T2&1.00&0.82&0.40&0.13&0.05&0.00\\\hline
2.50&H1&T1&0.00&0.00&0.00&0.00&0.00&0.03\\
      &&T2&0.02&0.02&0.00&0.07&0.00&0.07\\
    &H2&T1&0.02&0.02&0.00&0.00&0.00&0.00\\
      &&T2&0.02&0.02&0.00&0.00&0.00&0.00\\
    &H3&T1&0.02&0.02&0.00&0.00&0.00&0.00\\
      &&T2&0.18&0.00&0.00&0.07&0.05&0.03\\
    &H4&T1&0.96&0.18&0.00&0.00&0.00&0.00\\
      &&T2&1.00&0.68&0.10&0.07&0.00&0.03\\\hline
\end{tabular*}
\end{table*}

\subsection{Simulation experiment 2}
\begin{figure}[hptb]
\begin{center}
\subfigure[$\alpha=0.25$, H1 ]{\includegraphics[width=0.35\columnwidth]{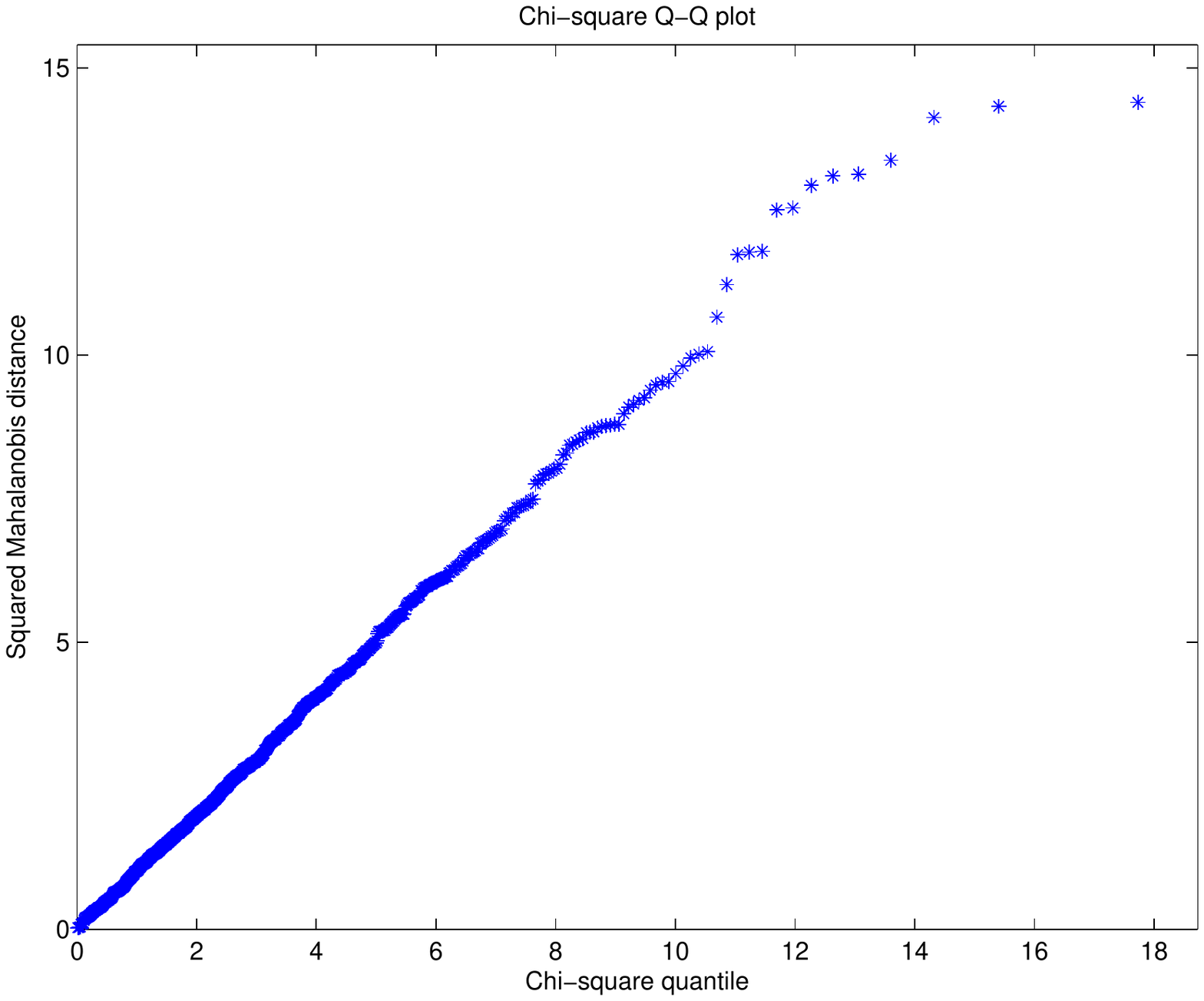}}
\hspace{0.1cm}
\subfigure[$\alpha=0.25$, H2 ]{\includegraphics[width=0.35\columnwidth]{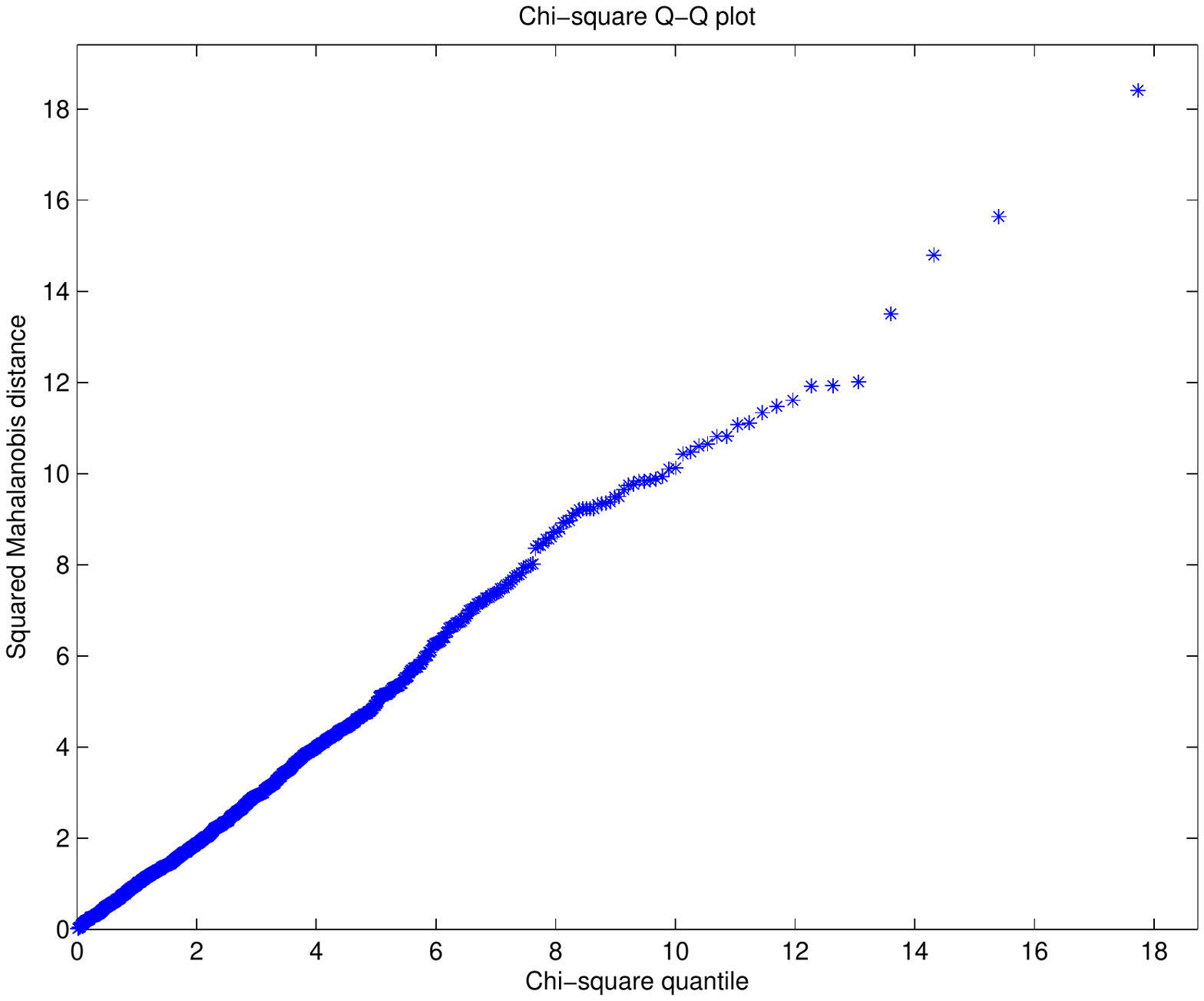}}
\hspace{0.1cm}
\subfigure[$\alpha=0.25$, H3 ]{\includegraphics[width=0.35\columnwidth]{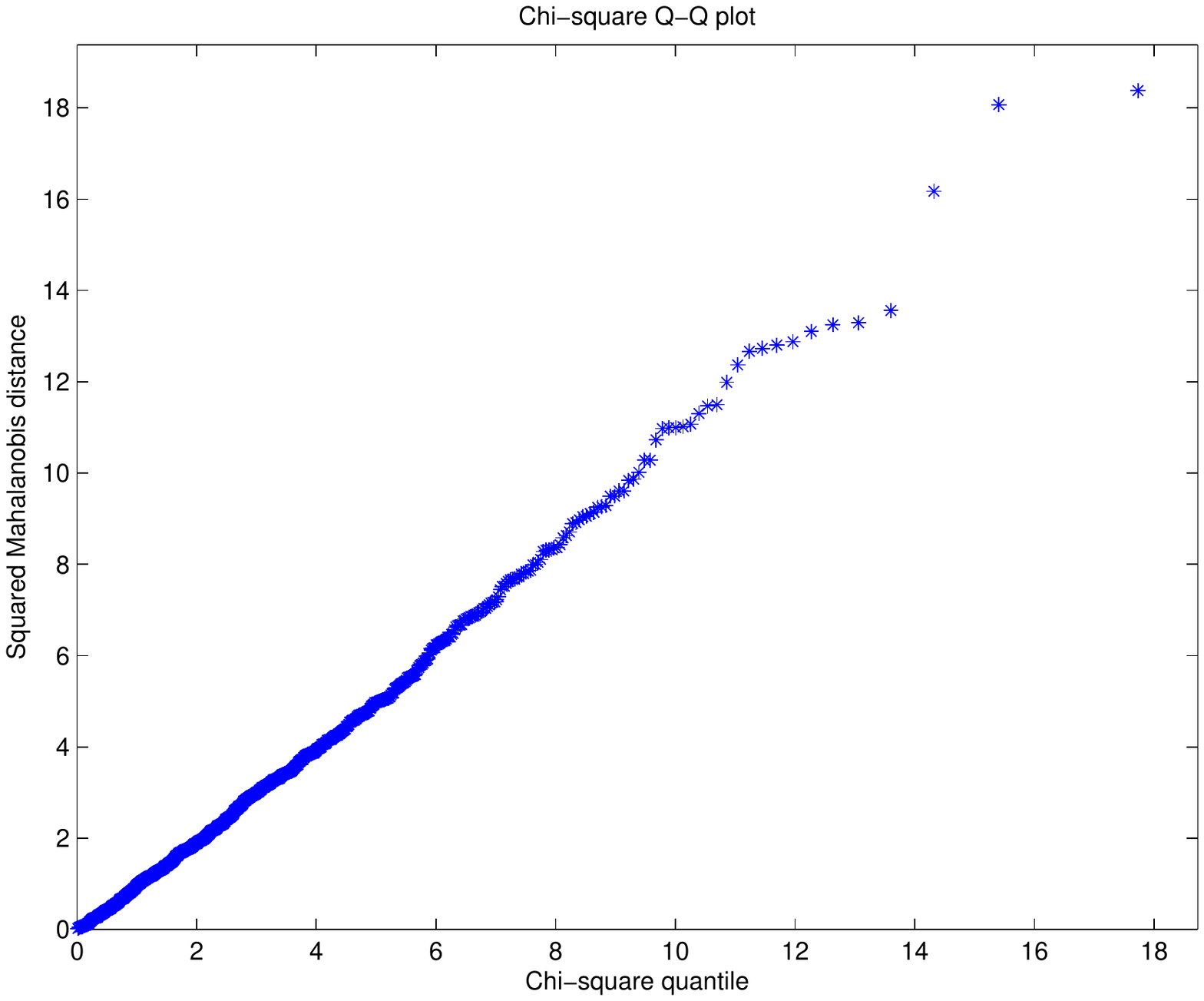}}
\hspace{0.1cm}
\subfigure[$\alpha=0.25$, H4 ]{\includegraphics[width=0.35\columnwidth]{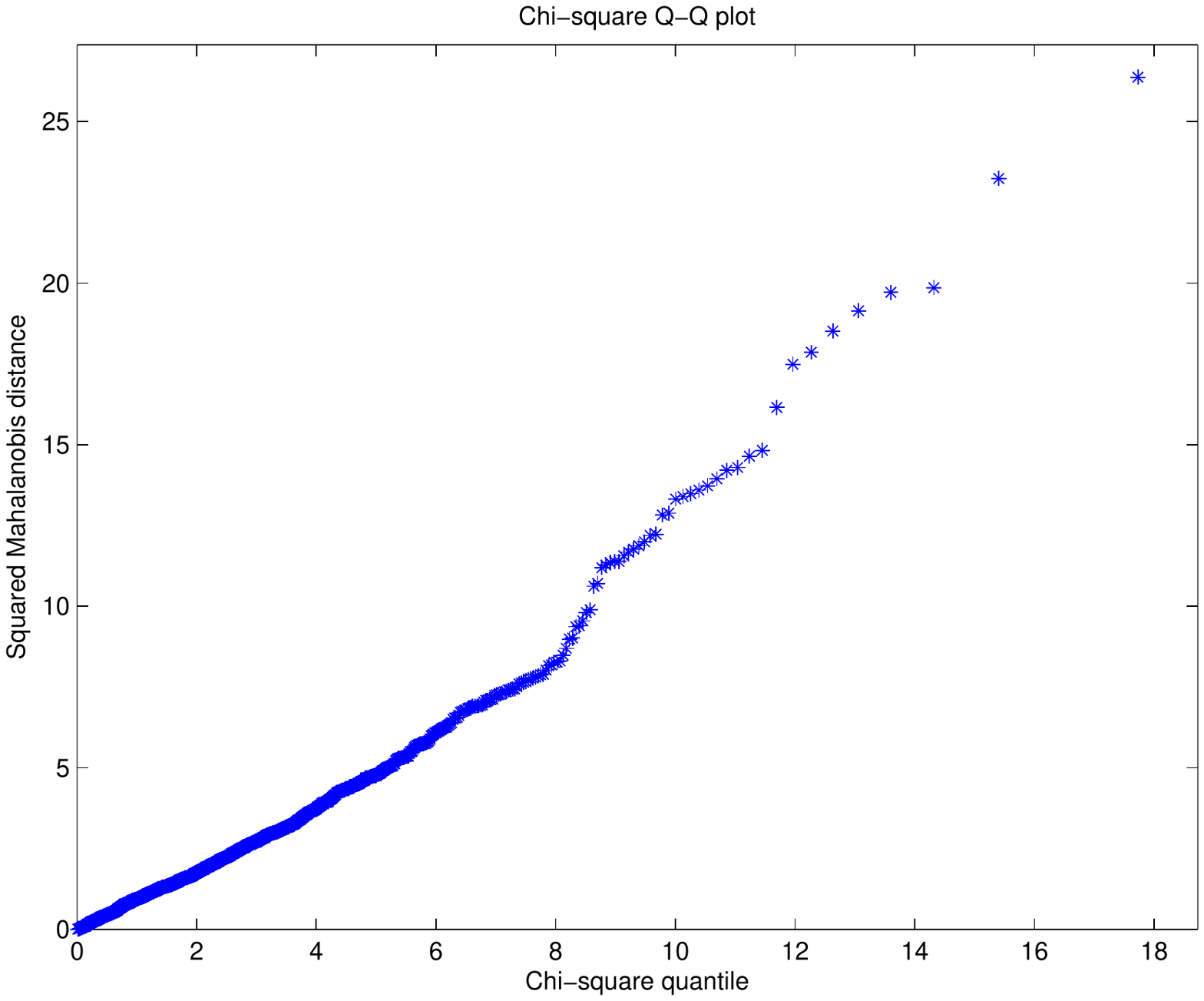}}
\subfigure[$\alpha=0.45$, H1 ]{\includegraphics[width=0.35\columnwidth]{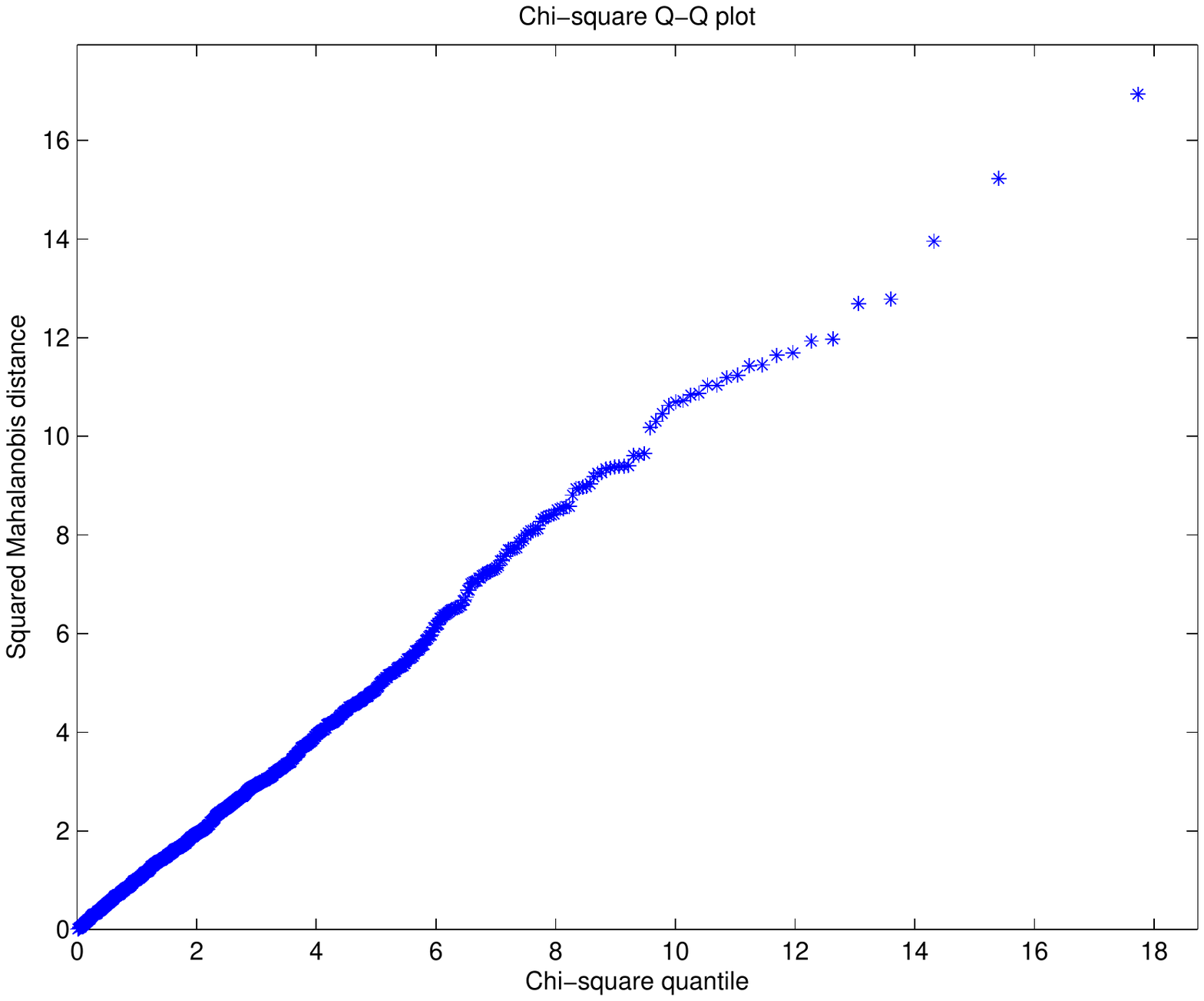}}
\hspace{0.1cm}
\subfigure[$\alpha=0.45$, H2 ]{\includegraphics[width=0.35\columnwidth]{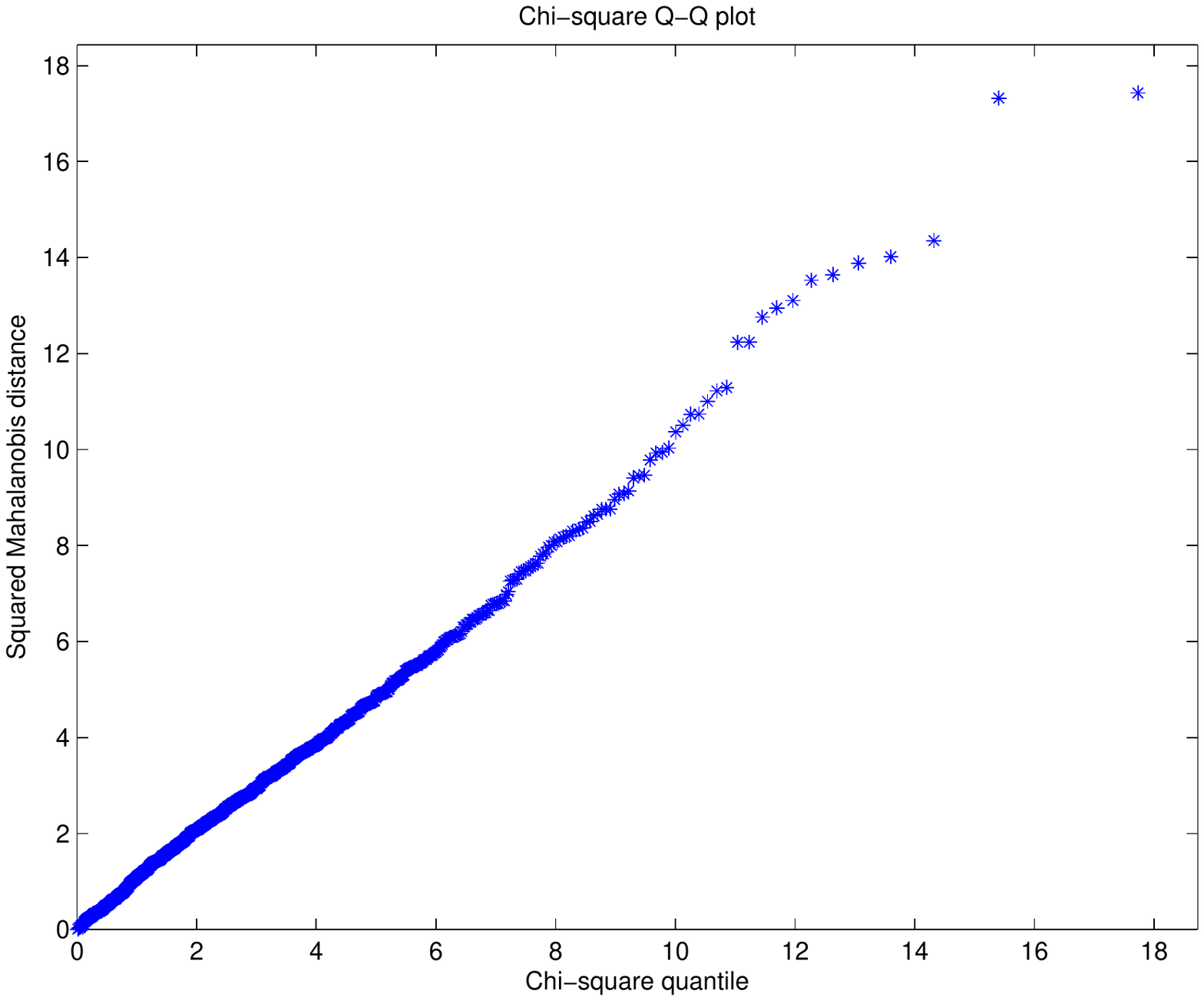}}
\hspace{0.1cm}
\subfigure[$\alpha=0.45$, H3 ]{\includegraphics[width=0.35\columnwidth]{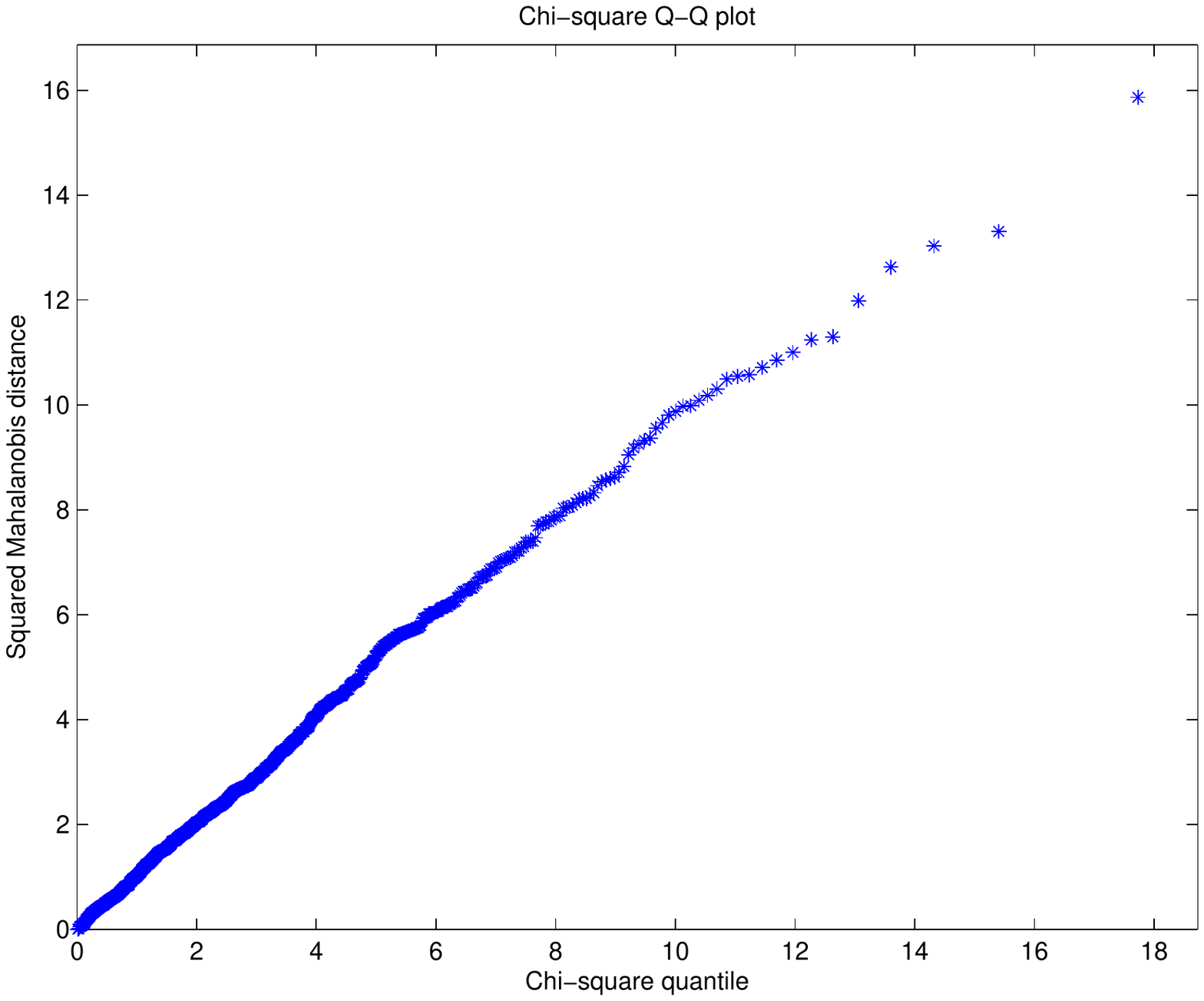}}
\hspace{0.1cm}
\subfigure[$\alpha=0.45$, H4 ]{\includegraphics[width=0.35\columnwidth]{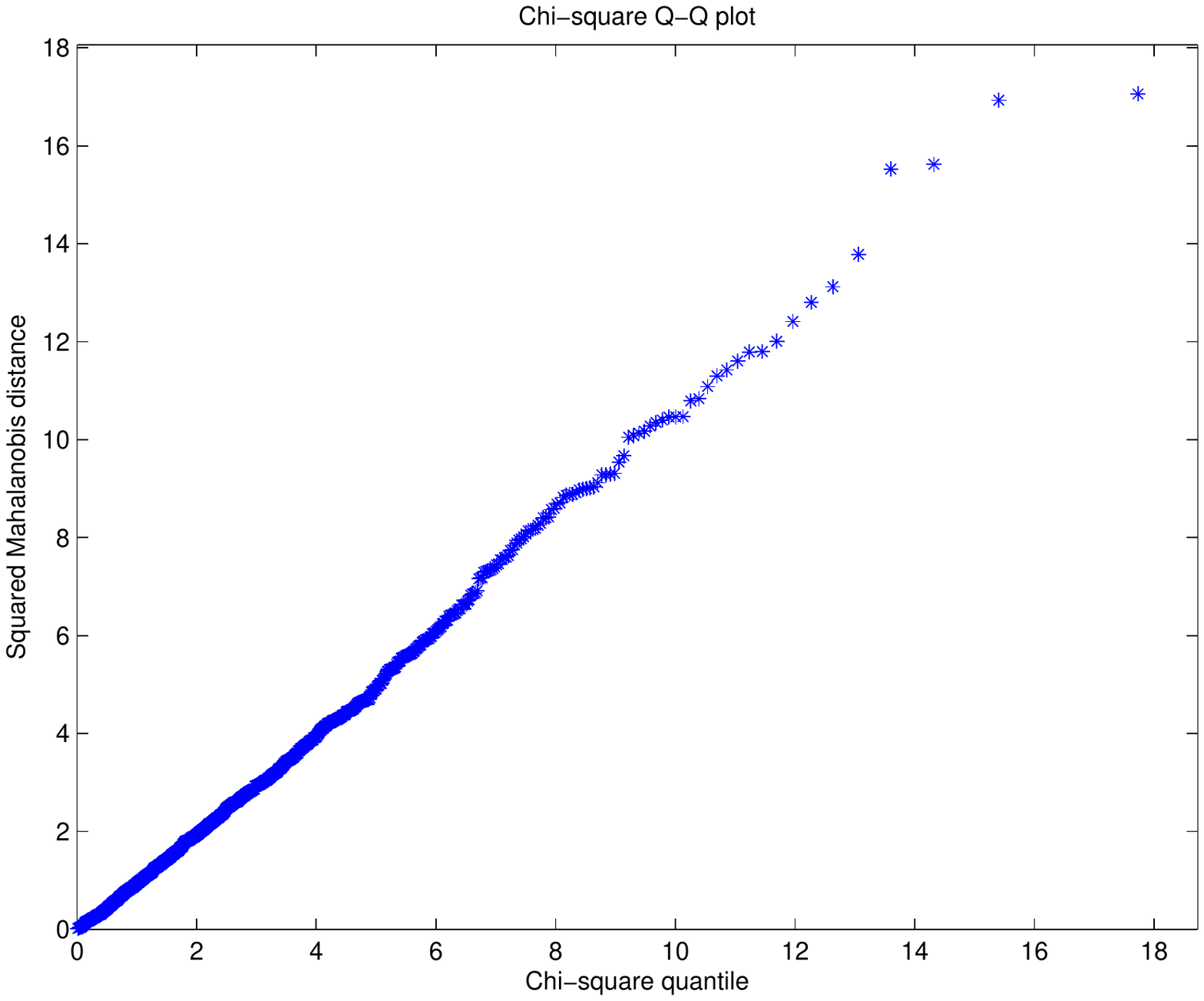}}
\caption{Chi-square Q-Q plot of the squared Mahalanobis distances of the simulated random vector ($\hat{\zeta}_{T}^{A}$, $\hat{\zeta}_{T}^{B}$,  $\hat{\zeta}_{T}^{\varphi}$), $T=30000$.}\label{Fig.7}
\end{center}
\end{figure}

\begin{figure}[hptb]
\begin{center}
\subfigure[$\alpha=0.25$, H1 ]{\includegraphics[width=0.35\columnwidth]{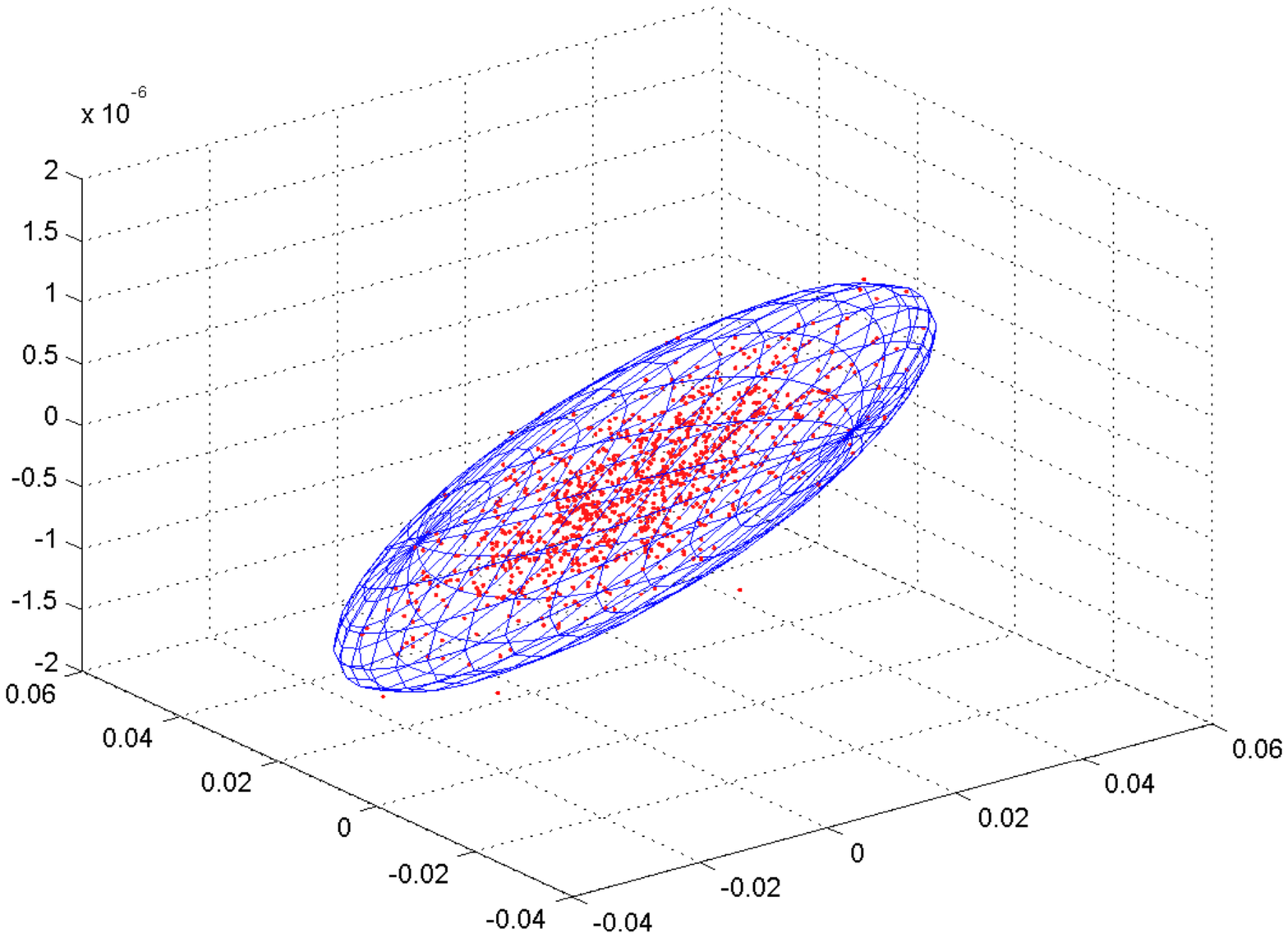}}
\hspace{0.1cm}
\subfigure[$\alpha=0.25$, H2 ]{\includegraphics[width=0.35\columnwidth]{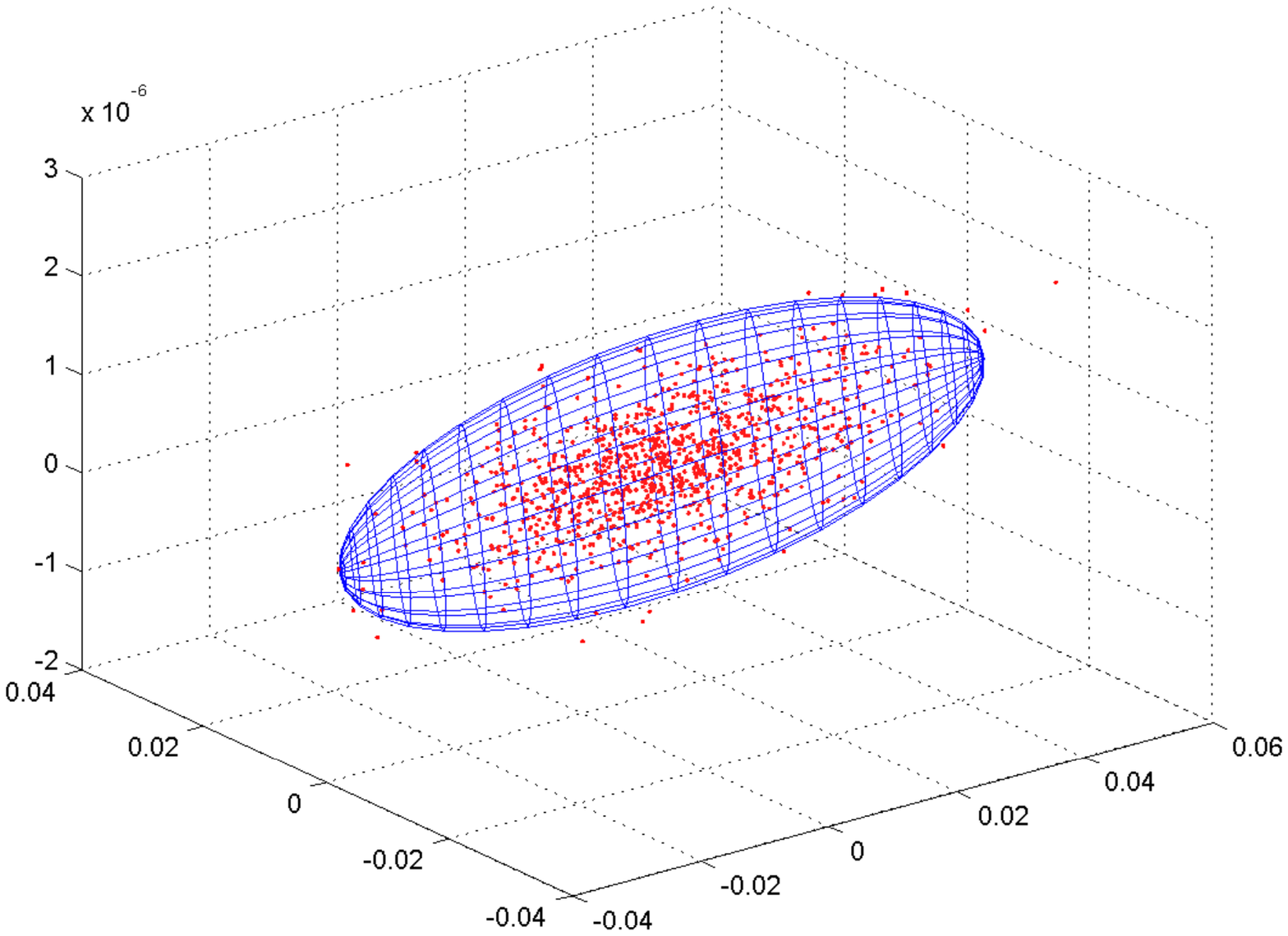}}
\hspace{0.1cm}
\subfigure[$\alpha=0.25$, H3 ]{\includegraphics[width=0.35\columnwidth]{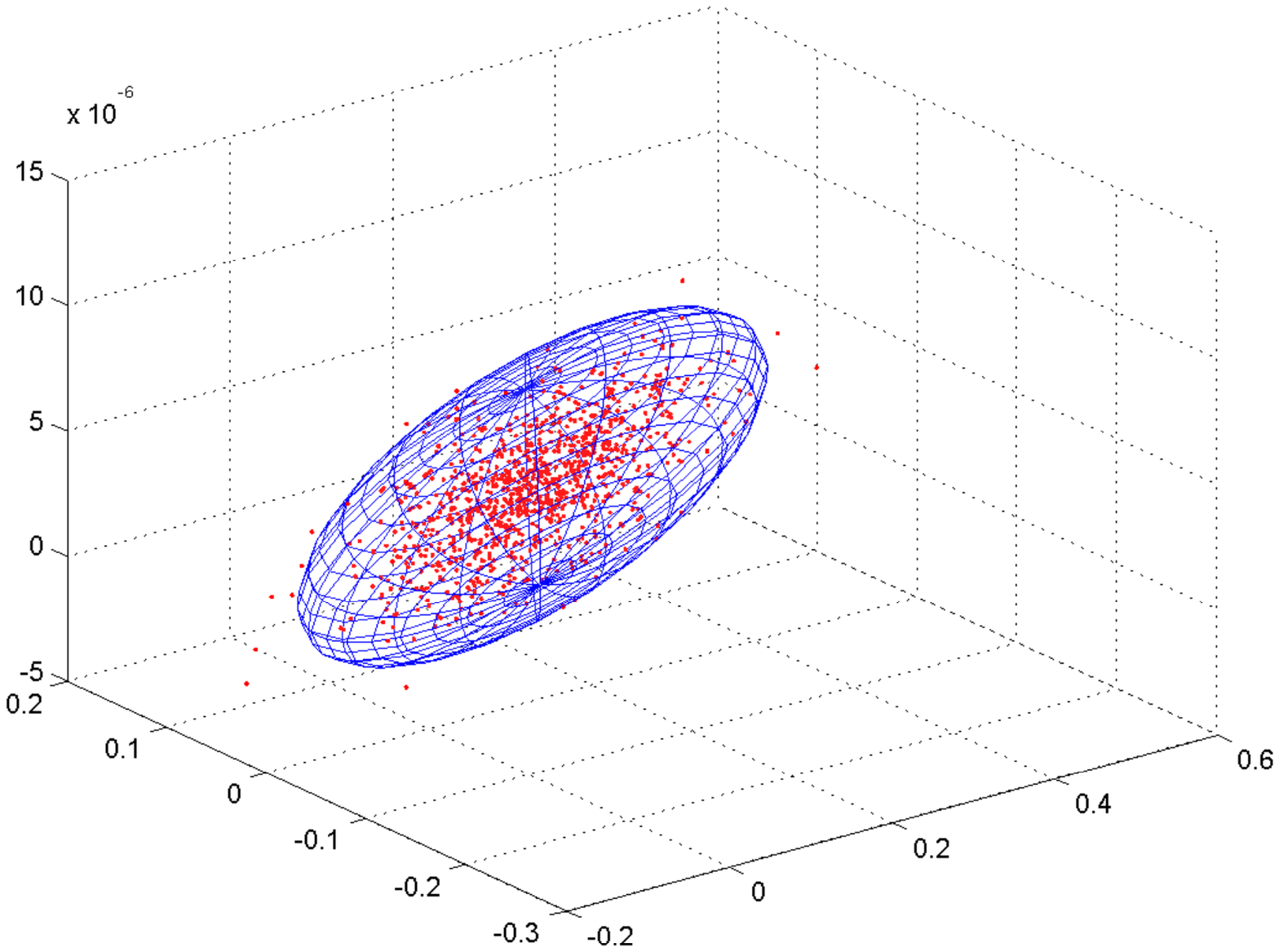}}
\hspace{0.1cm}
\subfigure[$\alpha=0.25$, H4 ]{\includegraphics[width=0.35\columnwidth]{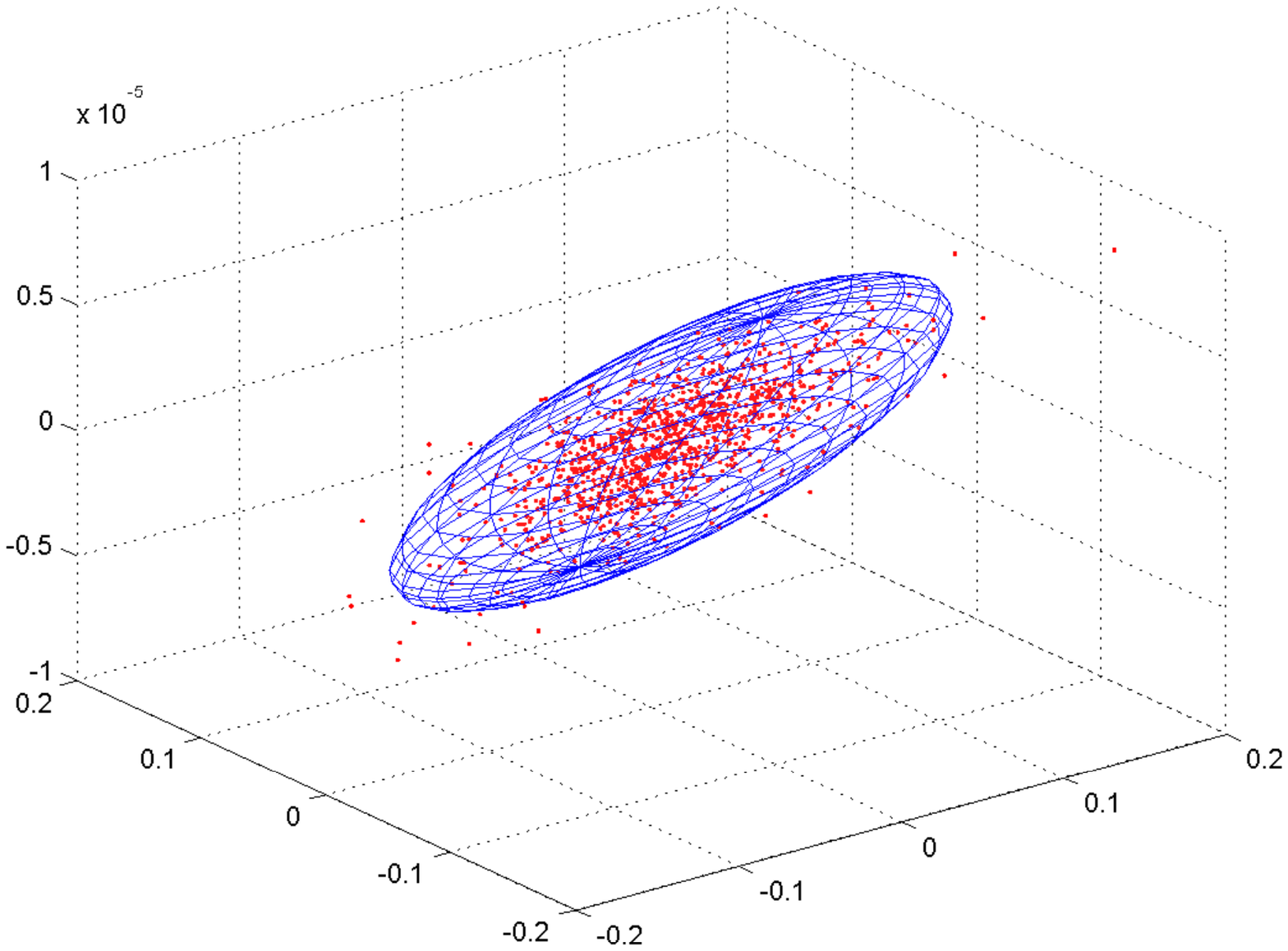}}
\subfigure[$\alpha=0.45$, H1 ]{\includegraphics[width=0.35\columnwidth]{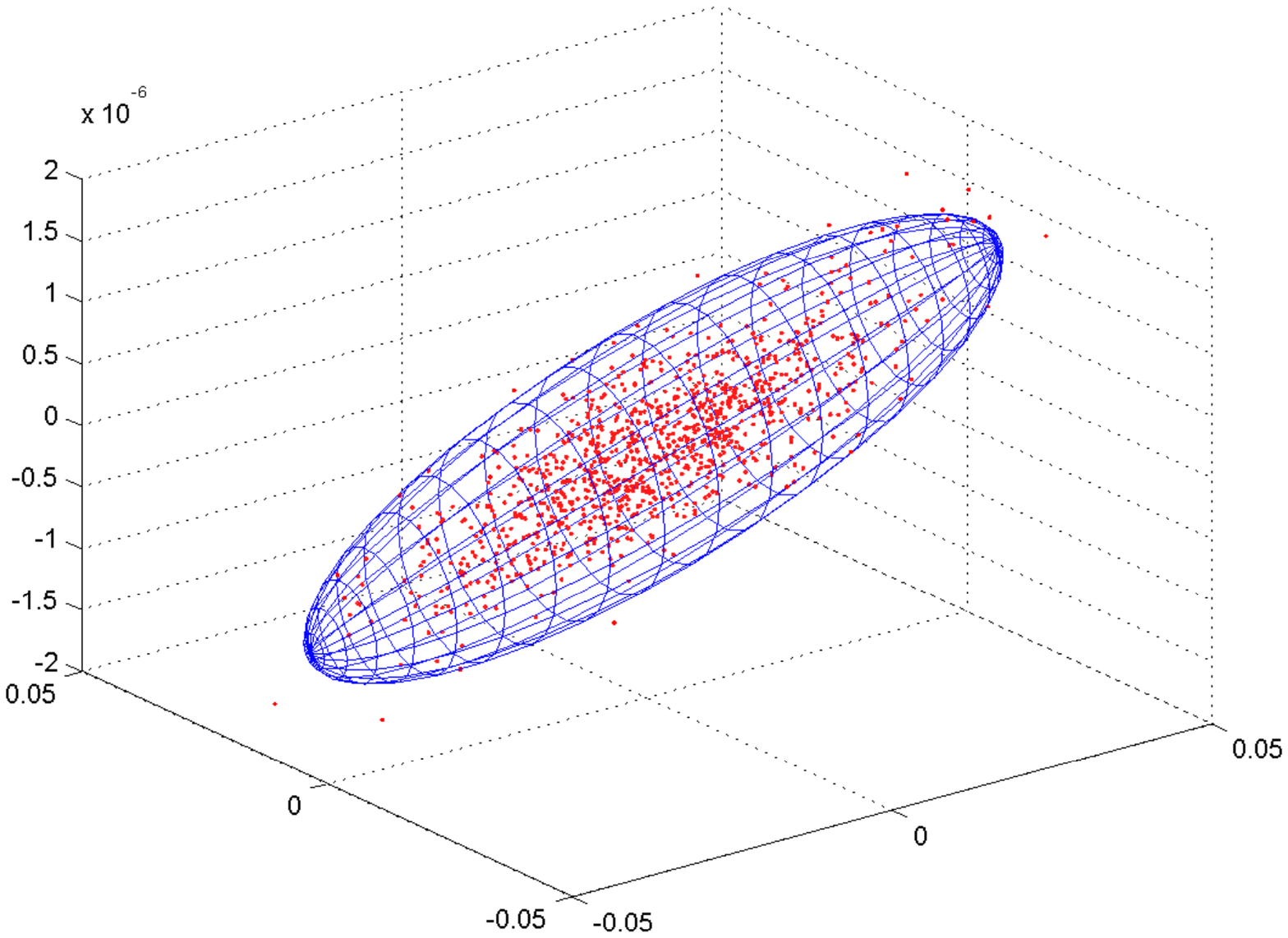}}
\hspace{0.1cm}
\subfigure[$\alpha=0.45$, H2 ]{\includegraphics[width=0.35\columnwidth]{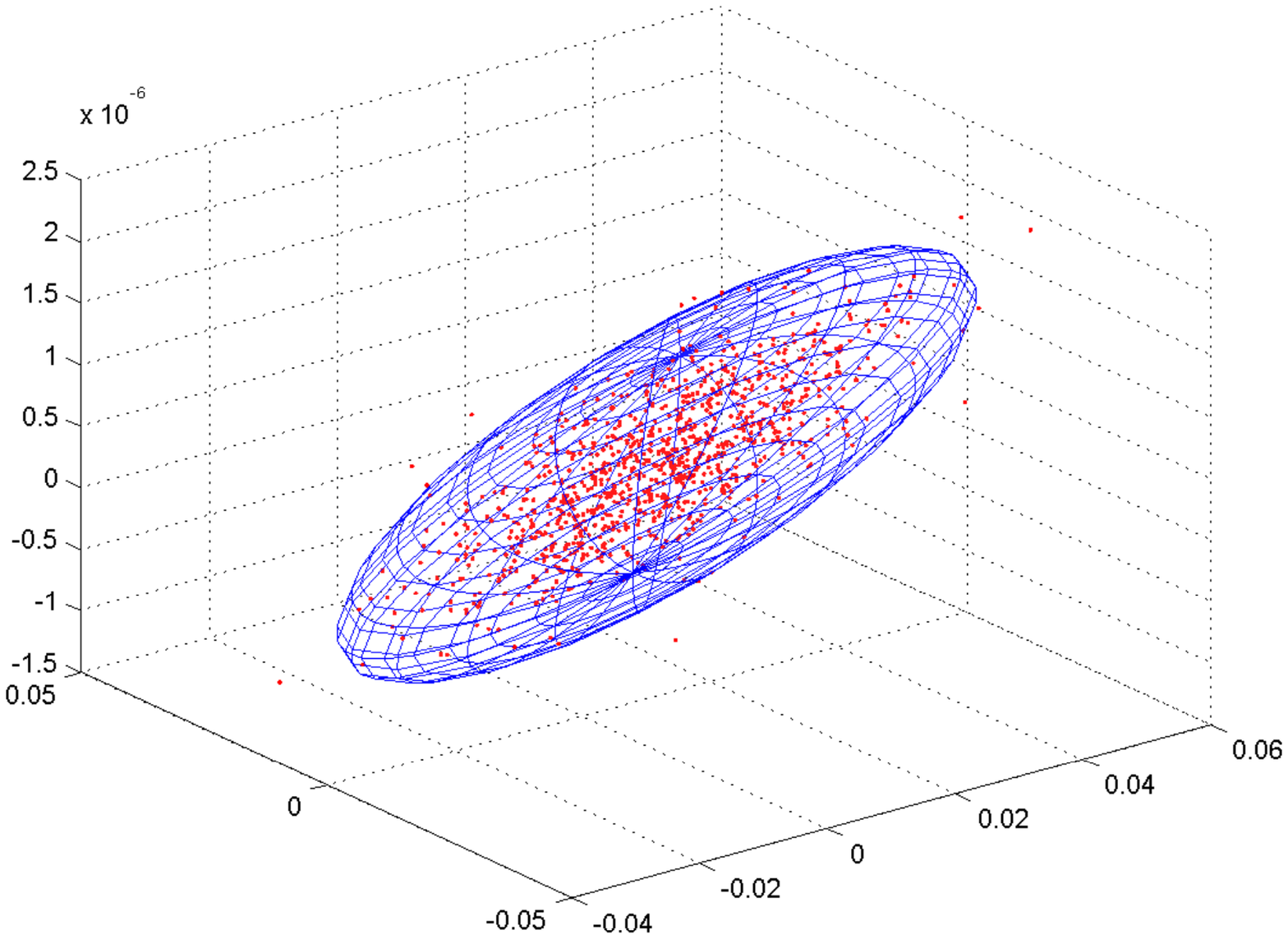}}
\hspace{0.1cm}
\subfigure[$\alpha=0.45$, H3 ]{\includegraphics[width=0.35\columnwidth]{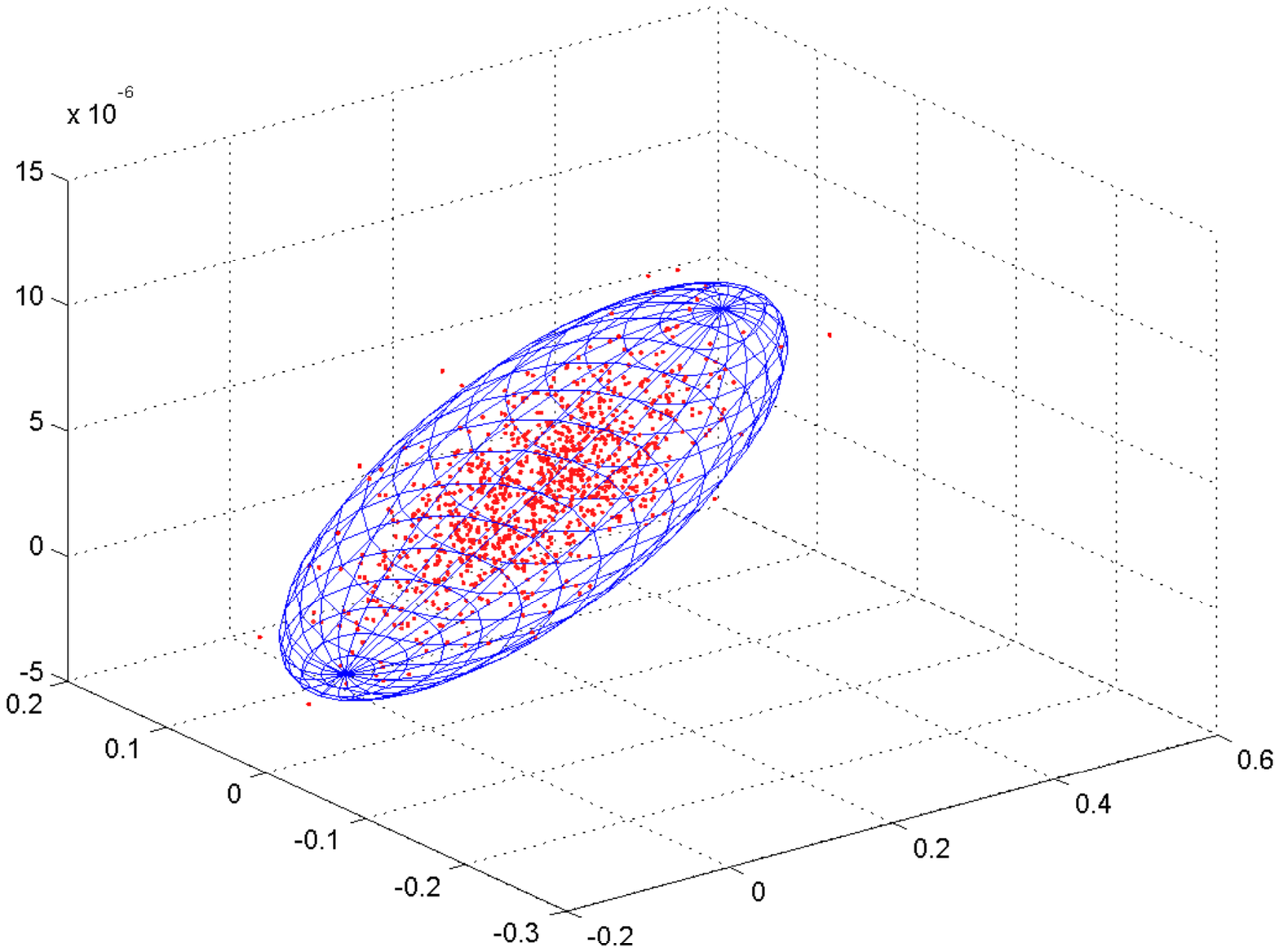}}
\hspace{0.1cm}
\subfigure[$\alpha=0.45$, H4 ]{\includegraphics[width=0.35\columnwidth]{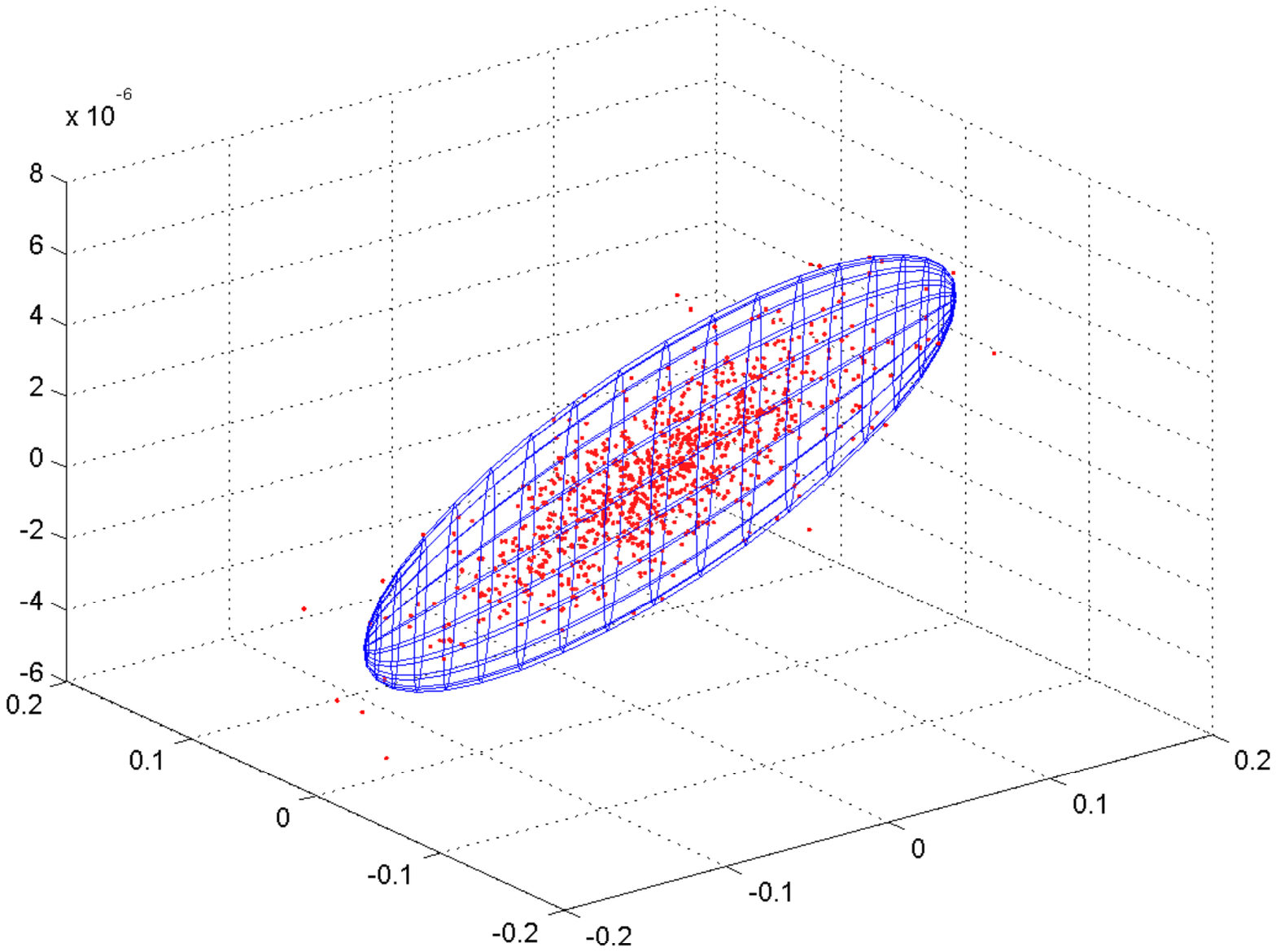}}

\end{center}
\caption{97\% Probability Contours of MVN and scatter plot for ($\hat{\zeta}_{T}^{A}$, $\hat{\zeta}_{T}^{B}$,  $\hat{\zeta}_{T}^{\varphi}$) values, $T=30000$.}\label{Fig.8}
\end{figure}

\begin{figure}[hptb]
\begin{center}
\includegraphics[width=0.35\columnwidth]{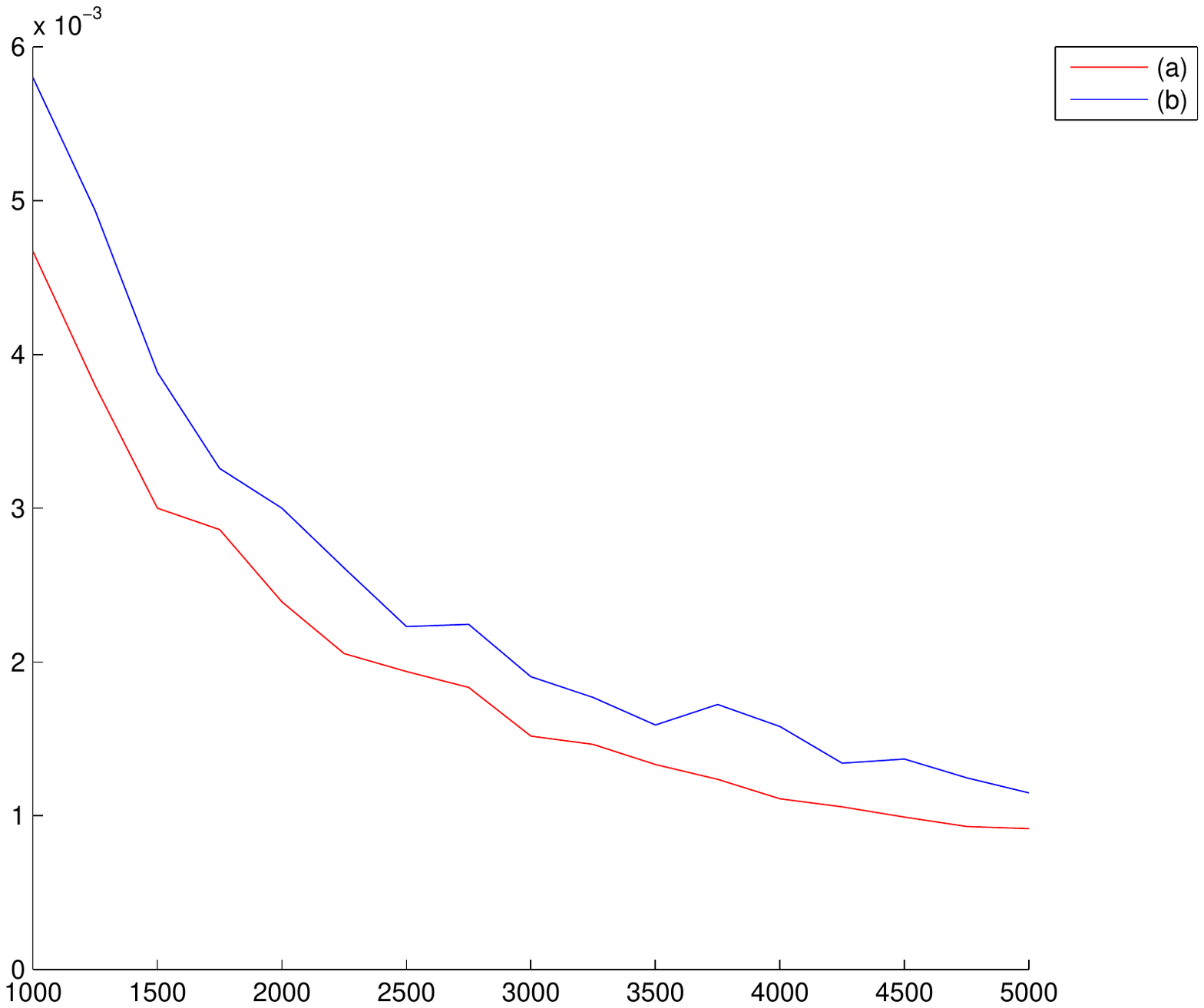}
\hspace{0.1cm}
\includegraphics[width=0.35\columnwidth]{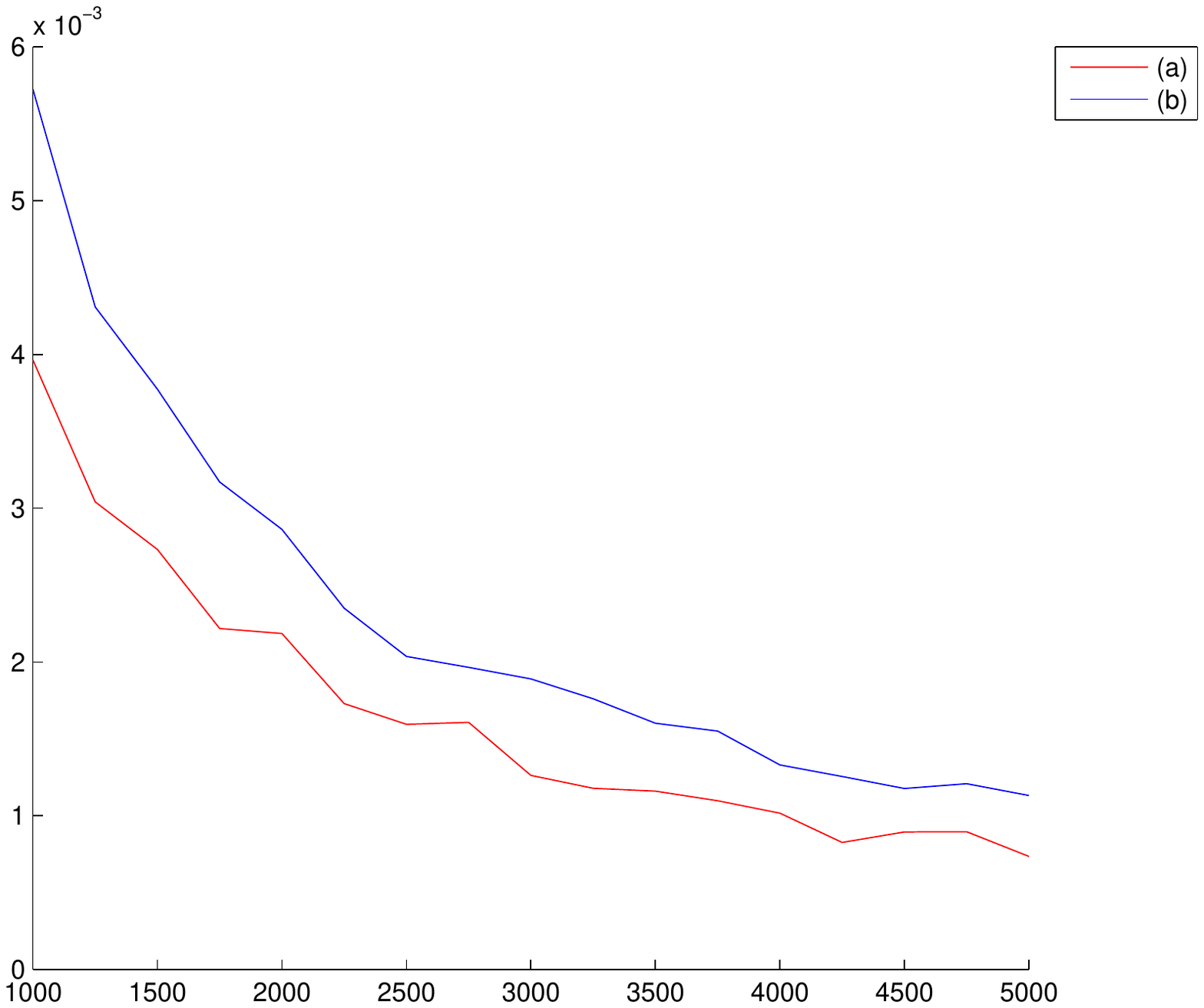}
\hspace{0.1cm}
\includegraphics[width=0.35\columnwidth]{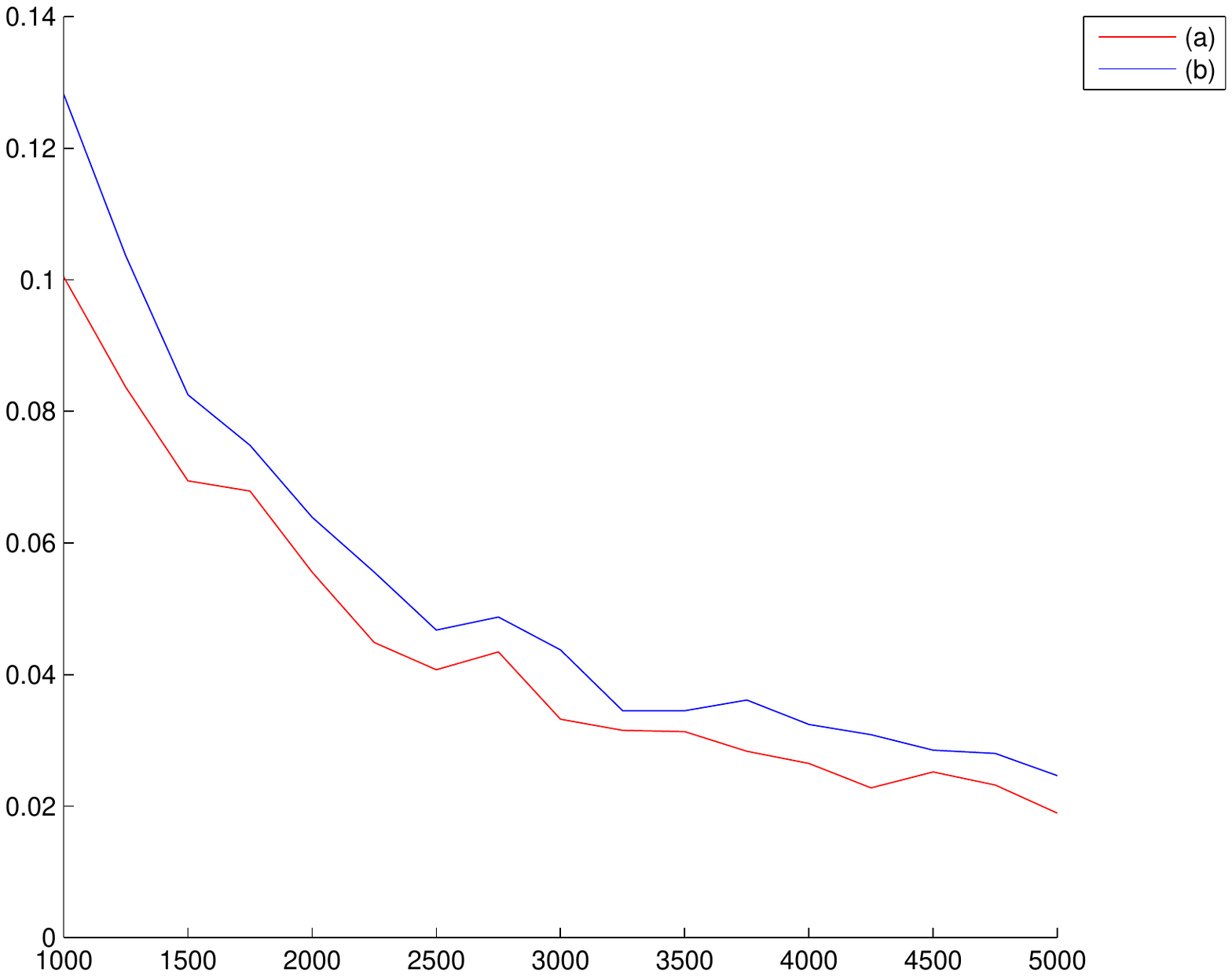}
\hspace{0.1cm}
\includegraphics[width=0.35\columnwidth]{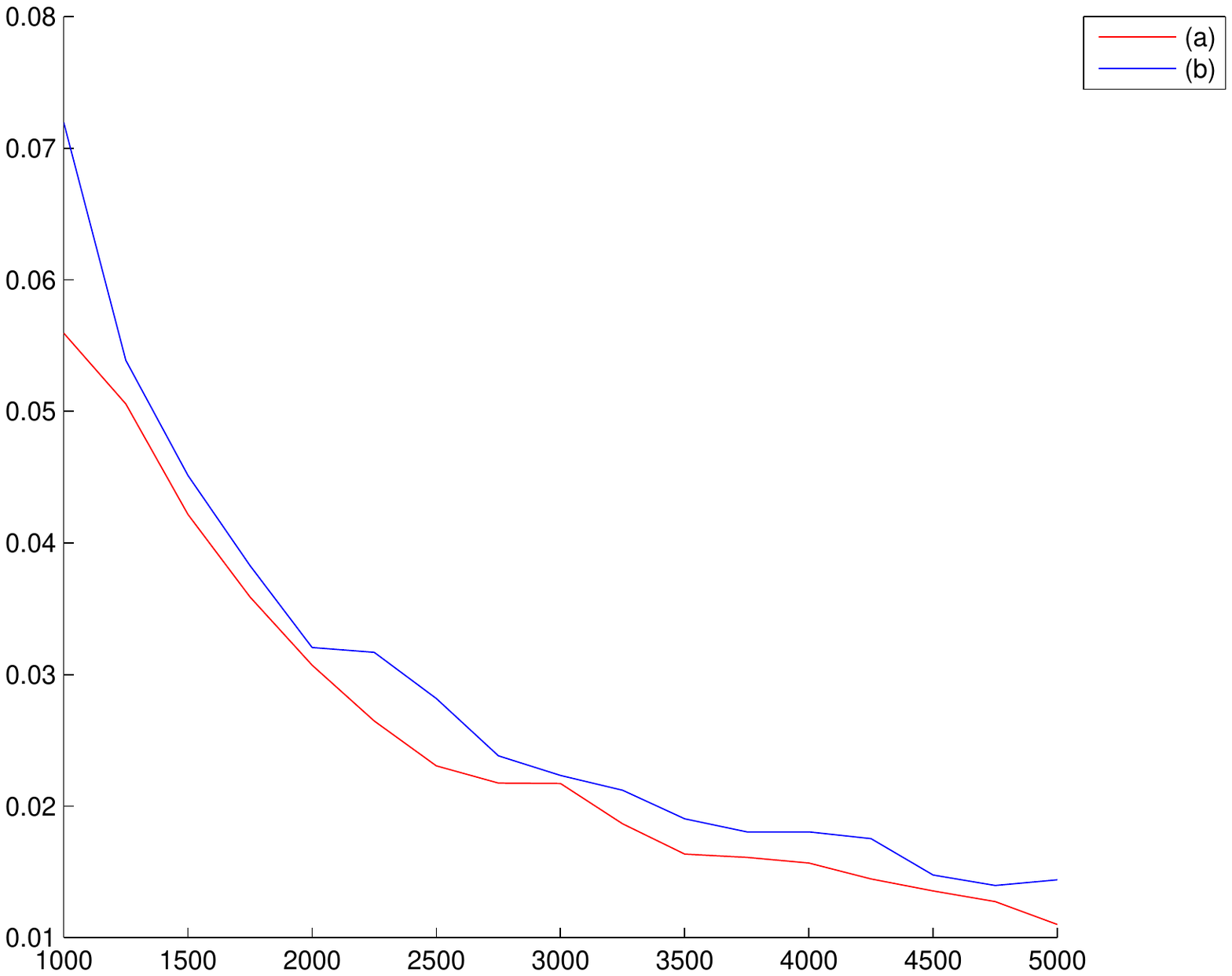}
\hspace{0.1cm}
\end{center}
\caption{Variance of  $\hat{A}_{T}$, $T \in [1000, 5000]$, with discretization step size 250, for cases:  (a) $\alpha=0.25$, (b)  $\alpha=0.45$,case H1 (top-left), case H2 (top-right), case H3 (bottom-left) and case H4 (bottom-right).}\label{Fig.10}
\end{figure}

\begin{figure}[hptb]
\begin{center}
\includegraphics[width=0.35\columnwidth]{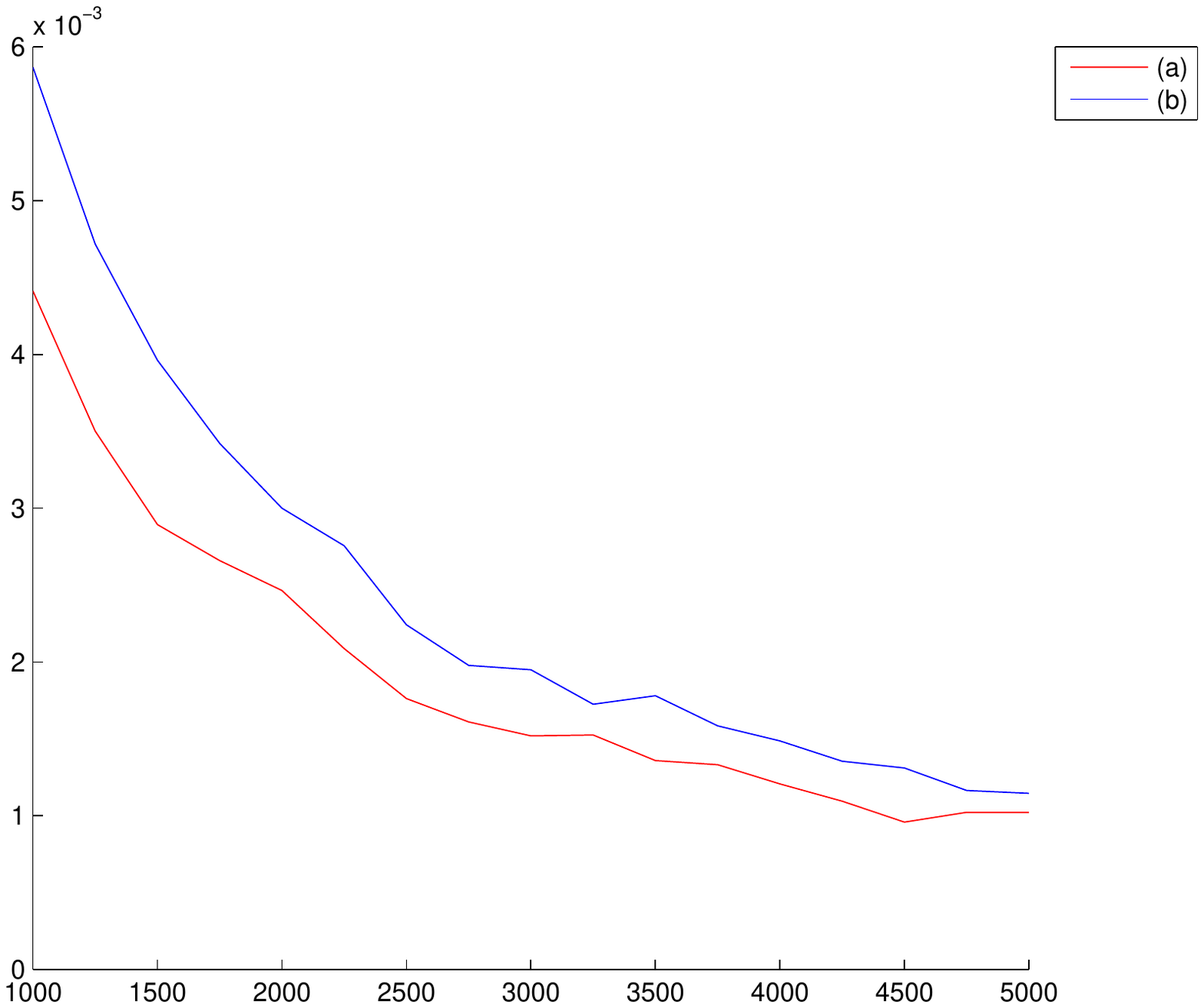}
\hspace{0.1cm}
\includegraphics[width=0.35\columnwidth]{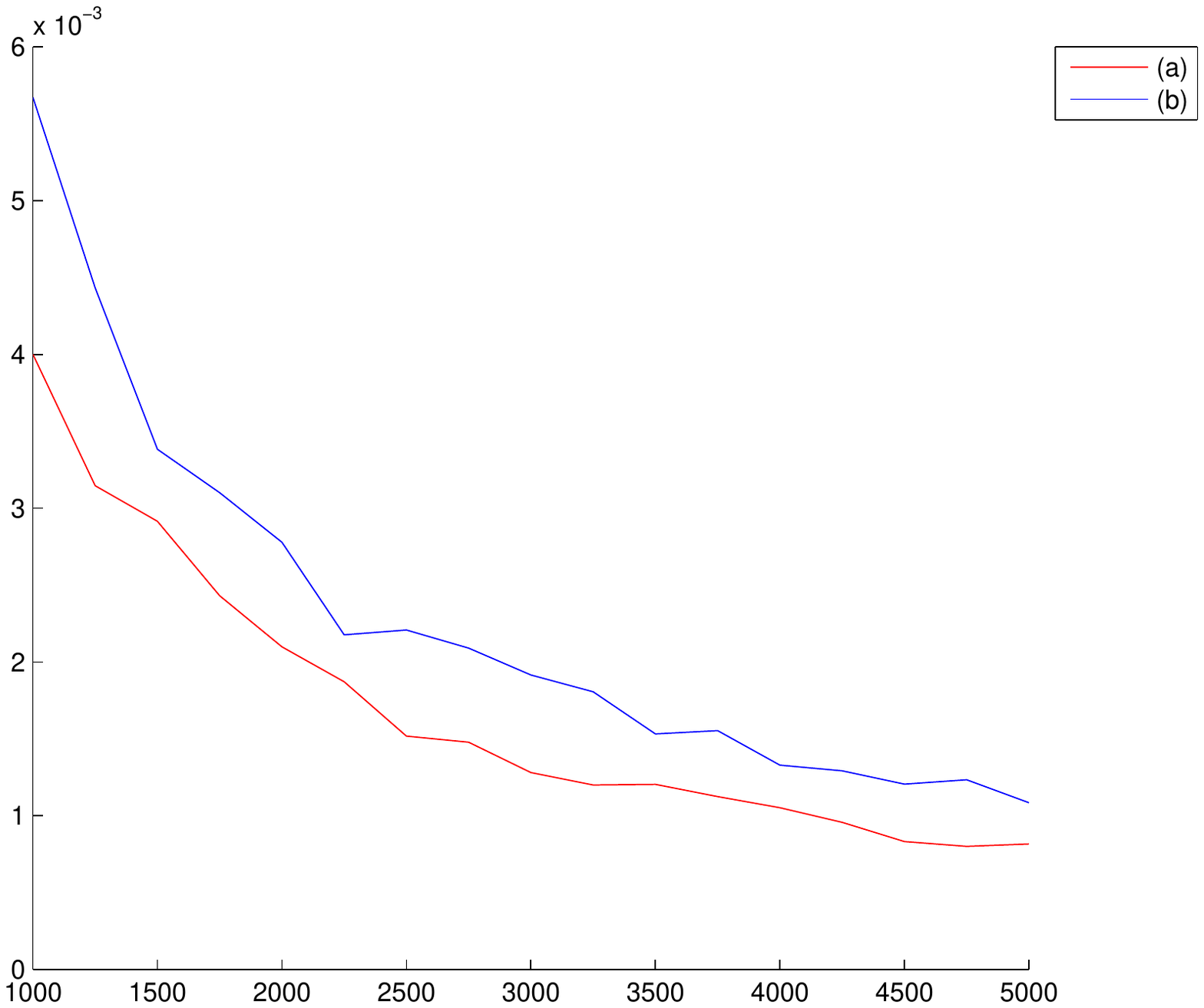}
\hspace{0.1cm}
\includegraphics[width=0.35\columnwidth]{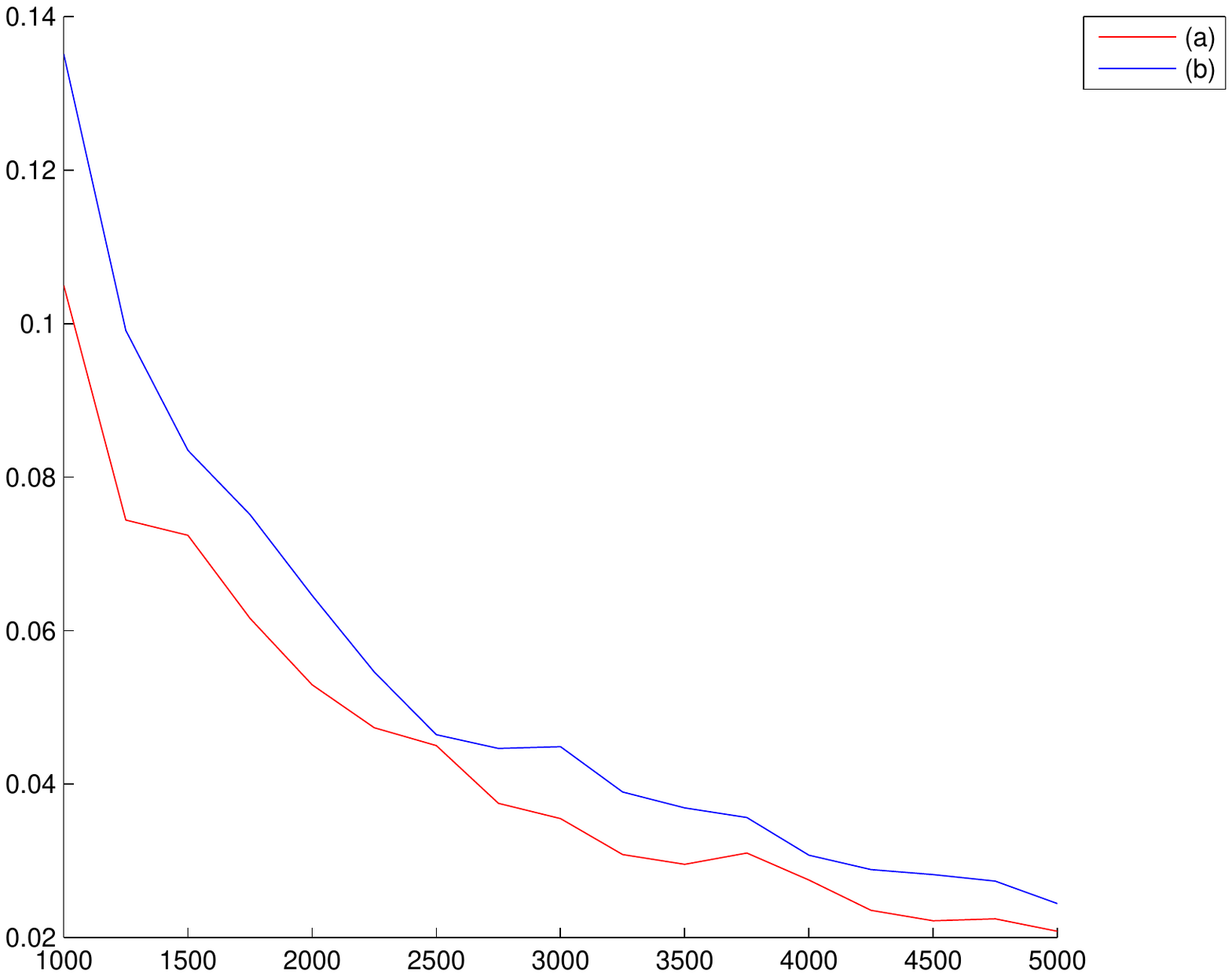}
\hspace{0.1cm}
\includegraphics[width=0.35\columnwidth]{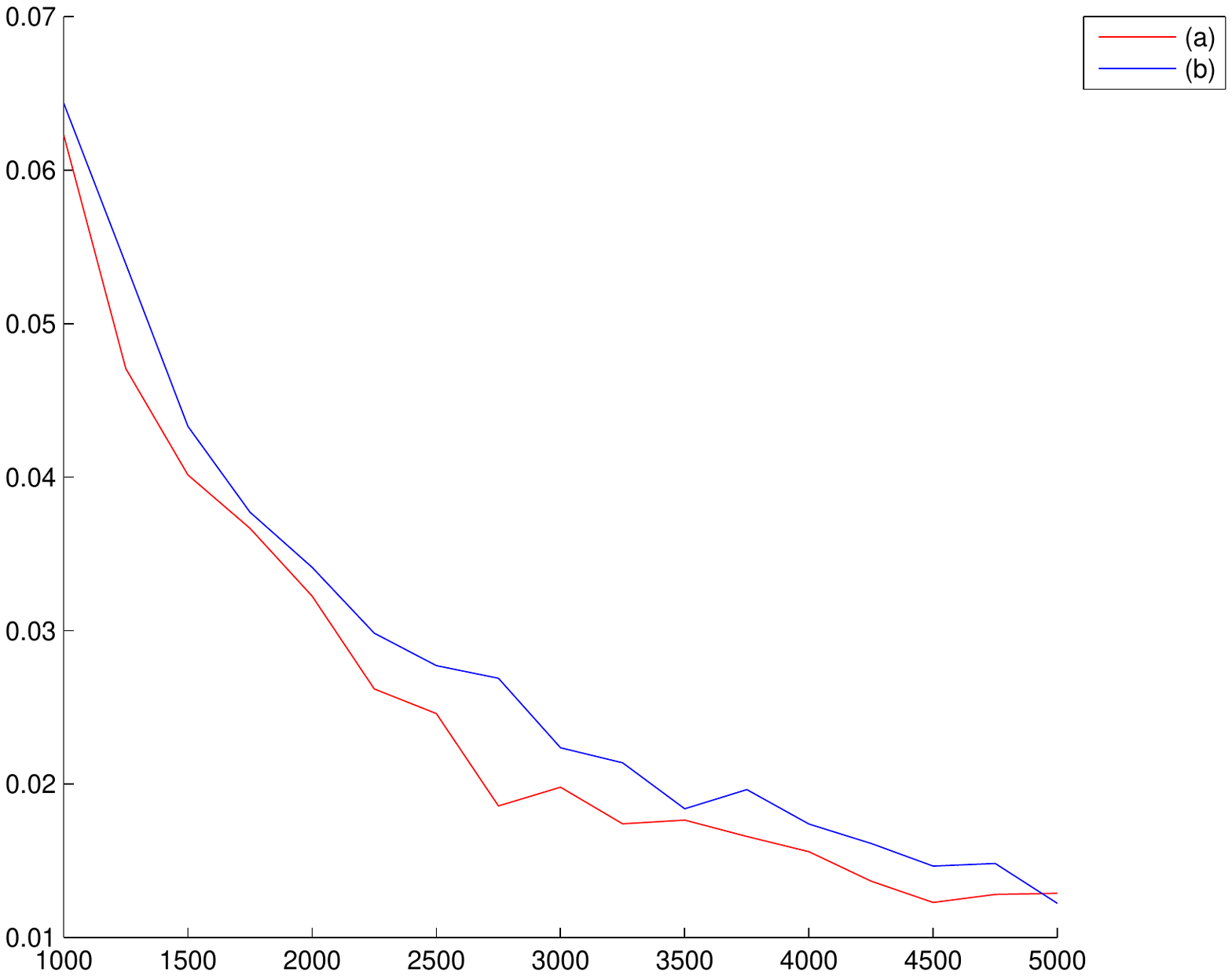}
\hspace{0.1cm}
\end{center}
\caption{Variance of  $\hat{B}_{T}$, $T \in [1000, 5000]$, with discretization step size 250, for cases: (a) $\alpha=0.25$, (b)  $\alpha=0.45$,case H1 (top-left), case H2 (top-right), case H3 (bottom-left) and case H4 (bottom-right).}\label{Fig.11}
\end{figure}

\begin{figure}[hptb]
\begin{center}
\includegraphics[width=0.35\columnwidth]{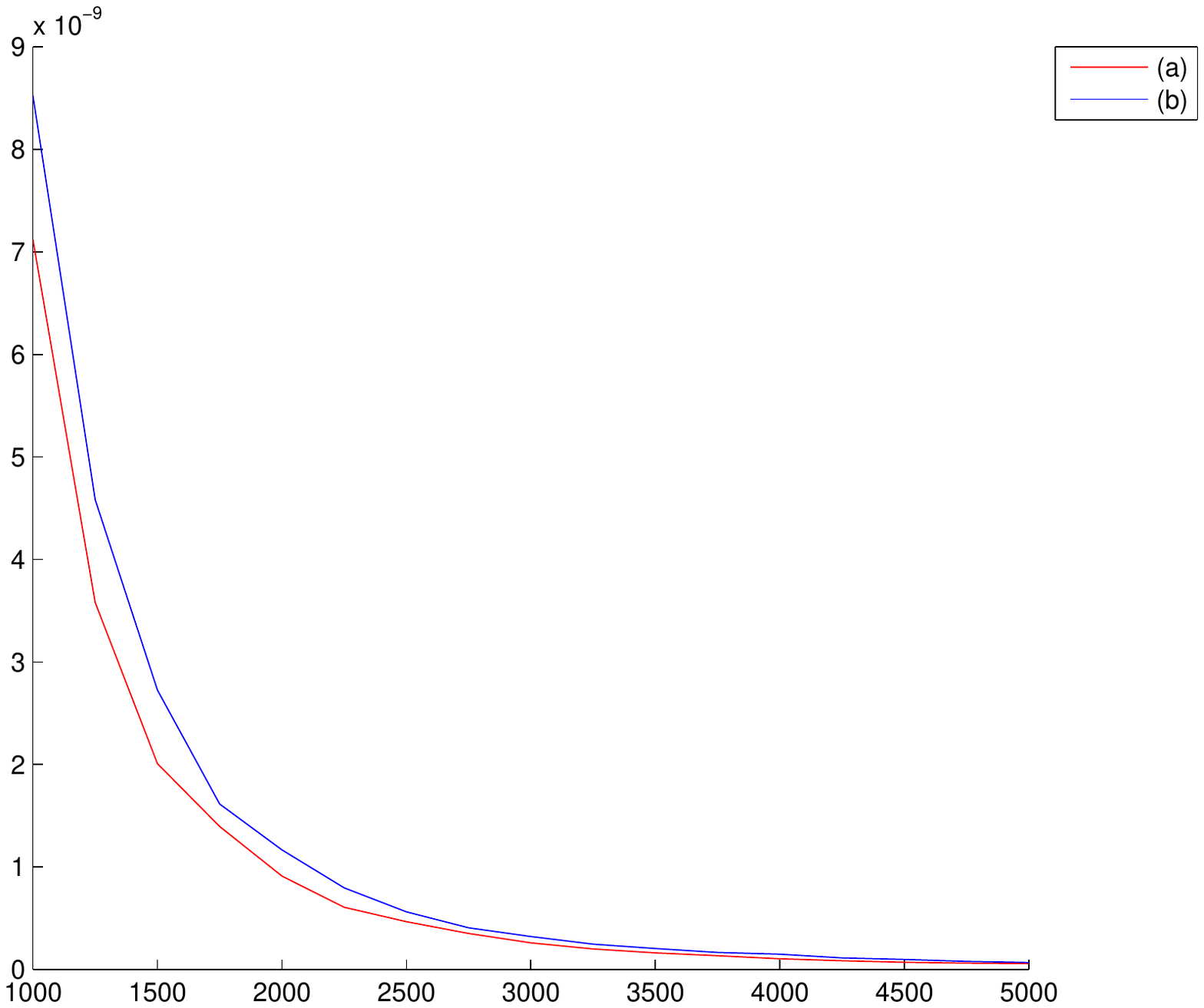}
\hspace{0.1cm}
\includegraphics[width=0.35\columnwidth]{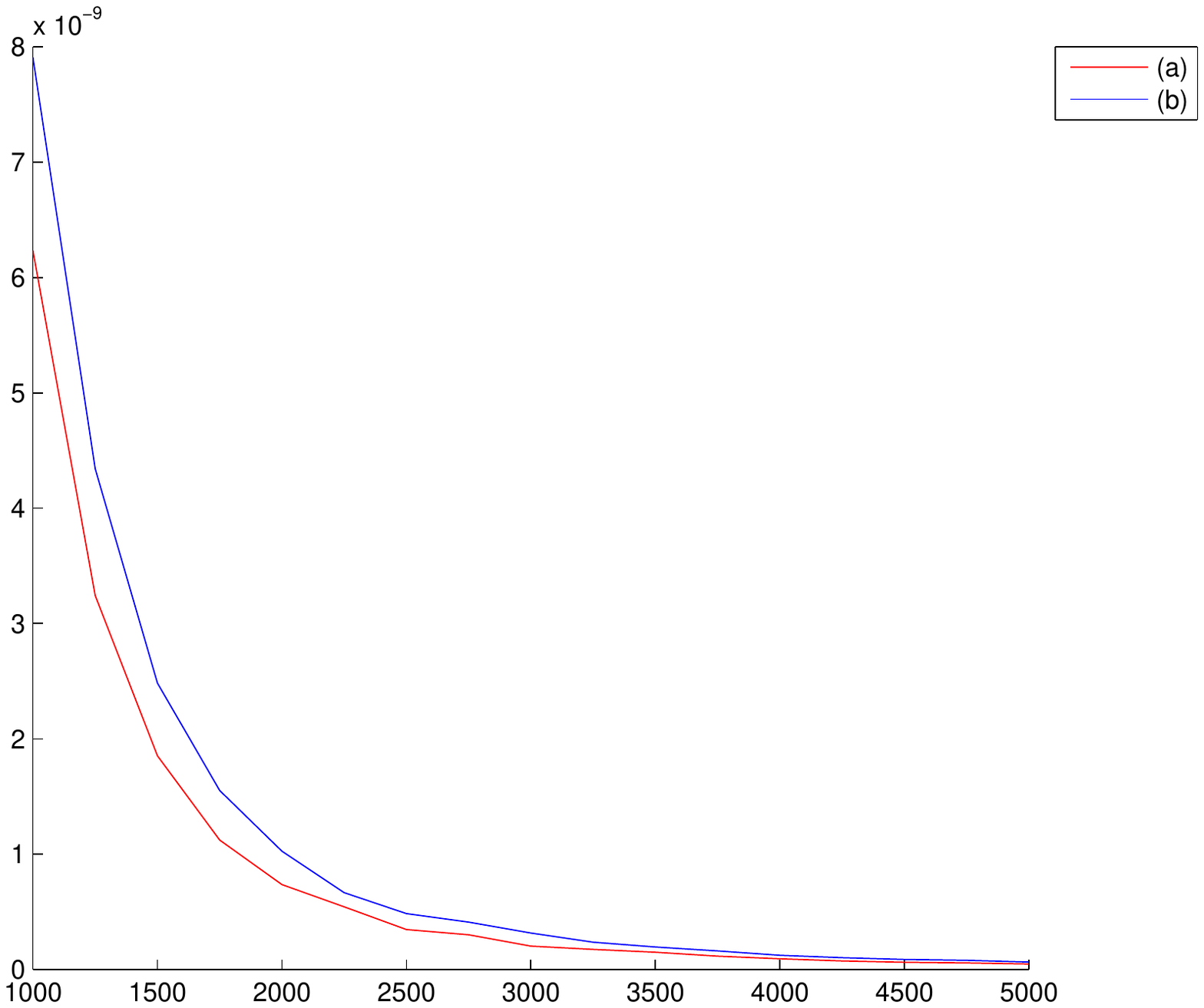}
\hspace{0.1cm}
\includegraphics[width=0.35\columnwidth]{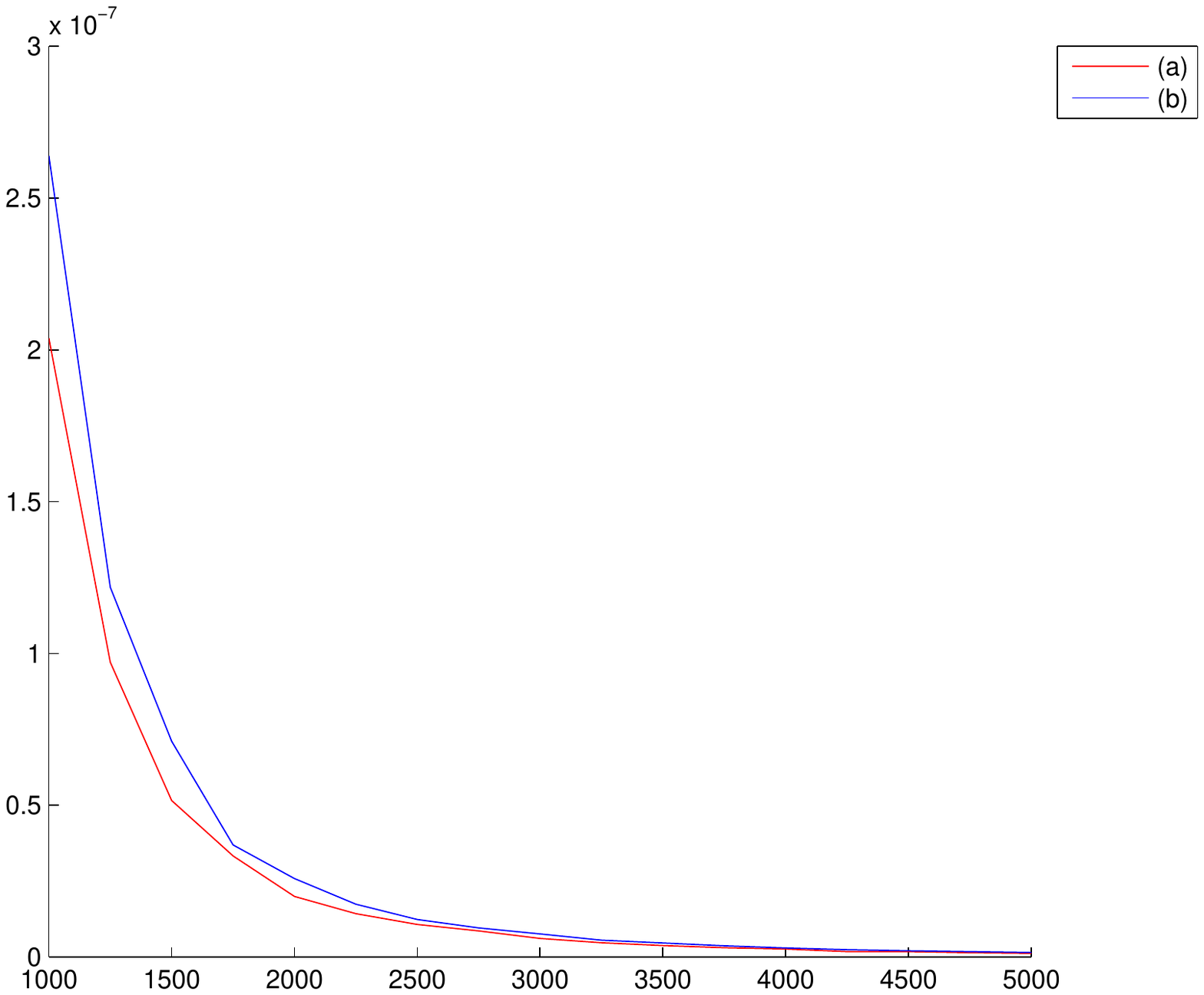}
\hspace{0.1cm}
\includegraphics[width=0.35\columnwidth]{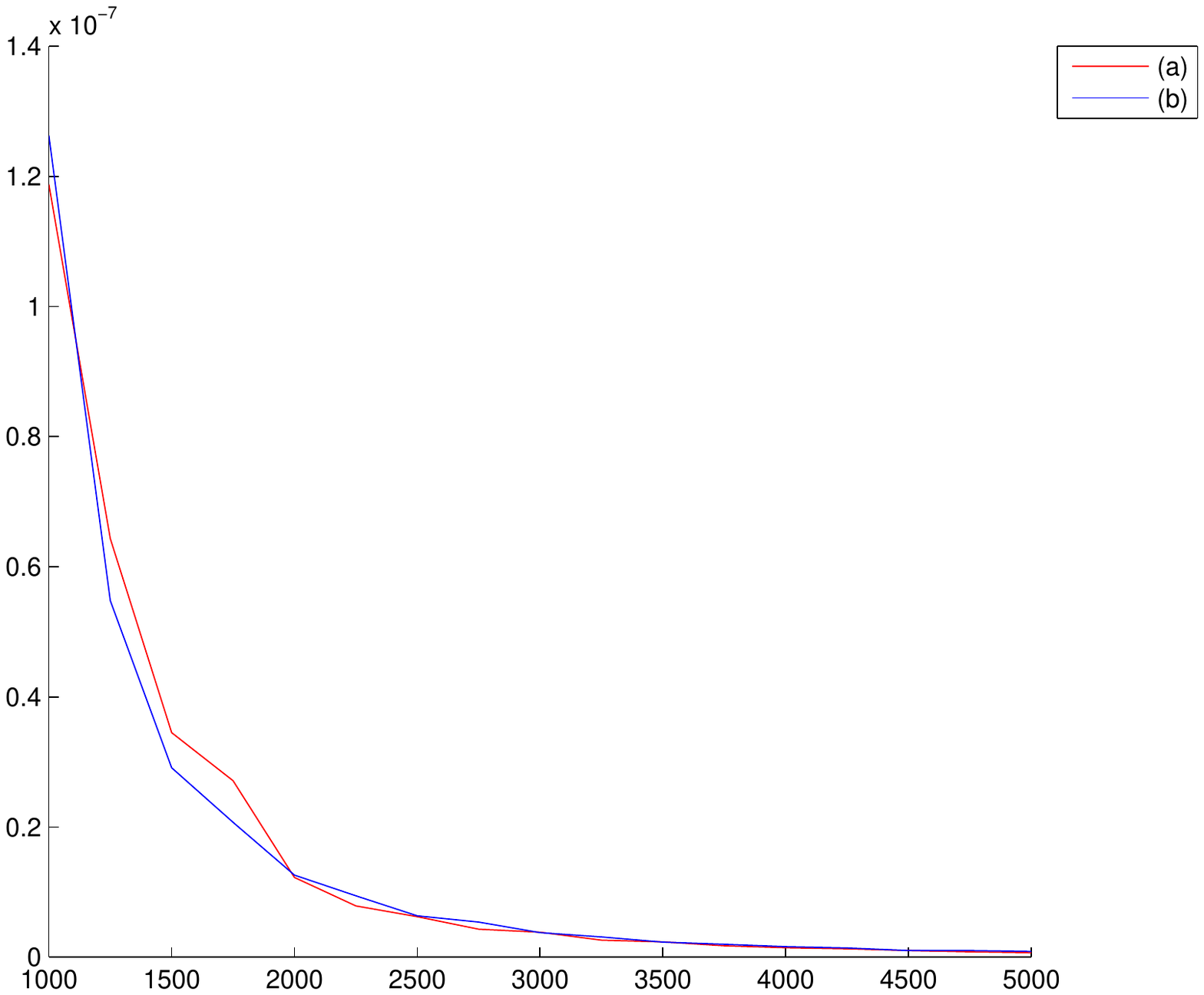}
\hspace{0.1cm}
\end{center}
\caption{Variance of $\hat{\varphi}_{T}$, $T \in [1000, 5000]$, with discretization step size 250, for cases: (a) $\alpha=0.25$, (b)  $\alpha=0.45$, case H1 (top-left), case H2 (top-right), case H3 (bottom-left) and case H4 (bottom-right).}\label{Fig.12}
\end{figure}

This subsection is aimed at proving asymptotic normality of $\theta$ in model (\ref{mod1}) under assumptions \textbf{A1}, \textbf{A2}, \textbf{A3}, \textbf{A4'} and \textbf{A5} by using simulation. We have considered model (\ref{mod1}) with $A=1$, $B=1$ and $\varphi=0.6$.
The generation of the random vectors $\{\xi(t),t=0,\ldots,T\},$  is performed from a MVN with zero mean vector, and covariance matrix (\ref{covA}) with the values of  $\varkappa=0.5,$ $\alpha=0.25,0.45.$ As in the previous section, we have tested MVN with different tools, Figure \ref{Fig.7} shows Chi-square Q-Q plot of ($\hat{\zeta}_{T}^{A}$, $\hat{\zeta}_{T}^{B},$ $\hat{\zeta}_{T}^{\varphi}$), $T=30000$, for different combinations of the selected parameter values. The 97 \% probability contours of a MVN and the scatter plot of ($\hat{\zeta}_{T}^{A}$, $\hat{\zeta}_{T}^{B},$ $\hat{\zeta}_{T}^{\varphi}$), $T=30000$, values are represented in Figure \ref{Fig.8}. Finally, rejection rates of Henze-Zirkler's and Doornik-Hansen MVN tests are calculated for 50 samples of size 1000 of the simulated random vectors ($\hat{\zeta}_{T}^{A}$, $\hat{\zeta}_{T}^{B}$ $\hat{\zeta}_{T}^{\varphi}$) for different $T$ values. The results are similar to the ones in the previous subsection. Under conditions \textbf{A1}, \textbf{A2}, \textbf{A3}, \textbf{A4'} and \textbf{A5}, normality and consistency (Figures \ref{Fig.10}-\ref{Fig.12}) of the sample estimator can be affirmed in most of the cases considered. For cases \textbf{H3} and \textbf{H4}  and low values of $\alpha$, $T$ must be increased to obtain a lower rejection ratio.

\begin{table}
\centering
\caption{Rejection rates of Henze-Zirkler's (T1) and Doornik-Hansen (T2) MVN tests applied to the simulated random vectors ($\hat{\zeta}_{T}^{A}$, $\hat{\zeta}_{T}^{B}$ $\hat{\zeta}_{T}^{\varphi}$) for different $T$ values. }
\label{Tabla.2}
\begin{tabular}{ccccccccc}%{\textwidth}{@{\extracolsep{\fill}}ccccccccc}
\hline
&&&\multicolumn{6}{c}{$T$} \\\cline{4-9}
$\alpha$&Case&Test &$1000$&$5000$& $10000$& $15000$&$20000$& $30000$ \\\hline

0.25&H1&T1&0.02&0.04&0.00&0.00&0.00&0.00\\
      &&T2&0.00&0.02&0.00&0.00&0.00&0.03\\
    &H2&T1&0.24&0.06&0.00&0.00&0.05&0.07\\
      &&T2&0.70&0.24&0.30&0.13&0.05&0.13\\
    &H3&T1&1.00&0.36&0.00&0.13&0.00&0.07\\
      &&T2&1.00&0.92&0.60&0.53&0.50&0.37\\
    &H4&T1&1.00&1.00&1.00&1.00&0.85&0.70\\
      &&T2&1.00&1.00&1.00&1.00&1.00&1.00\\\hline
0.45&H1&T1&0.02&0.02&0.00&0.00&0.00&0.00\\
      &&T2&0.00&0.02&0.00&0.13&0.00&0.00\\
    &H2&T1&0.06&0.04&0.00&0.00&0.00&0.00\\
      &&T2&0.06&0.04&0.00&0.00&0.00&0.00\\
    &H3&T1&0.36&0.02&0.00&0.00&0.00&0.03\\
      &&T2&0.78&0.12&0.00&0.00&0.10&0.03\\
    &H4&T1&1.00&0.82&0.20&0.27&0.10&0.03\\
      &&T2&1.00&1.00&0.80&0.60&0.30&0.17\\
 \hline
\end{tabular}
\end{table}

\clearpage

\section{Final Comments}

This paper studies the estimation of hidden periodicity in a nonlinear
regression model with stationary noise displaying cyclical dependence.
The problems of consistency, and Gaussian
limit distribution of the LSE, in the Walker sense, for
the harmonic regression model are addressed.  This kind of regression
constitutes an active research area, due to the existence of several open
problems and applications. In previous work, such as in \cite{Iv2014}, the
parameter range $\alpha m>1,$ $\alpha=\min_{j=0,\ldots,\kappa} \alpha_j,$ with $m$ is Hermite rank of $G$ (see, (\ref{3.1})), was considered. Here we have checked, by simulation experiments, that
 the Gaussian limit results hold for $0<\alpha <1/2$, $m=1$.

Specifically, the results proven in \cite{Iv2013,Iv2014} for the trigonometric regression function
(\ref{nonregr2}) for $\alpha>1/2$ have been confirmed by simulation.
Some experiments have been done to determine the validity of these results for $\alpha<1/2,$ under the assumption of non overlapping spectra. Consistency and asymptotic MVN of the LSE in the trigonometric regression has been verified. However, the convergence rate to the MVN  differs for each case included in the study.

\section*{Acknowledgements}

N.N. Leonenko and M.D. Ruiz-Medina partially supported by grant
of the European commission PIRSES-GA-2008-230804 (Marie Curie), projects
MTM2012-32674 of the DGI and the Australian Research Council grants
A10024117 and DP 0345577.

\end{document}